\documentclass[aps,prc,reprint,groupedaddress,nofootinbib]{revtex4-1}
\pdfoutput=1 
\usepackage{amsmath}
\usepackage{amssymb}
\usepackage{graphicx}
\usepackage{color}
\usepackage{soul}
\usepackage{float}
\usepackage{hyperref}
\usepackage{epsfig}
\usepackage{epstopdf}

\usepackage{natbib}

\newcommand{\beq}{\begin{equation}}
\newcommand{\eeq}{\end{equation}}
\newcommand{\bea}{\begin{eqnarray}}
\newcommand{\eea}{\end{eqnarray}}

\begin{document}
\title{Far-from-equilibrium search for the QCD critical point}

\author{Travis Dore}
\affiliation{Illinois Center for Advanced Studies of the Universe, Department of Physics, University of Illinois at Urbana-Champaign, Urbana, IL 61801, USA}
\author{Emma McLaughlin}
\affiliation{Department of Physics, Columbia University, 538 West 120th Street, New York, NY 10027, USA}
\author{Jacquelyn Noronha-Hostler}
\affiliation{Illinois Center for Advanced Studies of the Universe, Department of Physics, University of Illinois at Urbana-Champaign, Urbana, IL 61801, USA}

\date{\today}
\
\begin{abstract}
 Initial conditions for relativistic heavy-ion collisions may be far from equilibrium (i.e. there are large initial contributions from the shear stress tensor and bulk pressure) but it is expected that on  very short time scales the dynamics converge to a universal attractor that defines hydrodynamic behavior.  Thus far, studies of this nature have only considered an idealized situation at LHC energies (high temperatures $T$  and vanishing baryon chemical potential $\mu_B=0$) but, in this work, we investigate for the first time how far-from-equilibrium effects may influence experimentally driven searches for the Quantum Chromodynamic critical point at RHIC. We find that the path to the critical point is heavily influenced by far from equilibrium initial conditions where viscous effects lead to  dramatically different  $\left\{T,\mu_B\right\}$ trajectories through the QCD phase diagram. We compare hydrodynamic equations of motion with shear and bulk coupled together at finite $\mu_B$ for both DNMR and phenomenological Israel-Stewart equations of motion and discuss their influence on potential attractors at finite $\mu_B$ and their corresponding $\left\{T,\mu_B\right\}$ trajectories. 
\end{abstract}

\maketitle

\section{Introduction}

One of the major thrusts of the nuclear physics community is to map out the phase diagram of Quantum Chromodynamics (QCD) - specifically, the transition between a hadron gas and a deconfined state of matter composed of strongly interacting quarks and gluons, known as the Quark-Gluon Plasma (QGP).  While it is known from first principle Lattice QCD calculations \cite{Aoki:2006we,Borsanyi:2010bp,Bazavov:2011nk} that a cross-over phase transition existed in the early universe and in high-energy nuclear collisions, only conjectures and effective models provide indications that a real phase transition (i.e. first or second order) may exist at large baryon chemical potentials \cite{Halasz:1998qr,Stephanov:1998dy,Stephanov:1999zu,Dexheimer:2009hi,Critelli:2017oub,Fan:2016ovc,Fu:2019hdw,Motornenko:2019arp,Annala:2019puf,Tan:2020ics}. This phase transition would be separated from the cross-over by a critical point.  Nuclear physicists are searching for evidence of such a critical point in heavy-ion collisions and astrophysicists are searching at much lower temperatures and larger baryon densities for evidence of phase transitions in neutron star mergers \cite{Bedaque:2014sqa,Alford:2013aca,Dexheimer:2014pea,Benic:2014jia,Montana:2018bkb,Most:2018eaw,Tan:2020ics,Zha:2020gjw}. 

The crucial observable to search for the QCD critical point is the study of susceptibilities of baryon number (i.e net-proton fluctuations, which are measured by STAR \cite{Adam:2020unf} and HADES \cite{Adamczewski-Musch:2020slf}) because higher-order susceptibilities are increasingly sensitive to the correlation length and are, thus, expected to diverge at the critical point  \cite{Stephanov:2011pb}.  However, direct comparisons to experimental data are complicated because of finite volume, lifetime, size effects, and acceptance cuts \cite{Bzdak:2012ab,Bzdak:2012an,Bzdak:2013pha,Hippert:2015rwa,Steinheimer:2016cir,Bluhm:2016byc,Sombun:2017bxi,Hippert:2017xoj,Nouhou:2019nhe}.  Therefore, the best tool to search for the QCD critical point would be event-by-event relativistic viscous hydrodynamics that includes three conserved charges: baryon number, strangeness, and electric charge and that also incorporates stochastic fluctuations at the critical point. In such a fully dynamical framework, one could take into account all acceptance cuts and finite volume/size/lifetime effects. While a large number of theoretical efforts are underway to create such a model \cite{Auvinen:2013sba,Steinheimer:2014pfa,Monnai:2016kud,Auvinen:2017fjw,Stephanov:2017ghc,Shen:2017bsr,Nahrgang:2018afz,Akamatsu:2018olk,Kanakubo:2019ogh,Du:2019obx,Fotakis:2019nbq,Martinez:2019rlp,Martinez:2019jbu,Moreau:2019vhw,Soloveva:2019xph,An:2019csj,Bluhm:2020mpc}, no such framework is currently at one's disposal. Many of the needed advancements are outlined in \cite{Rao:2019vgy}.  In fact, even very simplistic studies of this baryon dense region are still in their infancy and have not gone through the same rigorous studies to constrain initial conditions \cite{Giacalone:2017uqx}, the equation of state \cite{Pratt:2015zsa,Moreland:2015dvc,Alba:2017hhe,Auvinen:2018uej}, and transport coefficients \cite{Noronha-Hostler:2015qmd,Niemi:2015qia,Bernhard:2019bmu} that have already been performed at the $\mu_B\sim 0$ region of the QCD phase diagram.  

Already at $\mu_B\sim 0$, a large amount of uncertainty remains when describing the initial state shortly after two heavy-ions collide and only more recently have theorists \cite{Heller:2015dha,Heller:2015dha,Buchel:2016cbj,Heller:2016rtz,Spalinski:2017mel,Romatschke:2017acs,Romatschke:2017vte,Behtash:2017wqg,Strickland:2017kux,Denicol:2017lxn,Blaizot:2017ucy,Casalderrey-Solana:2017zyh,Florkowski:2017olj,Heller:2018qvh,Rougemont:2018ivt,Denicol:2018pak,Almaalol:2018ynx,Casalderrey-Solana:2018uag,Behtash:2018moe,Behtash:2019txb,Strickland:2018ayk,Kurkela:2018wud,Strickland:2019hff,Kurkela:2019set,Jaiswal:2019cju,Denicol:2019lio,Brewer:2019oha,Almaalol:2020rnu,Berges:2020fwq,Bemfica:2020xym,NunesdaSilva:2020bfs} begun to systematically study the effects of far-from-equilibrium behavior at $\mu_B\sim 0$.  Looking towards the baryon dense region, there has not yet been a single study of the influence of a far-from-equilibrium initial state on the search for the critical point and, in fact, most hydrodynamical models have assumed only ideal hydrodynamic equations of motion \cite{Aguiar:2007zz,Petersen:2008dd,Steinheimer:2009nn,Steinheimer:2011ea} with just a handful of models that incorporate viscosity or diffusion in the last couple of years \cite{Karpenko:2013wva,Rougemont:2015ona,Feng:2018anl,Du:2019obx,Denicol:2018wdp,Batyuk:2017sku,Fotakis:2019nbq}.  In the baryon dense region, we are unaware of any initial conditions that include an initialized shear stress tensor or bulk pressure (although this may be possible using SMASH \cite{Weil:2016zrk}, URQMD \cite{Steinheimer:2007iy}, or NEXUS \cite{Werner:1993uh} but these are currently  coupled to ideal hydrodynamic models). Therefore, there is no real understanding of how far-from-equilibrium initial conditions would influence the ability of different beam energies to approach the QCD critical point (musings that it may affect the search for the critical point can be found in \cite{Romatschke:2016hle}).

While astrophysical searches for a first order phase transition also utilize relativistic hydrodynamics \cite{Most:2018eaw,Most:2019onn,Zha:2020gjw} (in this context coupled to general relativity), the current models do not incorporate shear and bulk viscosities, nor do they have diffusion currents due to conserved charges (such as baryon number).  Current efforts are underway to incorporate bulk viscosity into such models \cite{Haensel:2000vz,Alford:2010gw,Alford:2017rxf,Alford:2019qtm,Bemfica:2019cop,Bemfica:2017wps}. We note that if the initial contribution from bulk viscosity is large immediately after the two neutron stars collided, similar issues when determining their trajectories through the QCD phase diagram will arise. 

Typically, in most studies of large baryon density effects in heavy-ion collisions there is an underlying assumption that the QGP is a nearly perfect fluid so that there is almost no entropy production.  If one assumes that entropy is not produced at all, then the ratio of total entropy to baryon number ($S/N_B$) is fixed throughout the collision, the subsequent expansion, and cooling throughout the phase diagram\footnote{This is the same underlying assumption made when using partial chemical equilibrium for hadronic decays, e.g. \cite{Bebie:1991ij}.}.  These trajectories, known as isentropes, have been studied in a number of recent papers \cite{Gunther:2017sxn,Bellwied:2018tkc,Parotto:2018pwx,Noronha-Hostler:2019ayj,Monnai:2019hkn,Stafford:2019yuy} and are used extensively to understand equilibrium properties of QCD at large baryon densities.   However, since hydrodynamic models require both shear and bulk viscosity to reproduce experimental data, entropy production must occur and deviations from the isentropic trajectories are expected.  This may be exacerbated at large $\mu_B$ since a number of studies have suggested the viscosity could increase in this region \cite{Demir:2008tr,Denicol:2013nua,Kadam:2014cua,Monnai:2016kud,Auvinen:2017fjw,Martinez:2019bsn}.  Thus, large deviations from isentropic trajectories may be possible, especially at the critical point.  

In this paper, we perform the first study of the effects of far-from-equilibrium initial conditions (arising from a fully initialized shear stress tensor and bulk pressure) on the search for the QCD critical point. We find large deviations from isentropic trajectories, especially near the critical point. We show this in both Israel-Stewart and DNMR hydrodynamic equations of motion and elaborate on difficulties in ensuring positive entropy production throughout the evolution.  Furthermore, we find that Israel-Stewart and DNMR do not traverse the QCD phase diagram in the same manner, which implies that the specific way such approaches implement second order corrections matters for the evolution of the baryon rich fluid. This means that special attention must be paid when selecting the hydrodynamic equations of motion in the presence of phase transitions. The sensitivity to the initial conditions indicates that they play a crucial role in determining the trajectory of the QGP through the QCD phase diagram and significant efforts must be made to constrain initial conditions before a fully dynamical model can properly describe heavy-ion collisions at finite baryon densities. Furthermore, if there are significant event-by-event fluctuations in the initial conditions of shear and bulk, certain events may pass through the critical points while others can miss it entirely (even starting from the same initial energy density and baryon density). 

This paper is organized as follows. In Sec.\ \ref{sec:eqs} we outline our hydrodynamical model, the transport coefficients, and equation of state. In Sec.\ \ref{sec:TmuB} we calculate the $\left\{T,\mu_B\right\}$ trajectories across the QCD phase diagram for fixed $\rho$ and $\varepsilon$ in Sec.\ \ref{sec:TmuBerho} and fixed freeze-out point in Sec.\ \ref{sec:TmuBFO}.  The indirect effects of the critical point on shear viscosity are shown in Sec.\ \ref{sec:shear}.  Then, the potential existence of  attractors for shear and bulk channels are discussed in Sec.\ \ref{sec:att}. Sec.\ \ref{sec:CP} discusses the influence of a critically scaled bulk viscosity in our results. Our conclusions in Sec.\ \ref{sec:conclusions} explore the consequences of our calculations on the search for the QCD critical point (potential consequences to neutron star mergers are also discussed).  In Appendix \ref{sec:dbetas} we study the influence of the $\dot{\beta}$ terms present in   Israel-Stewart theory to the evolution of the fluid at large baryon chemical potentials.

\section{Hydrodynamical Setup}\label{sec:eqs}

In the past years, a significant effort has been made to incorporate at least one conserved charge (baryon density) and more recently two (strangeness) in event-by-event relativistic viscous hydrodynamics codes
\cite{Du:2019obx,Denicol:2018wdp,Batyuk:2017sku,Fotakis:2019nbq}.  Additionally, transport coefficients can also depend on $\left\{T,\mu_B\right\}$ \cite{Demir:2008tr,Denicol:2013nua,Kadam:2014cua,Rougemont:2015ona,Monnai:2016kud,Rougemont:2017tlu,Rougemont:2017tlu,Auvinen:2017fjw,Martinez:2019bsn} and they are also sensitive to the presence of critical fluctuations \cite{Son:2004iv}, which should  influence final state observables. In the following we only consider the effects from one conserved charge (baryon number) but we point out that a more realistic description of trajectories on the QCD phase diagram would require effects from the conservation of baryon number, strangeness, and electric charge, which would severely complicate the type of analysis done here.

In this first study of how the viscous fluid traverses the QCD phase diagram we use a simplistic, highly symmetric Bjorken flow \cite{Bjorken:1982qr} picture where the hydrodynamic equations of motion are greatly simplified \cite{Muronga:2003ta}. We use two different formulations of relativistic viscous hydrodynamics, DNMR \cite{Denicol:2012cn} and Israel-Stewart \cite{Israel:1979wp}, in order to determine how assumptions regarding the derivation of the equations of motion, and their choices of second-order transport coefficients, affect the evolution of the baryon rich viscous fluid. 

Both DNMR and Israel-Stewart are based on the idea that the dissipative currents, such as the shear-stress tensor $\pi^{\mu\nu}$ and bulk scalar $\Pi$, evolve according to relaxation equations that describe how such quantities deviate from their relativistic Navier-Stokes values. Using hyperbolic coordinates with the metric $g_{\mu\nu} = \text{diag}(1,-1,-1,-\tau^2)$, the underlying symmetries of Bjorken flow imply that all dynamical quantities depend only on the proper time $\tau = \sqrt{t^2-z^2}$. Furthermore, in Bjorken flow the state of the fluid is described by only 4 dynamical variables: the proper energy density $\varepsilon(\tau)$, the baryon number density $\rho(\tau)$, $\Pi(\tau)$, and $\pi^\eta_\eta(\tau)$ (where $\eta$ here stands for the spacetime rapidity). For   DNMR the equations of motion in Bjorken flow become \cite{Denicol:2012cn,Bazow:2016yra} \begin{eqnarray}
    \dot{\epsilon}&=&-\frac{1}{\tau}\left[\epsilon+p+\Pi-\pi^{\eta}_{\eta}\right]\\
      \tau_{\pi}\dot{\pi}^{\eta}_{\eta}+\pi^{\eta}_{\eta}&=&\frac{1}{\tau}\left[\frac{4\eta}{3} - \pi^\eta_\eta\left(\delta_{\pi\pi} + \tau_{\pi\pi} \right) + \lambda_{\pi\Pi}\Pi \right]\\
 \tau_{\Pi}\dot{\Pi}+\Pi&=&-\frac{1}{\tau}\left(\zeta + \delta_{\Pi\Pi}\Pi+\frac{2}{3}\lambda_{\Pi\pi}\pi^{\eta}_{\eta}\right)\\
    \dot \rho &=& -\frac{\rho}{\tau}\label{eqn:rhoBevo}
\end{eqnarray}
where $\dot\epsilon=d\epsilon/d\tau$, $p$ is the equilibrium pressure defined by the equation of state, $\zeta$ is the bulk viscosity, and the remaining second order transport coefficients are taken from \cite{Denicol:2014vaa}. We note that in Bjorken flow the particle diffusion contribution vanishes and, thus, the baryon density equation can be readily solved to give $\rho(\tau)=\rho_0 (\tau_0/\tau)$, where $\rho_0$ and $\tau_0$ are the initial baryon density and time, respectively.

We make the point of including second order transport coefficients terms that couple the shear and bulk contributions (e.g. $\lambda_{\pi\Pi}$ and $\lambda_{\Pi\pi}$) since there should be a nontrivial coupling between the two \cite{Denicol:2014mca}. The transport coefficients for DNMR used in this paper are defined as follows:
\begin{eqnarray}
    \tau_\pi &=& \frac{5\ \eta}{\epsilon + p}\\
   \tau_{\Pi}&=&\frac{\zeta}{15(\epsilon+p)\left(\frac{1}{3}-c_s^2\right)^2}\label{eqn:tauPI}\\
    \lambda_{\pi \Pi} &=& \frac{6}{5} \tau_\pi\\
    \delta_{\pi\pi} &=& \frac{4}{3}\tau_\pi\\
    \tau_{\pi\pi} &=& \frac{10}{7}\tau_\pi\\
    \lambda_{\Pi \pi} &=& \frac{8}{5}\left(\frac{1}{3} - c_s^2\right)\\
    \delta_{\Pi \Pi} &=& \frac{2}{3}.
\end{eqnarray}
where the speed of sound squared is $c_s^2 = dp/d\epsilon$ (computed at constant entropy). Given $\eta/(\epsilon+p)$ and $\zeta/(\epsilon+p)$ as functions of $T$ and $\mu_B$, all the second order transport coefficients (such as the bulk and shear relaxation times, $\tau_\Pi$ and $\tau_\pi$, respectively) can be readily obtained. For the Israel-Stewart case, the energy density and baryon density evolution remain the same (as they stem from the conservation laws) while the relaxation equations for shear-stress and bulk viscous pressure evolution are given by
 \begin{eqnarray}
    \tau_\pi \dot\pi^\eta_\eta + \pi^\eta_\eta &=& \frac{4\eta}{3\tau} - \frac{\eta T \pi^\eta_\eta}{2} \left( \frac{\beta_\pi}{\tau} + \dot \beta_\pi \right)\label{eqn:ISsh}\\
    \tau_\Pi \dot\Pi + \Pi &=& -\frac{\zeta}{\tau} - \frac{\zeta T \Pi}{2} \left(\frac{\beta_\Pi}{\tau}  + \dot \beta_\Pi\right)\label{eqn:ISbu}
\end{eqnarray}
where we defined
\begin{eqnarray}
    \beta_\pi &=& \frac{\tau_\pi}{2 \eta T}\\
    \beta_\Pi &=& \frac{\tau_\Pi}{\zeta T}.
\end{eqnarray}
 When the Israel-Stewart equations were first derived in \cite{Israel:1979wp}, the terms in Eq.\ (\ref{eqn:ISsh}) and Eq.\ (\ref{eqn:ISbu}) that contain $\dot \beta_\pi$ and $\dot \beta_\Pi$ were left off, since these derivatives were presumed to be small on the scales they were interested. This is certainly not true in heavy-ions where early in the expansion these terms can be quite large. Thus, to gauge the importance of these terms and also the possibility of needing to include higher order terms in the power counting scheme of DNMR \cite{Denicol:2012cn}, we will also make comparisons with and without including the $\dot\beta$ terms. This comparison is shown in Appendix \ref{sec:dbetas}.  However, for the rest of the main text we will only show results comparing DNMR and Israel-Stewart including the $\dot \beta$ terms because they play an important role in the system's evolution.
 
\begin{figure}
    \centering
    \includegraphics[width =\linewidth]{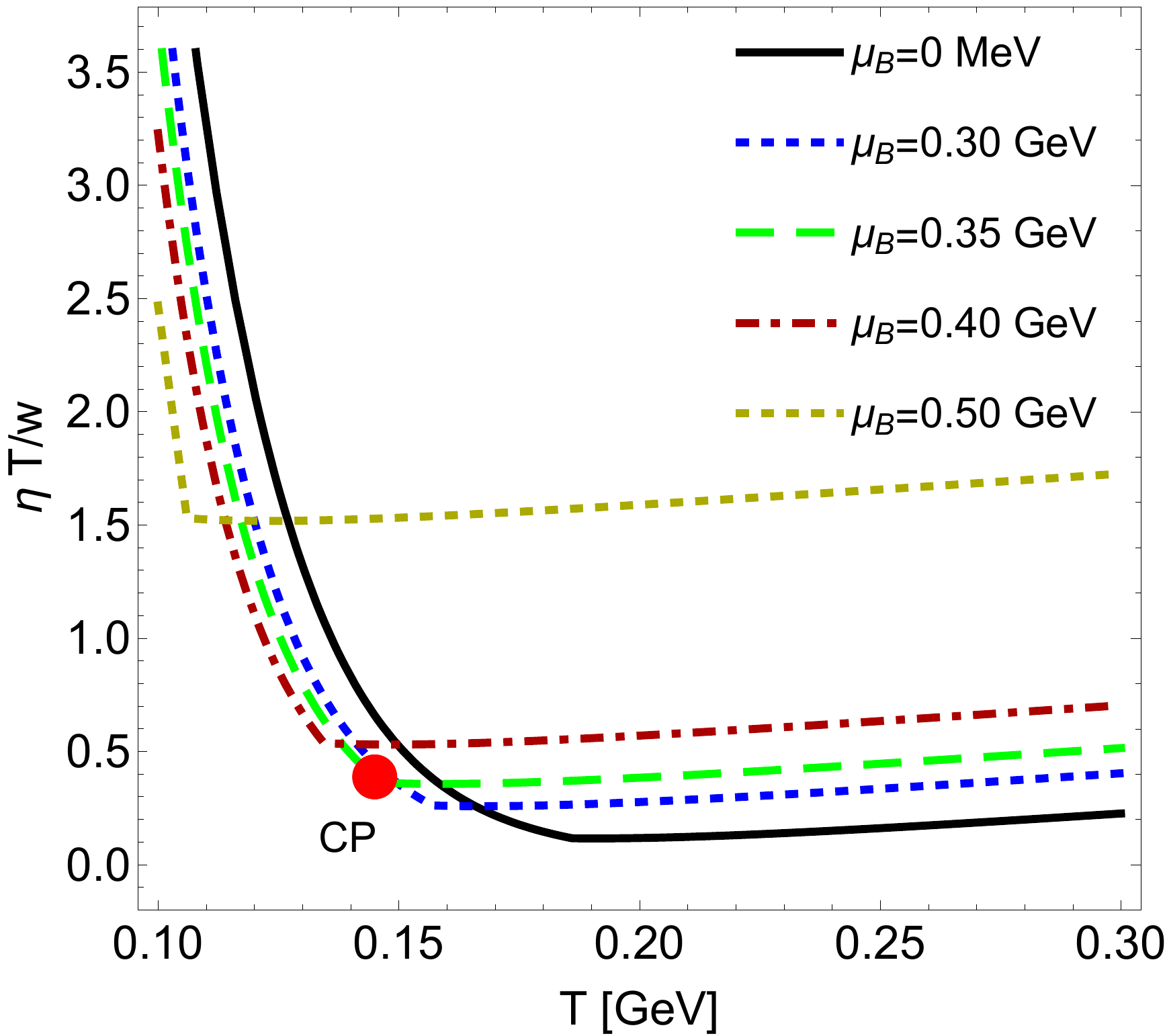}
    \caption{(Color online) Our phenomenological $\eta T/w(T,\mu_B)$ across the phase diagram. The critical point is shown in red (note, no critical scaling was included in the shear viscosity).}
    \label{fig:etaTw}
\end{figure}

 The shear viscosity used in this paper was derived from an excluded hadron resonance gas model similarly to what was done in \cite{NoronhaHostler:2012ug}. Then, this hadronic shear viscosity was coupled to a simplistic parameterized QGP phase (based on the parameterization in \cite{Christiansen:2014ypa,Dubla:2018czx}) and was matched at $T\sim0.195$ GeV at $\mu_B=0$, similar to \cite{Christiansen:2014ypa,Rougemont:2017tlu,Dubla:2018czx}). The finite $\mu_B$ behavior is determined by the change in $\eta T/w$ (where $w=\epsilon+p$ is the enthalpy) in the excluded volume hadron resonance gas model and the switching temperatures between the hadron resonance gas where the QGP phase is adjusted to match the critical point at finite $\mu_B$. The variation of $\eta T/w(T,\mu_B)$ is shown in Fig.\ \ref{fig:etaTw} for various values of $\mu_B$. Note that no critical behavior is incorporated in the shear viscosity.  Rather, the $\mu_B$ dependence is driven entirely by the matching to the hadron resonance gas at lower and lower values of the temperature with increasing $\mu_B$. Generally, lower temperatures lead to a large shear viscosity and, therefore, this quantity increases with increasing $\mu_B$.  A forthcoming paper will appear shortly about this work with further details.

 For the bulk viscosity, two different parameterizations were used, both of which are scaled up from one that is in the same ballpark as the $\zeta/s$ extracted from Bayesian analysis \cite{Bernhard:2016tnd,Bernhard:2019bmu} that is also consistent with that from holographic models \cite{Finazzo:2014cna,Rougemont:2017tlu} and quasi-particle models \cite{Alqahtani:2017mhy,Almaalol:2018gjh}. This base parameterization of the bulk viscosity is given by
 \begin{eqnarray}\label{eqn:zetanorm}
     \frac{\zeta T}{w} =36\times \frac{1/3 - c_s^2}{8\pi}
 \end{eqnarray}
 where the factor of 36 is included to obtain a maximum $\frac{\zeta T}{w}\sim 0.2$ similar to the maximum value employed in certain hydrodynamic simulations \cite{Ryu:2015vwa}.
 Given that this quantity depends on $c_s^2$, there is at least some sensitivity to the critical point (since the critical point has a vanishing $c_s^2$).

 As previously mentioned, in the Bjorken picture the baryon density evolution is trivial, as seen in Eq.\ (\ref{eqn:rhoBevo}).  This is because the baryon diffusion can only be included in less symmetrical evolution dynamics, which we will consider in a future work. However, the non-trivial time evolution of the energy density due to viscous effects as well as the non-trivial mapping of $\{\epsilon,\rho\}\rightarrow\left\{T,\mu_B\right\}$ due to the equation of state lead to  unique trajectories in the QCD phase diagram. These trajectories are necessarily off of the isentropes calculated in equilibrium, such as those from Lattice QCD, and should be associated with some amount of entropy production.
  
To close the hydrodynamic equations of motion we use the Lattice QCD-based equation of state (EOS) from \cite{Parotto:2018pwx} that is coupled to a parameterized 3D Ising model.  This equation of state allows us to test the influence of a critical point on the $T-\mu_B$ trajectories.  Since we do not, in fact, know the location (or even the existence) of the QCD critical point, the results are simply to test the qualitative influence of the critical point.  Thus, we only consider one readily available parameterization of the EOS from  \cite{Parotto:2018pwx} where the critical point is located at $\{T,\mu_B\} = \{143,350\}$ MeV. In this EOS the critical point always lies on the chiral phase transition line, which is currently known up to $\mathcal{O}(\mu_B^2)$:
\begin{equation}\label{eq:trline}
T = T_0 + \kappa_2 \, T_0 \left( \frac{\mu_B}{T_0} \right)^2 + {\cal O} (\mu_B^4),
\end{equation}
where we use $T_0=0.155$ GeV and the central value of $\kappa_2= 0.0153$ from \cite{Borsanyi:2020fev}.

At this point in time we do not have the necessary framework to include critical fluctuations (see  \cite{Stephanov:1999zu,Jiang:2015hri,Mukherjee:2016kyu,Stephanov:2017ghc,Nahrgang:2018afz,Akamatsu:2018vjr,An:2019csj}).  The only contribution from criticality arises in the EOS itself and the influence on the parameterized bulk viscosities because of either a sharp dip in the speed of sound at the critical point or large increase in bulk viscosity due to the critical scaling.

One final remark on the limitations on the EOS derived in \cite{Parotto:2018pwx} is in order.  Because the 3D Ising model is coupled to the Lattice QCD reconstructed EOS up to $\mathcal{O}(\mu_B^4)$, the absolute maximum that we can reasonably extend the EOS out to in $\mu_B$ along the phase transition is $\mu_B\sim 450$ MeV. Beyond this point, pathologies begin to appear in the EOS.  At high temperatures we have a slightly higher reach and we can extend the phase diagram out to $\mu_B\sim 600$ MeV.  However, because a number of trajectories that pass through the critical point begin at relatively low temperatures but high $\mu_B$ (and the time evolution is nearly flat in $T$), we are limited in the phase space that we can explore our initial conditions.  This is especially problematic for the Israel-Stewart equations of motion, which appear to prefer these type of trajectories.

\section{$T-\mu_B$ trajectories across the QCD phase diagram}\label{sec:TmuB}

\begin{figure}
    \centering
    \includegraphics[width =\linewidth, height = 4cm]{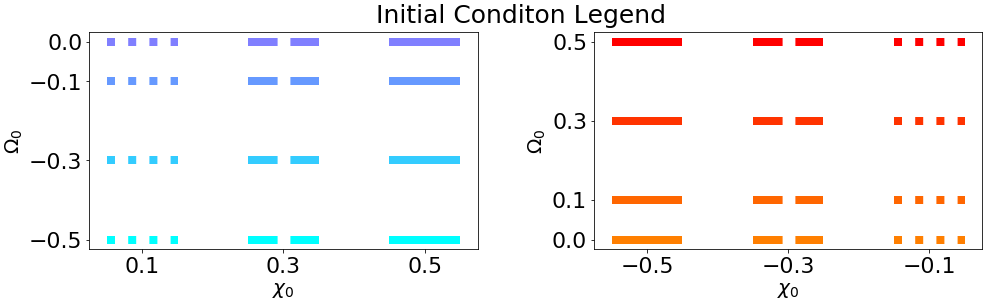}
    \caption{(Color online) Figure showing how to distinguish between different initial conditions in the various plots presented in this work.}
    \label{fig:piLeg}
\end{figure}

Up until this point there have been two main approaches to studying the evolution of a hot and  baryon rich QGP through the QCD phase diagram.  On one hand, a significant part of the community assumes that the system can be described as an ideal fluid such that one can follow Lattice QCD-computed isentropes (where the total entropy to baryon number ratio is fixed throughout the expansion $S/N_B=const$) throughout the QCD phase diagram. In order to determine the correct $S/N_B$ ratio, one determines it from freeze-out properties (typically comparing net-charge fluctuations at freeze-out) and works backwards from the freeze-out point to extract these trajectories, see \cite{Ejiri:2005uv,Schmid:2008sy,Bellwied:2016cpq,Noronha-Hostler:2019ayj,Motornenko:2019arp} for recent examples of this approach. 

\begin{figure}
    \centering
    \includegraphics[width=\linewidth]{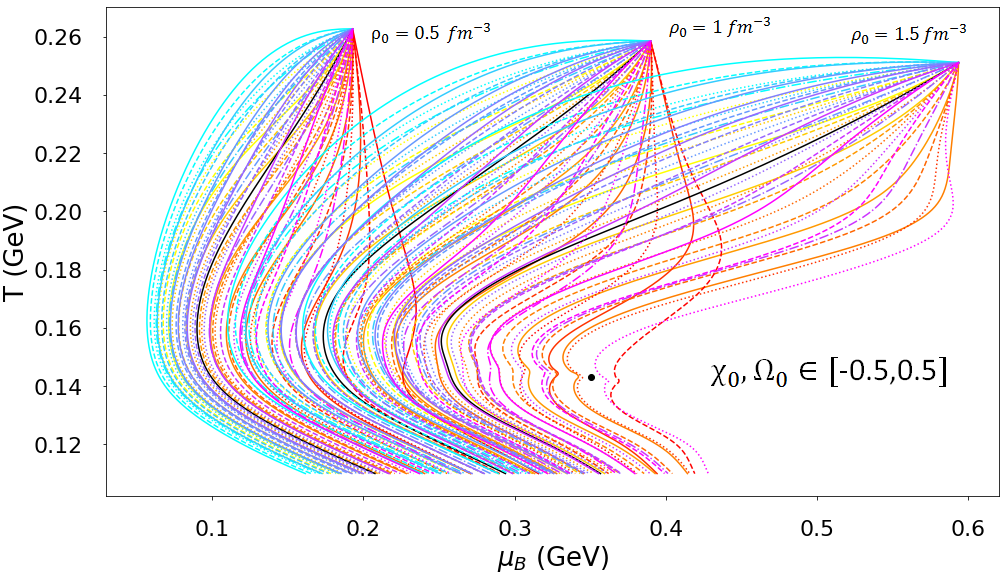}
    \caption{Trajectories produced using DNMR equations of motion, for the same initial energy density, with $\rho_0 \in \{0.5,1,1.5\}$fm$^{-3}$ , and with $\chi_0,\Omega_0 \in \{-0.5,0.5\}$.}
    \label{fig:drama}
\end{figure}

The second approach has been to study full scale 3+1 dimensional hydrodynamic simulations such as in \cite{Ivanov:2005yw,Steinheimer:2007iy,Shen:2018pty,Moreau:2019vhw} and concentrate on the central cells passage through the QCD phase diagram.  In  \cite{Steinheimer:2007iy} ideal hydrodynamic equations of motion were used and, unsurprisingly, the  $T-\mu_B$ evolution of the central cell closely followed that of isentropes.  However, in \cite{Shen:2018pty} full viscous simulations (but assuming the initialization $\pi^{\mu\nu}=\Pi=0$) were used and cells from the center certainly pass through a wide swath of the phase diagram throughout the hydrodynamic evolution.  As far as we know, there has yet to be a study on the influence of viscosity (or better put, entropy production) on the $T-\mu_B$ trajectories. Nor are we aware of any initial conditions that initialize the full energy momentum tensor ($T^{\mu\nu}$) at finite baryon densities that are coupled to viscous hydrodynamic codes and, thus, explore the influence of far-from-equilibrium behavior on the $T-\mu_B$ evolution.

One has no reason to believe that initial conditions at the beam energy scan should be close to equilibrium (in fact, very little is known about initial conditions at the beam energy scan and they have not gone through nearly as many rigorous checks as what has been performed at LHC energies \cite{Giacalone:2017uqx} and the idea of BSQ eccentricities is still being developed \cite{Martinez:2019rlp,Martinez:2019jbu}). Thus, it is necessary to include this systematic uncertainty in our calculations.  For this study we perturb the initial state between $\chi_0=\pm 0.5$ and $\Omega_0=\pm 0.5$, where we define 
\begin{eqnarray}
    \chi &\equiv& \pi^{\eta}_{\eta}/(\epsilon+p)\label{eqn:chi}\\
    \Omega &\equiv& \Pi/(\epsilon+p)\label{eqn:om}
\end{eqnarray}
which are, respectively, the inverse Reynolds numbers, $Re^{-1}$ for shear and bulk viscosity \cite{Denicol:2012cn}. We then systematically run a large number of initial baryon densities. Because we run a large number of trajectories, we have employed a color scheme to denote the initial conditions used in our model, as shown in Fig.\ \ref{fig:piLeg}. In Sec.\ \ref{sec:entropy}, we touch on some physical constraints in allowed choices for the initial $Re^{-1}$ for both shear and bulk.

\subsection{Trajectories for fixed initial $\rho$ and $\varepsilon$}\label{sec:TmuBerho}

To demonstrate how strong of an effect that far-from-equilibrium behavior can have from the initial conditions on the trajectory through the QCD phase diagram, we pick three different initial conditions in $\rho_0$ at a fixed initial energy density and then vary $\chi_0$ and $\Omega_0$, as shown in Fig.\ \ref{fig:drama}. At the lowest baryon density of
$\rho_0=0.5\;fm^{-3}$ 
we already see a wide spread in the $\left\{T,\mu_B\right\}$ trajectories and around the chiral phase transition they cover a swath in baryon chemical potential of about $\Delta \mu_B\sim 200$ MeV.  Thus, even far from the critical point it is extremely important to know the initial conditions for $\chi$ and $\Omega$. One can also see an interesting dependence on the range of chemical potentials at the chiral phase transition, depending on the choice of $\rho_0$. For the intermediate baryon density initial condition of $\rho_0=1\;fm^{-3}$ we find a range of chemical potentials at the chiral phase transition to be even larger, on the order of $\Delta \mu_B\sim 250$ MeV.  However, at our maximum initial baryon density of $\rho_0=1.5\;fm^{-3}$ we begin to see a bend in all the trajectories and what may even be some hints of an attraction towards the critical region.  The chiral phase transition range in initial chemical potential range is smaller than for $\rho_0 =1\; fm^{-3}$, and is again $\Delta \mu_B\sim 200$ MeV. We also do not obtain trajectories that pass far to the right of the critical point.  Unfortunately, we cannot explore this trend further because of the limits of our EOS.

\begin{figure}[t]
\centering

\begin{tabular}{c}
\includegraphics[width=\linewidth]{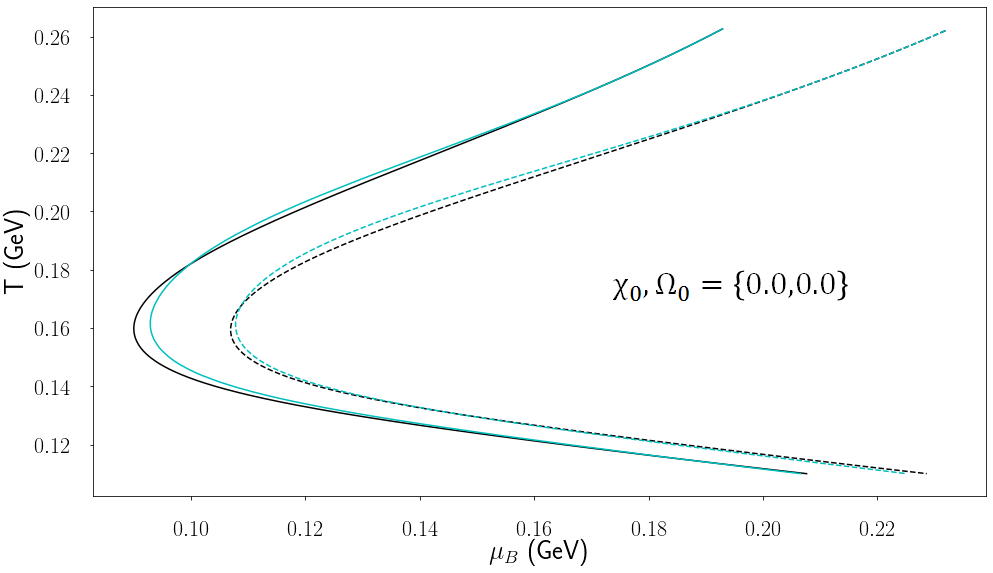} \\
\includegraphics[width=\linewidth]{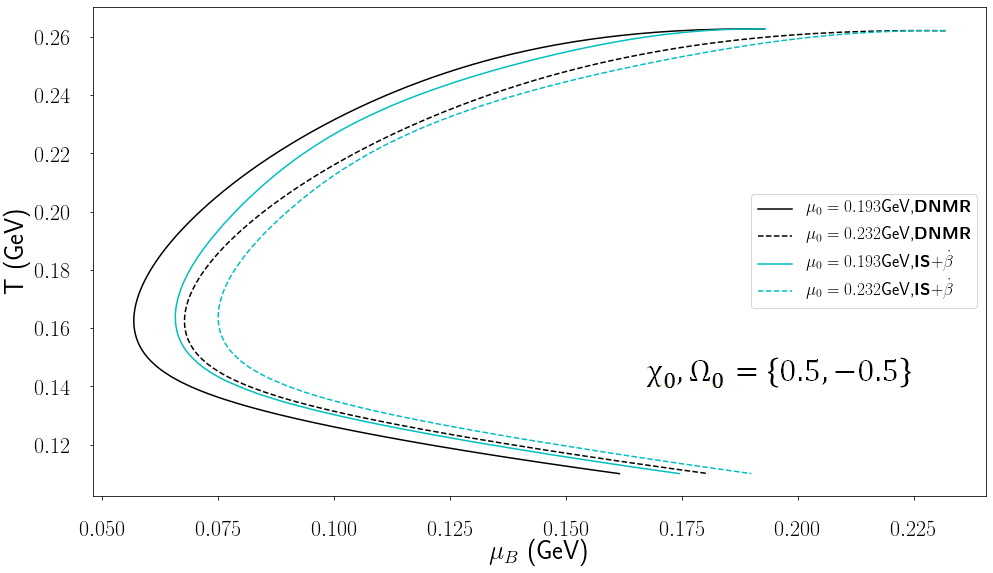} 
\end{tabular}
\caption{Here we compare the $T-\mu_B$ trajectories for DNMR and Israel-Stewart equations of motion, with different initial conditions. The legend is the same in both figures.}
\label{fig:compareEOM}

\end{figure}


In Fig.\ \ref{fig:compareEOM}, we directly compare the phase diagram trajectories generated by hydrodynamic runs of the same initial conditions, comparing DNMR and Israel-Stewart equations of motion. When the initial conditions for the shear stress tensor and bulk pressure are all set to zero, then the two trajectories are relatively similar to each other (although not identical!).
However, if we consider far-from-equilibrium initial conditions, specifically $\chi_0=0.5$ and $\Omega_0=-0.5 $.
The differences between IS and DNMR and are very pronounced, especially where the trajectories cross the chiral phase transition where DNMR appears to freeze-out at a lower $\mu_B$ compared to IS.  A more interesting comparison can be made when one finds the range of initial conditions which lead to the same freeze-out point, that is, a degeneracy in the final state thermodynamics when attempting to trace back to the initial state. This is the approach that we shall take for the rest of this paper.


\subsection{Trajectories for a fixed freeze-out}\label{sec:TmuBFO}

While the initial state is certainly unknown at the beam energy scan, freeze-out has been well studied with both thermal fits \cite{Becattini:2005xt,Becattini:2012xb,Cleymans:2005xv,Torrieri:2006yb,Andronic:2011yq,BraunMunzinger:2003zd,Andronic:2005yp,Vovchenko:2015idt} and fluctuations of conserved charges \cite{Alba:2014eba,Karsch:2010ck,Garg:2013ata,Borsanyi:2013hza,Borsanyi:2014ewa,Bellwied:2018tkc,Bellwied:2019pxh,Braun-Munzinger:2020jbk}.  Some tension still exists between the freeze-out estimates in terms of $T$ and $\mu_B$ from thermal fits versus fluctuations, although reasonable agreement exists when two separate freeze-out temperatures are used for light and strange hadrons \cite{Alba:2020jir}.  Therefore, in this study we require that our hydrodynamic evolution must pass approximately\footnote{Since we only include one conserved charge the isentropes are slightly different.} through the light hadron freeze-out point from \cite{Alba:2014eba} and can then determine the range in initial conditions that lead to that point.  

\begin{figure}[t]
    \centering
    
    \begin{tabular}{c}
    \includegraphics[width=\linewidth]{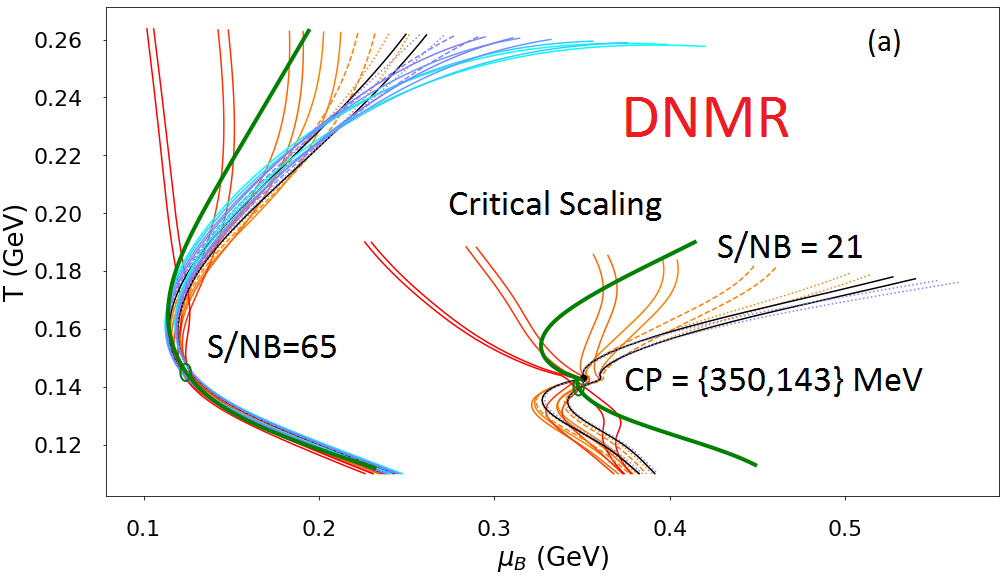}\\

    \includegraphics[width=\linewidth]{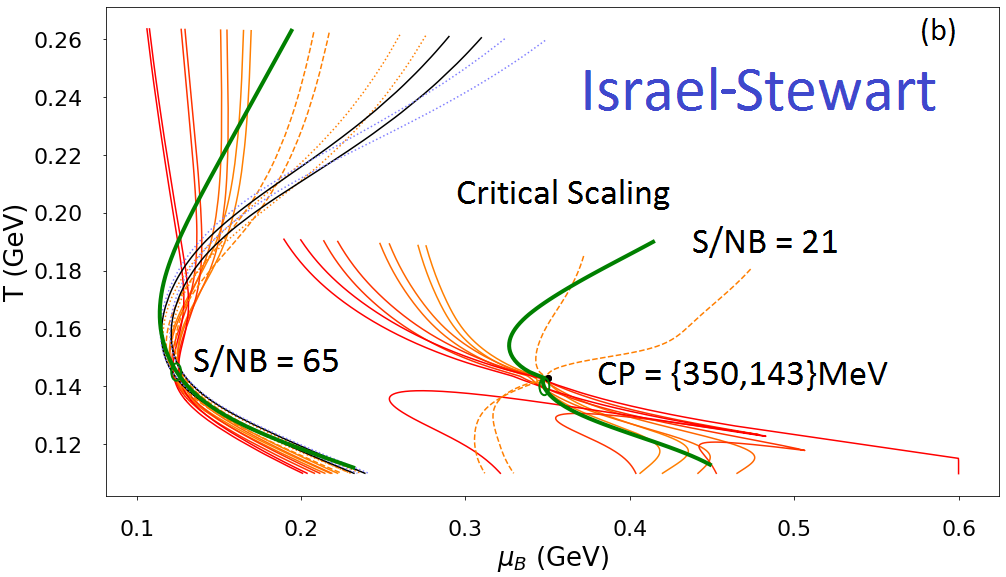}

    \end{tabular}

    \caption{Trajectories in the QCD phase diagram for different hydrodynamic equations of motion. The green lines are isentropes and are the same in each figure. The freeze-out region is shown as a green circle centered along the freeze-out point on the isentrope.}
    \label{fig:Runs}

\end{figure}

In Fig.\ \ref{fig:Runs} we study the intermediate beam energy of $\sqrt{s_{NN}}=27$ GeV as well as a hypothetical lower beam energy which would have an isentrope that passes through the critical point, for DNMR and Israel-Stewart equations of motion. The freeze out region is defined at some point along the green isentrope lines by choosing a reasonable temperature at which to freeze out at. We then select on hydrodynamic trajectories that pass through a circular region centered on the freeze out point, with a radius of $2.5$ MeV (we motivate this value by the approximate order of magnitude of the error bars on the extracted freeze-out points from thermal fits and fluctuations).

At some point we expect standard relativistic hydrodynamics to breakdown sufficiently close to the critical point.  However, the point where this occurs is still unclear.  Additionally, due to the limited influence the critical point may have due to finite time and volume effects, as a first step one may run hydrodynamics at the critical point without critical fluctuations.  

We find that regardless of the choice of equations of motion, contributions coming from viscous effects play an important role in determining the trajectory of the system through the phase diagram. 
For our range of $\chi_0$ and $\Omega_0$, the possible initial conditions that lead to the same freeze-out conditions are wide-spread in chemical potential for the same initial energy density, as shown in Fig.\ \ref{fig:Runs}. In the Israel-Stewart case, away from the critical point, the initial chemical potential can lie in a range of nearly $\Delta \mu_B\sim 200$ MeV and still make it to the same freeze-out region. Closer to the critical point, the range increases to $\sim 300$ MeV. The DNMR trajectories have the same characteristics, only the initial conditions appear to converge closer to the isentrope (solid green line) more quickly at least far from the critical point. It is interesting to note that trajectories that go through or near the critical point accept a larger range of initial conditions. This is, again, indicative of some attractive like behavior, specific to this EOS, and the question remains as to whether this behavior persists upon inclusion of the necessary critical
fluctuation framework previously mentioned. It should be the case that the behavior of the trajectories before entering the critical region will be the same. However, the dynamics within the critical region will surely be modified. 

The solid black lines in Fig.\ \ref{fig:Runs} are the points where the initial shear and bulk are set to zero but that the transport coefficients are still turned on i.e. $\pi^\eta_{\eta,0}=\Pi_0=0$.  One can see that for DNMR the effect of the transport coefficients alone is smaller (transport coefficients lead to initial conditions that start at $\Delta \mu_B\sim 50$ MeV larger than for the isentropes) than for Israel-Stewart, where we find that the effect of transport coefficients alone increases the initial baryon chemical potential by $\Delta \mu_B\sim 100$ MeV. This demonstrates that the trajectories through the QCD phase diagram are strongly dependent on the choice of second order hydrodynamic equations of motion. Since those theories only differ in the transient regime (given that both approaches have the same relativistic Navier-Stokes limit), our results indicate that transient hydrodynamic effects must be taken into account when determining the path traversed by the QGP on the QCD phase diagram.  

Finally, we find that the sign of the initial conditions plays a large role if the trajectories are to the left or the right of the isentropes.  Generally, values for the initial conditions with $\Pi\leq0$ and $\pi^{\eta}_\eta\geq0$ push the trajectories towards larger $\mu_B$ whereas initial conditions with $\Pi\geq0$ and $\pi^{\eta}_\eta\leq0$ push the trajectories to smaller $\mu_B$.

\subsection{Viscous Effects}\label{sec:shear}

 \begin{figure} 
      \centering

    \begin{tabular}{c}
    \includegraphics[width=\linewidth]{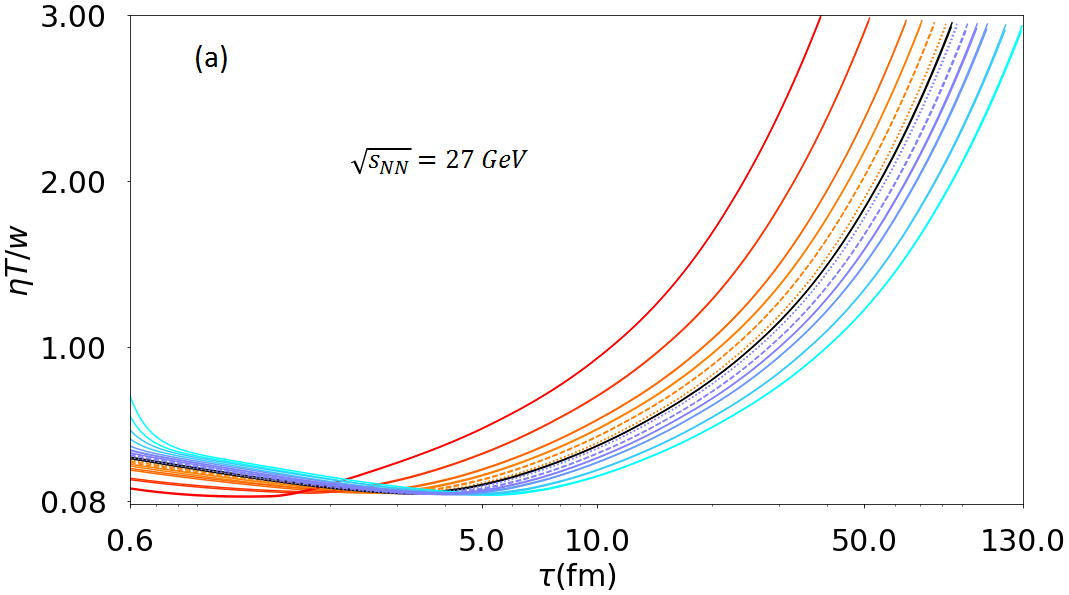}\\
    \includegraphics[width=\linewidth]{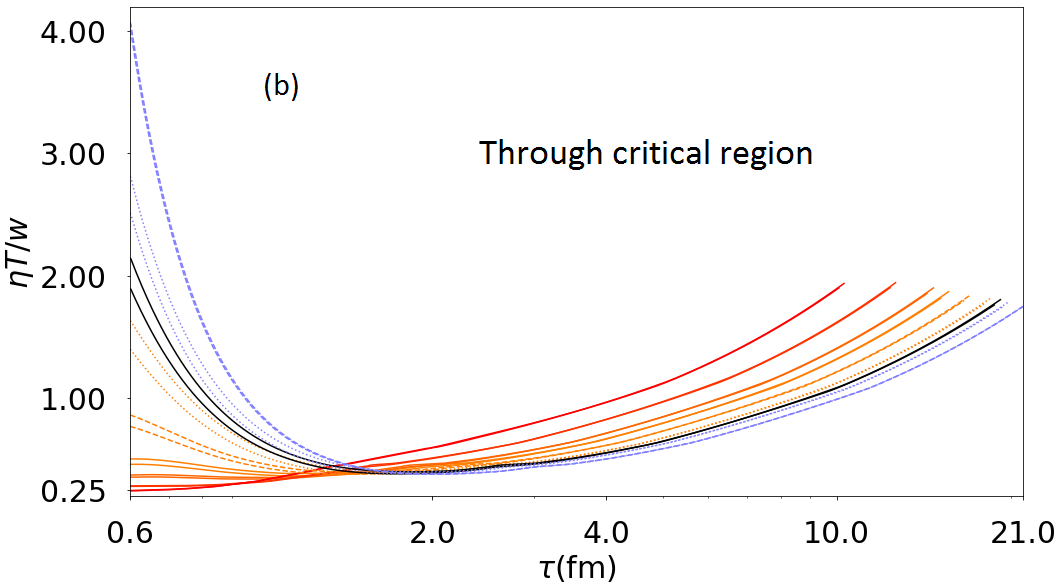}
    \end{tabular}
    \caption{Time evolution of $\frac{\eta T}{w}$ for different initial conditions in DNMR, close to the critical point (top) and away from the critical point (bottom).}\label{fig:shear}
 \end{figure}

Fig.\ \ref{fig:shear} plots different trajectories of our shear viscosity over enthalpy ratio for DNMR equations of motion for trajectories both far from and near to the critical point. We note that our construction of shear viscosity does not incorporate any critical scaling since it does not scale as strongly with the correlation length \cite{Son:2004iv}.  The time evolution of $\eta T/w$ varies with the choice in the initial $\pi$ and $\Pi$, which sends the hydrodynamical expansion along different trajectories. Since $\eta T/w$ depends on both $T$ and $\mu_B$, different values of $\eta T/w$ as a function of time are probed depending on the initial conditions. 
 
 We then compare the bulk viscosity in Eq.\ (\ref{eqn:zetanorm}) to its critically scaled form proposed in \cite{Monnai:2016kud}. The form of this bulk viscosity is then
 \begin{eqnarray}\label{eqn:zetaCS}
     \left(\frac{\zeta T}{w}\right)_{CS} =\frac{\zeta T}{w}
     \left[1 +\left(\frac{\xi}{\xi_0}\right)^3\right]
 \end{eqnarray}
 where $\xi$ is the correlation length and $\xi_0$ sets the scale for the critical region. When not including the critical component, we simply set $\xi$ to $0$.
 
  \begin{figure}[b]
      \centering

    \begin{tabular}{c}
     \includegraphics[width=\linewidth]{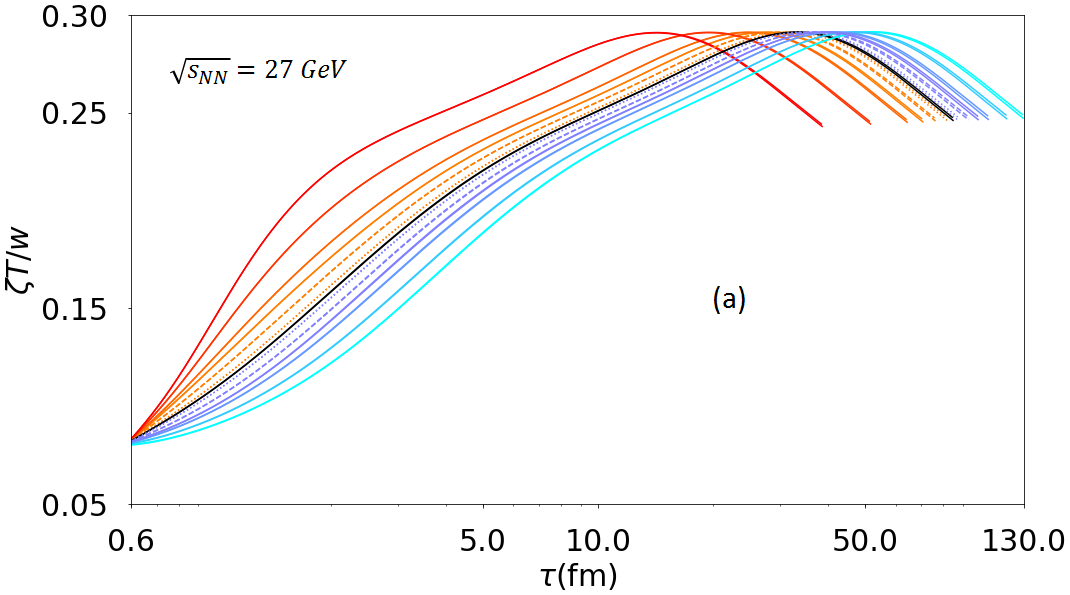}\\
     \includegraphics[width=\linewidth]{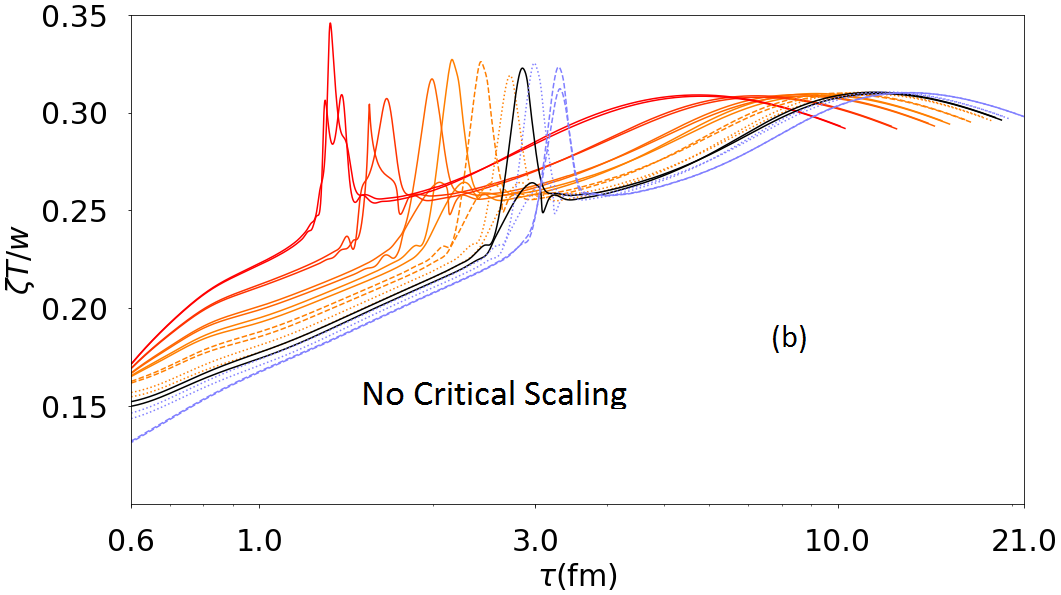}\\
    \includegraphics[width=\linewidth]{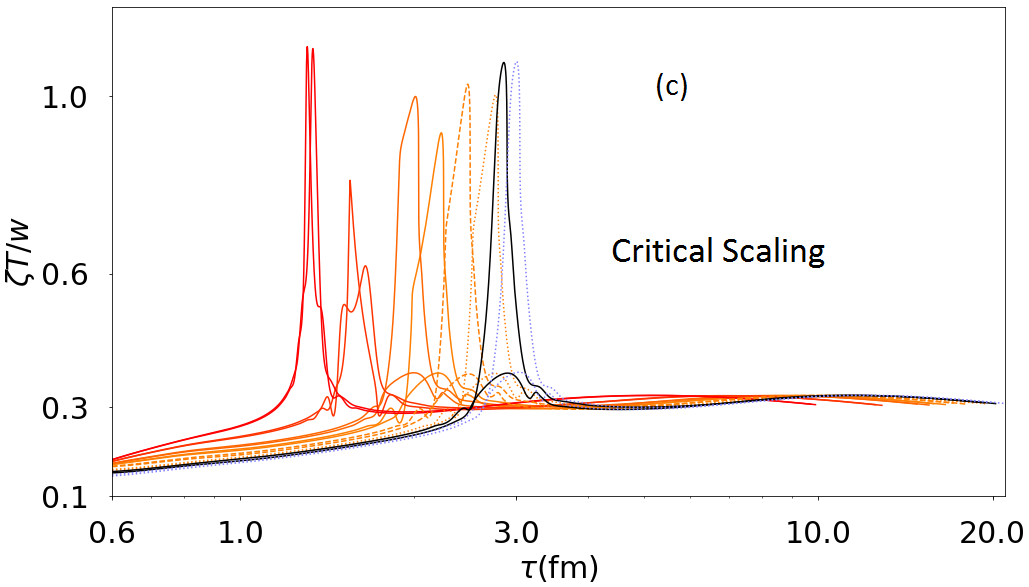}
    \end{tabular}
    \caption{Time evolution of $\frac{\zeta T}{w}$ for different initial conditions in DNMR, away from the critical point (top), near the critical point (middle), and critically scaled (bottom) }\label{fig:bulk}
 \end{figure}
 
Because the bulk viscosity depends on $c_s^2$, the non-trivial structure that arises in its dependence over time is due to the change of degrees of freedom. When plotted on trajectories close to $\mu_B=0$ (i.e. far from the critical point) a bump it seen as the quarks and gluons transition into hadrons, as seen in Fig.\ \ref{fig:bulk} (a). The different lines demonstrate how different trajectories probe different values of $\zeta T/w$ at different times.\ However, at the critical point the speed of sound goes to zero, which produces a spike in $\zeta T/w$ as one passes through it. In this paper we compare two scenarios, one where $\zeta T/w$  only scales with $c_s^2$ across the critical point, which is shown in Fig.\ \ref{fig:bulk} (b) and another where the correlation length affects the $\zeta T/w$, as shown in Fig.\ \ref{fig:bulk} (c).

When incorporating the critical scaling through the correlation length into the bulk viscosity, there is some freedom in choosing the scaling constant, $\xi_0$, such that $\zeta T/w$ is smaller outside the critical region, and much larger inside. In this work, it was chosen so that the peak in $\zeta T/w$ increases by a factor of 3 close to the critical point. The correlation length is calculated using a formula found in \cite{Monnai:2016kud} that calculates the equilibrium value as:
\begin{eqnarray}
    \xi^2 = \frac{1}{H_0} \left(\frac{\partial M(r,h)}{\partial h} \right)_r
\end{eqnarray}
As is done in \cite{Monnai:2016kud}, we use the linear paramterization model \cite{Schofield:1969,Zinn:2001}, but instead derive an expression to fifth order in $\theta$. This is consistent with the accuracy of our EOS. The expression to fifth order in $\theta$ is then:
\begin{multline}
    \left(\frac{\partial M(r,h)}{\partial h} \right)_r = 
    \frac{M_0}{H_0 R^{\beta(\delta-1)}}\left(\frac{1 + \theta^2(2\beta -1)}{2\beta\delta\theta\widetilde{h}+ \widetilde{h'}(1 - \theta^2)} \right)
\end{multline}
with
\begin{eqnarray}
    \widetilde{h} = \theta(1 + a \theta^2 + b \theta^4)
\end{eqnarray}
where the coefficients $a,b$ are accessible output from our EOS, and the critical exponents are taken as their mean field approximate values. We leave for future work the studying of consequences of changing the strength and shape of the critical region, which should change the peak in $\zeta T/w$, accordingly.

\begin{figure} 
      \centering

    \begin{tabular}{c}
    \includegraphics[width=\linewidth]{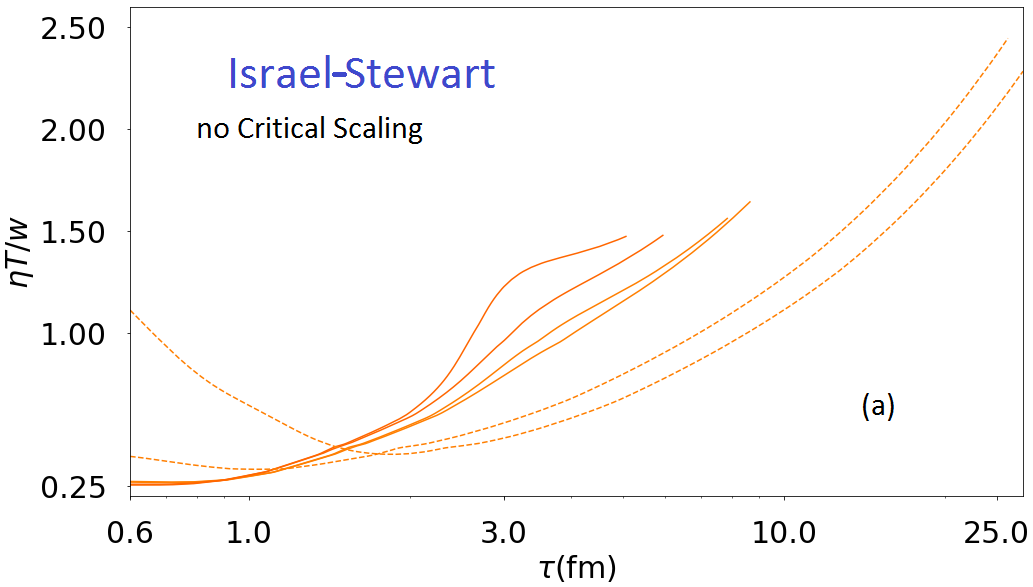}\\
    \includegraphics[width=\linewidth]{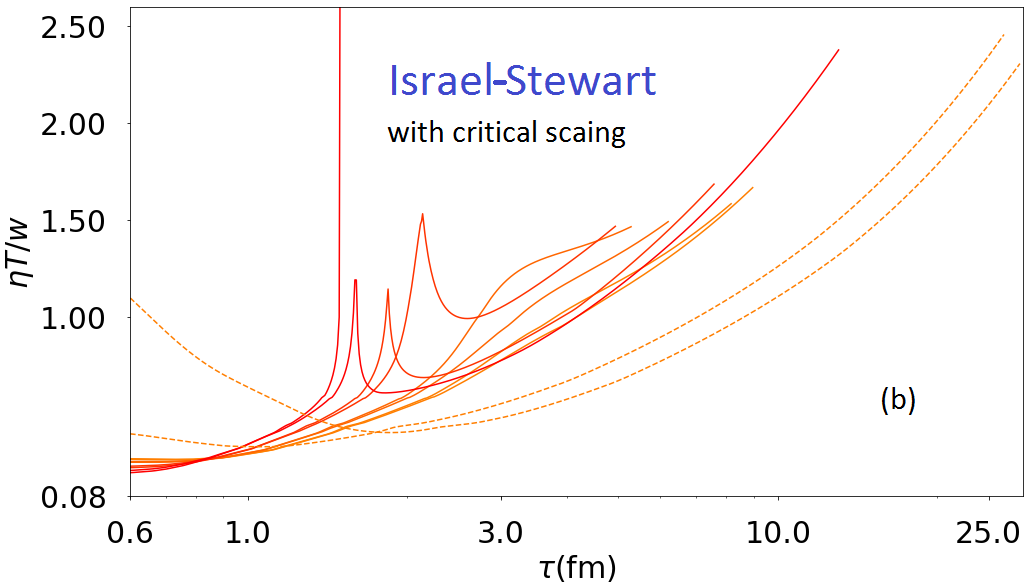}
    \end{tabular}
    \caption{Time evolution of $\frac{\eta T}{w}$ for different initial conditions in Israel-Stewart, with $\dot\beta$ terms that go through the critical region, without critical scaling (top) and with (bottom).}\label{fig:sheardbetas}
 \end{figure}
  
A crucial piece to understanding $\chi$ at the critical point in Israel-Stewart is to observe the $\eta T/w$ trajectories at the critical point, as shown in Fig.\ \ref{fig:sheardbetas}. Because of the rather non-trivial trajectories across the critical point for Israel-Stewart equations of motion, $\eta T/w$ inherits a non-trivial time dependence even though no critical behavior was built into the transport coefficient. One can see in Fig.\ \ref{fig:IS_Dbetas_Com} that many lines traverse the chemical potential in a complicated and non-trivial way (specifically the red line). It is this chemical potential dependence that produces the peak behavior seen in Fig. \ref{fig:sheardbetas}. Comparing the red lines with a spike to Fig.\ \ref{fig:IS_Dbetas_Com} we find that this caused by trajectories that begin at high $T$ and low $\mu_B$ that then pass through the critical point and continue onto low $T$ and high $\mu_B$ trajectories.  Eventually these lines end abruptly because they have reached the edge of our EOS. Also worth noting is the increased sensitivity of the shear viscosity to critical scaling shown in the bottom of Fig. \ref{fig:sheardbetas} compared to the non-critically scaled runs shown in the top.

In the case of the DNMR equation of motion, the $T-\mu_B$ trajectories behave much more smoothly, and thus no spike is seen in Fig.\ \ref{fig:shear}.

\section{Potential attractors}\label{sec:att}

\begin{figure*}
    \centering
    \begin{tabular}{c c}
    \includegraphics[width=0.5\linewidth]{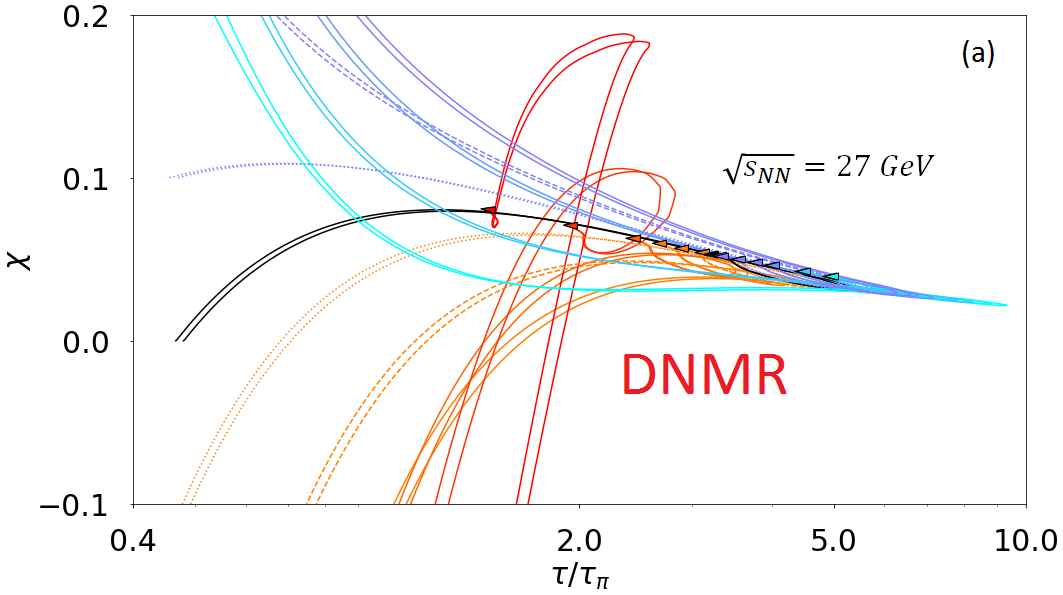} &  \includegraphics[width=0.5\linewidth]{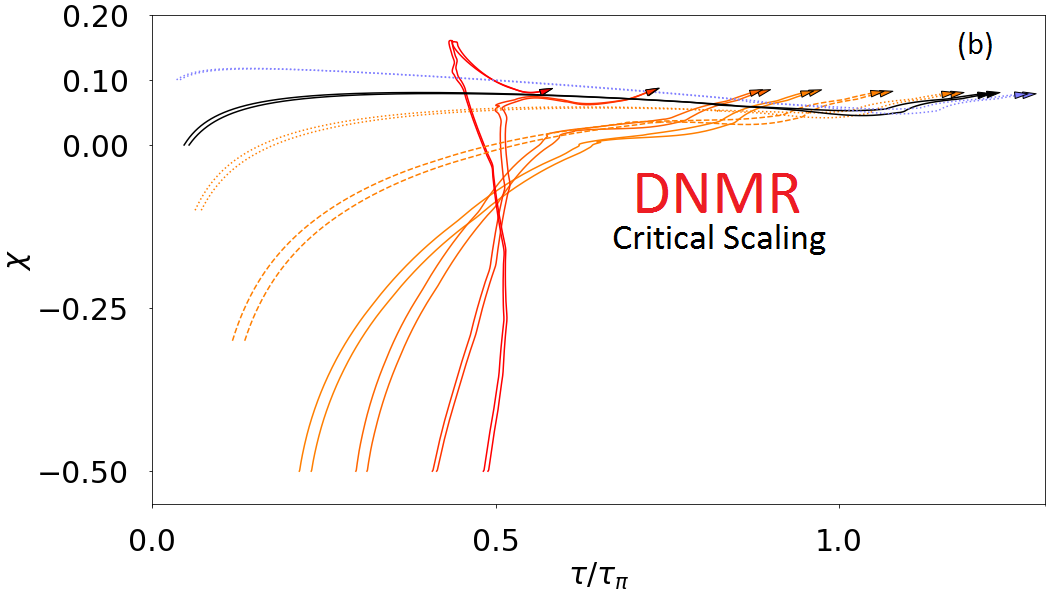} \\
    \includegraphics[width=0.5\linewidth]{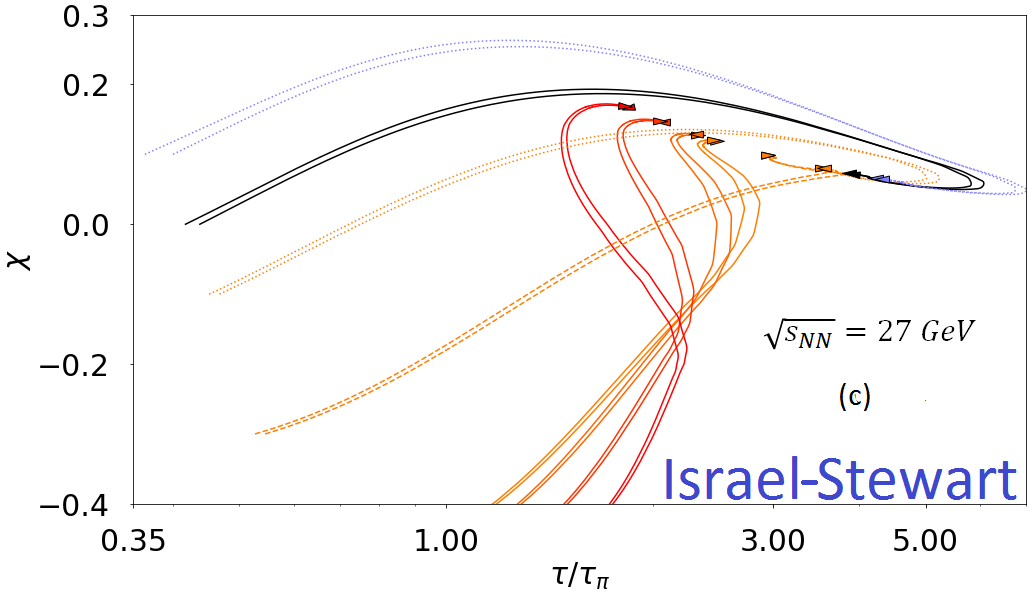} &  \includegraphics[width=0.5\linewidth]{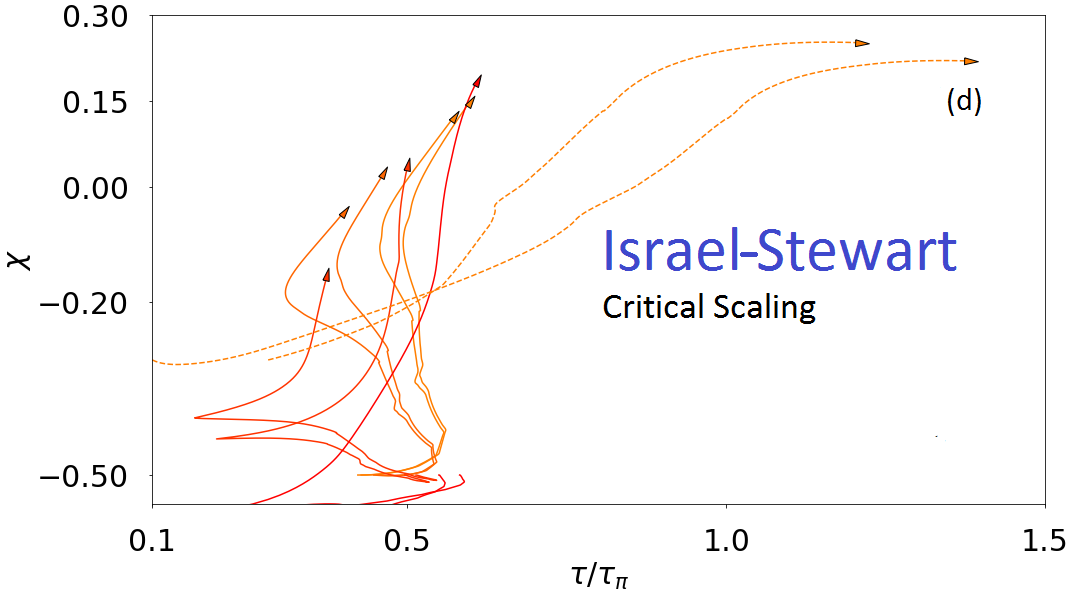}
    \end{tabular}
    \caption{Shear inverse Reynolds number $\chi=\pi^\eta_\eta/w$ trajectories far  from the critical point (left) and at the critical point (right) using DNMR (top) and Israel-Stewart (bottom) equations of motion. The solid black lines assume that the initial $Re^{-1}=0$ for both shear and bulk (the band demonstrates the width of our range of the freeze-out $\left\{T,\mu_B\right\}$. }
    \label{fig:chi}
\end{figure*}

\begin{figure*}
    \centering
    \begin{tabular}{c c}
    \includegraphics[width=0.5\linewidth]{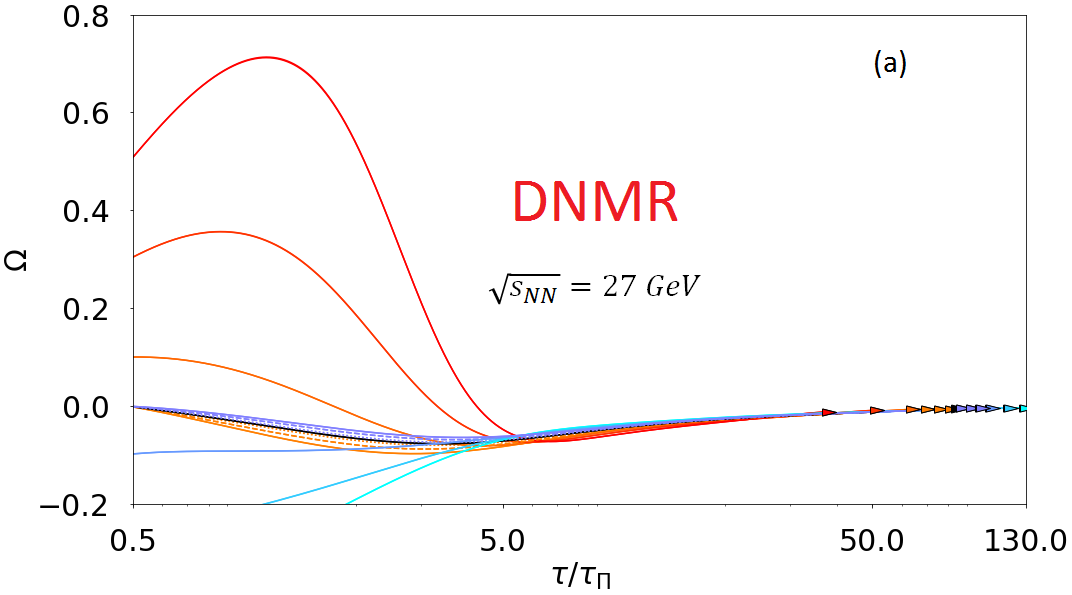} &  \includegraphics[width=0.5\linewidth]{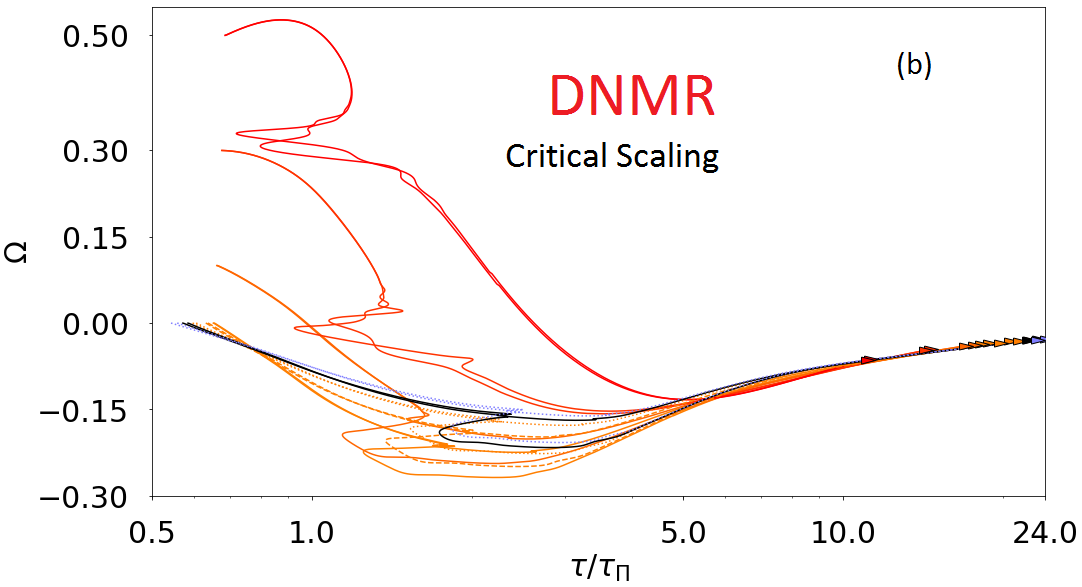} \\
    \includegraphics[width=0.5\linewidth]{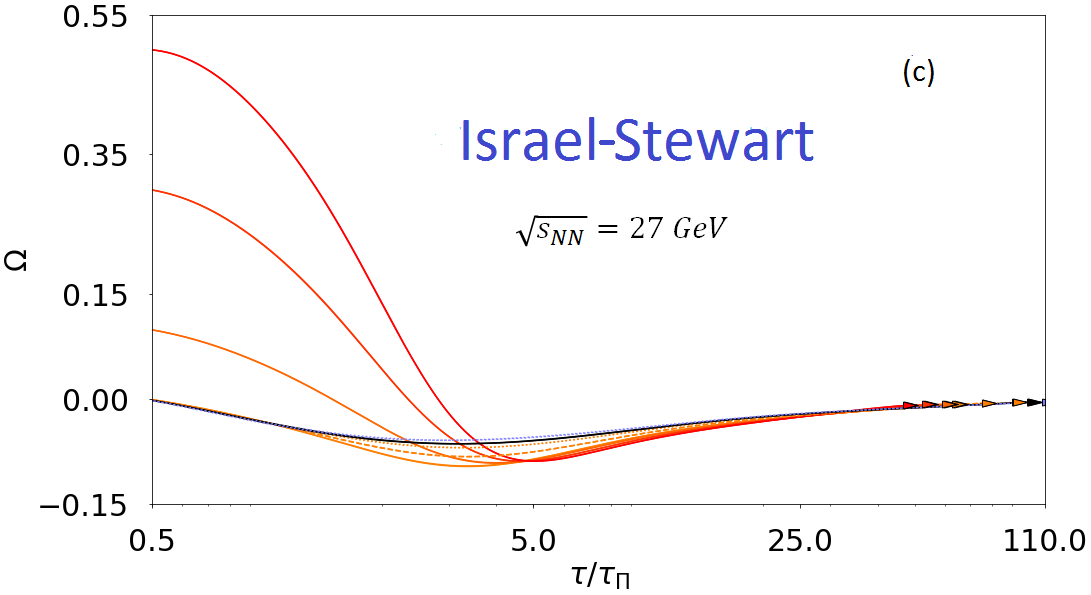} &  \includegraphics[width=0.5\linewidth]{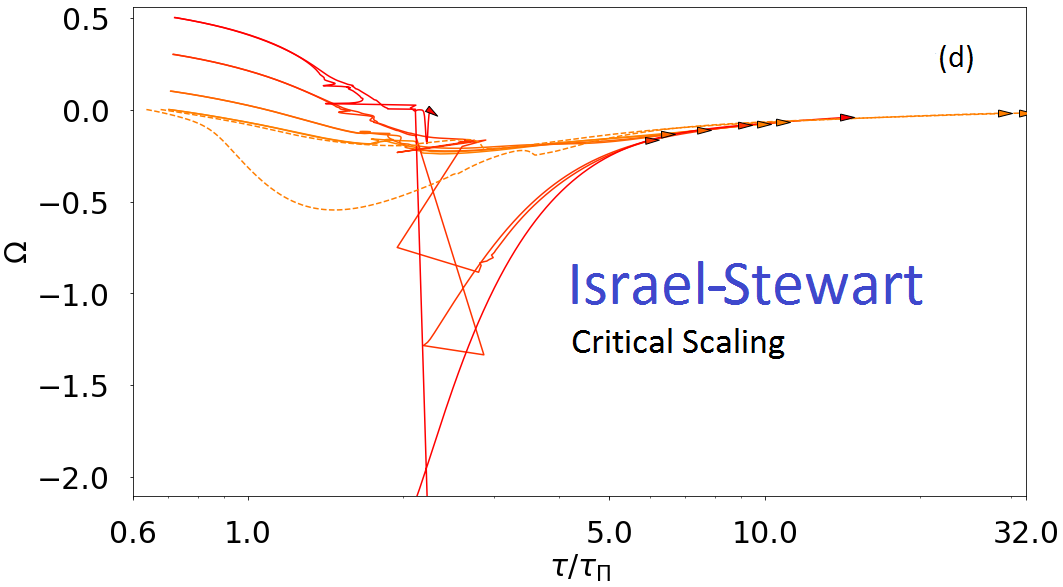}
    \end{tabular}
    \caption{Bulk inverse Reynolds number $\Pi/w$ trajectories far  from the critical point (left) and at the critical point (right) using DNMR (top) and Israel-Stewart (bottom) equations of motion. The solid black lines assume that the initial $Re^{-1}=0$ for both shear and bulk (the band demonstrates the width of our range of the freeze-out $\left\{T,\mu_B\right\}$.}
    \label{fig:omega}
\end{figure*}

In this paper we do not attempt to systematically investigate the presence of attractors for these rather non-trivial transport coefficients and complex equation of state.  However, we can check for a convergence of $\chi=\pi^\eta_\eta/(\epsilon+p)$ in Fig.\ \ref{fig:chi} and $\Omega=\Pi/(\epsilon+p)$ in Fig.\ \ref{fig:omega} on time scales normalized by their respective relaxation times. The points observed in Figs.\ \ref{fig:chi} and \ref{fig:omega} are those passing through the freeze-out for $\sqrt{s_{NN}}=27$ GeV and the critical point, respectively.

In Fig.\ \ref{fig:chi} the inverse Reynolds numbers for shear viscosity are shown both far from the critical point and at the critical point. The shape of $\chi$ over time is rather complicated because the minimum of $\eta T/w$ at the phase transition, which leads to this bending backwards in $\chi$ since $\tau_\pi$ depends on the shear viscosity (similar to what was found in \cite{Chattopadhyay:2019jqj}).  For both DNMR and Israel-Stewart we immediately note that due to the short lifetime of our hydrodynamic runs, none of our trajectories converge to a single line by freeze-out.  However, we also plot the direction of the derivative at the freeze-out point and it does appear that in all cases that an attractor could be reached if hydrodynamics would run for a longer period of time.  From now on, we will refer to this as a ``potential attractor" because we are not certain if this is an attractor but it certainly hints at one. 

One curious difference between DNMR and Israel-Stewart is that for DNMR the potential attractor appears to always sit on a nearly flat line in $\chi$.  However, for Israel-Stewart equations of motion the potential attractor line has a clear slant far from the critical point.  At the critical point the potential attractor for Israel-Stewart is even more bizarre in that it appears to be growing in $\chi$ and then potentially flattening out.  Unfortunately, we cannot investigate this further with our current EOS due to its limitations in $\mu_B$. In fact, due to the limitations in the EOS we are not even able to obtain the $\Pi_0=\pi^\eta_{\eta,0}=0$ curves because they would begin at much larger values of $\mu_B$.

The bulk pressure is more intuitive to understand and we find that despite a wide range of initial conditions (and multiple different combinations for the initial shear and bulk) that all curves quickly collapse onto a universal scaling behavior.  While the time scale may appear to be long, we note that this is because the bulk relaxation time is quite significant (due to the small bulk viscosity used here e.g. see Eq.\ (\ref{eqn:tauPI})).  

In Fig. \ref{fig:omega} we find that far from the critical point both equations of motion quickly converge to what appears to be an attractor, although it appears that Israel-Stewart takes longer to converge.  At the critical point we find that the DNMR equations of motion are more well-behaved and generally do not have large inverse Reynolds numbers even though the critically scaled $\zeta T/w$ is quite large.  On the other hand, the Israel-Stewart equations of motion diverge quite dramatically when passing through the critical point but, despite this effect, they manage to converge afterwards.

\section{Consequence of $\zeta T/w$ diverging due to the critical point}\label{sec:CP}

In the previous section, we always assumed that  $\zeta T/w$ scaled with the correlation length, according to Eq.\ (\ref{eqn:zetaCS}).  In this section we will compare this assumption to the regular $\zeta T/w$ that only scales with the speed of sound, as shown in Eq.\ (\ref{eqn:zetanorm}).  We note that outside of the critical region that our choice of the inclusion of critical scaling is irrelevant since this only affects $\zeta T/w$ near to the critical point.

\begin{figure*}
    \centering
    \begin{tabular}{c c}
    \includegraphics[width=0.5\linewidth]{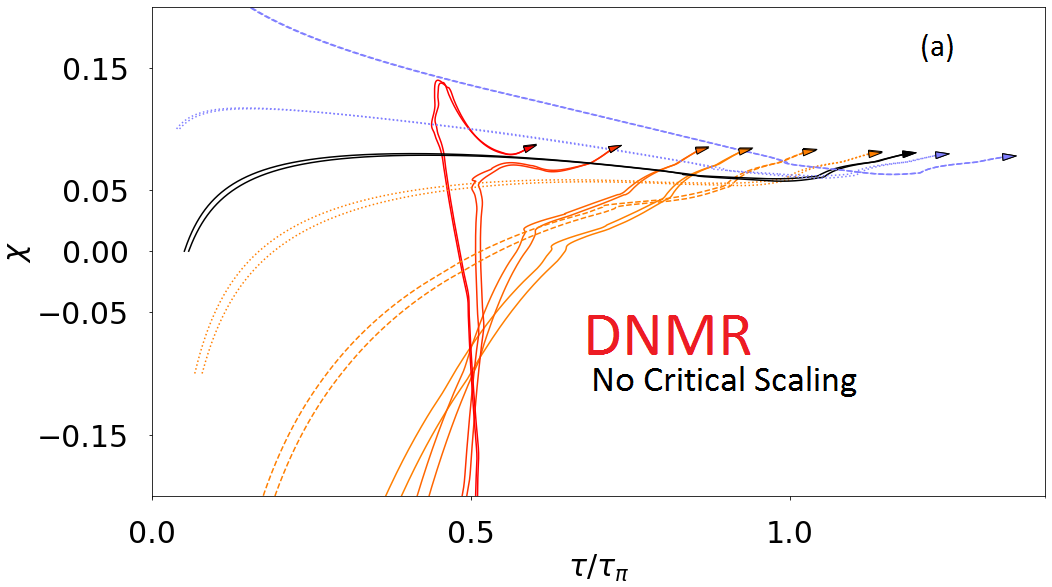} &  \includegraphics[width=0.5\linewidth]{chi_DNMR_CSB_Energy0_15_FO_0348,014_ShearResc.png} \\
    \includegraphics[width=0.5\linewidth]{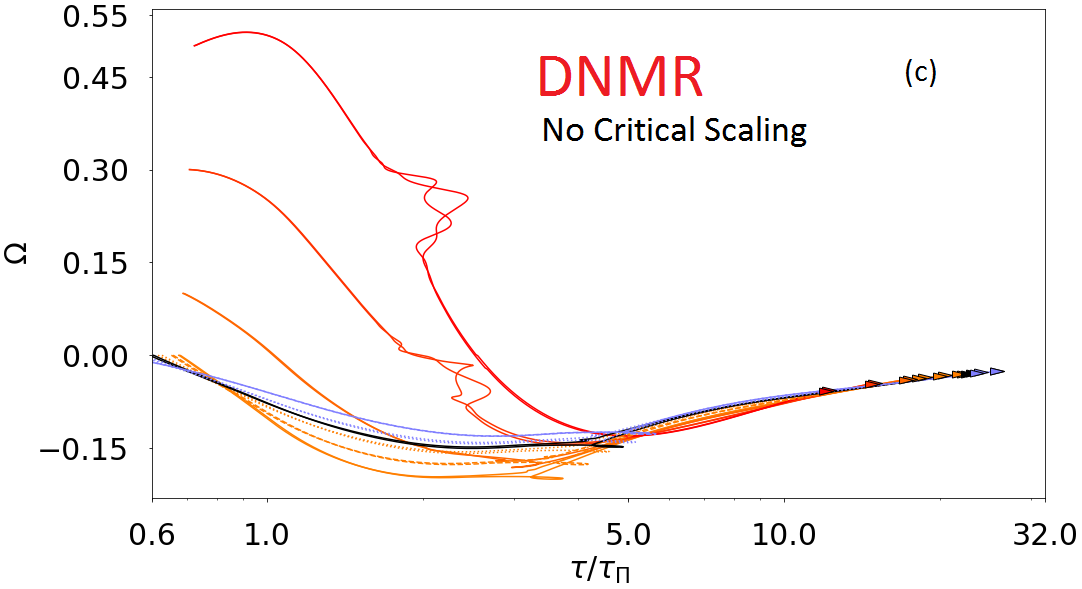} &  \includegraphics[width=0.5\linewidth]{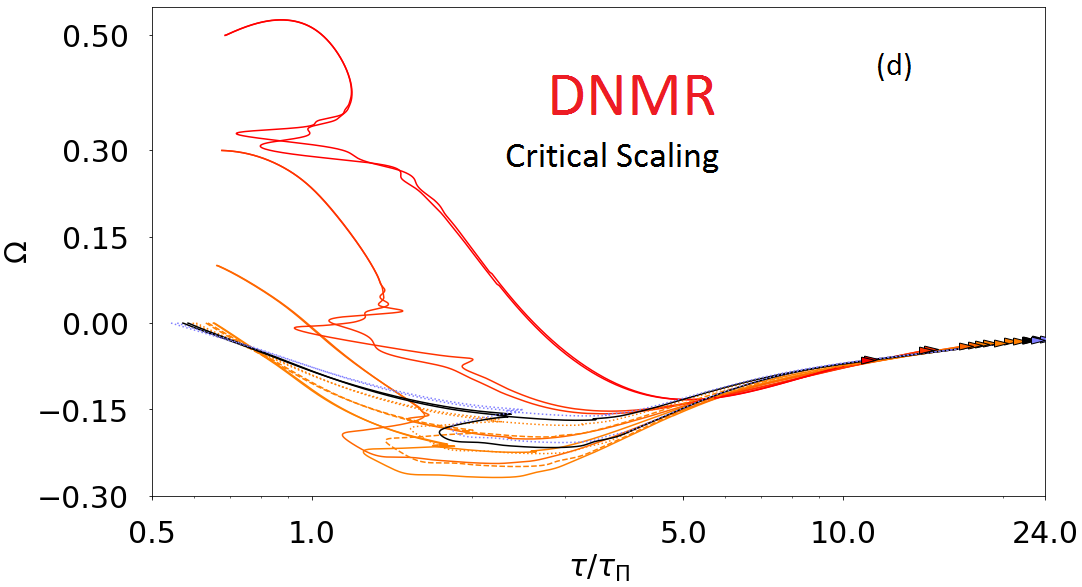}
    \end{tabular}
    \caption{Shear (top) and bulk (bottom) inverse Reynolds number trajectories at the critical point for DNMR equations of motion where either $\zeta T/w$ only scales with $c_s^2$ (left) or also scales with the correlation length (Right). The solid black lines assume that the initial $Re^{-1}=0$ for both shear and bulk (the band demonstrates the width of our range of the freeze-out $\left\{T,\mu_B\right\}$. }
    \label{fig:zetacritical}
\end{figure*}

In Fig.\ \ref{fig:zetacritical} we plot the inverse Reynolds numbers of both shear and bulk viscosity comparing with and without critical scaling of $\zeta T/w$. In the shear $Re^{-1}$ trajectories, we see very little difference if the bulk viscosity has critical scaling or not.  This is not entirely unexpected because while there are coupling terms between shear and bulk viscosity in DNMR, they are non-linear terms and, thus, they do not affect $\chi$ very strongly.  Additionally, the large peak  in $\zeta T/w$ only appears close to freeze-out and, therefore, the $\chi$ trajectory has already converged much closer to its potential attractor at that point.  

As expected, the bulk $Re^{-1}$  is more affected by critical scaling of $\zeta T/w$.  In fact, one can see quite clearly in the plots the point where the peak in $\zeta T/w$ is reached.  However, despite a brief interruption in the approach to the potential attractor in $\Omega$, the curves quickly fall on top of each other in both scenarios. It is clear from these results that the potential bulk attractor is quite large for heavy-ion collisions - likely because bulk only plays a role briefly around the phase transition. 
\begin{figure}[t]
    \centering
    
    \begin{tabular}{c}
    \includegraphics[width=\linewidth]{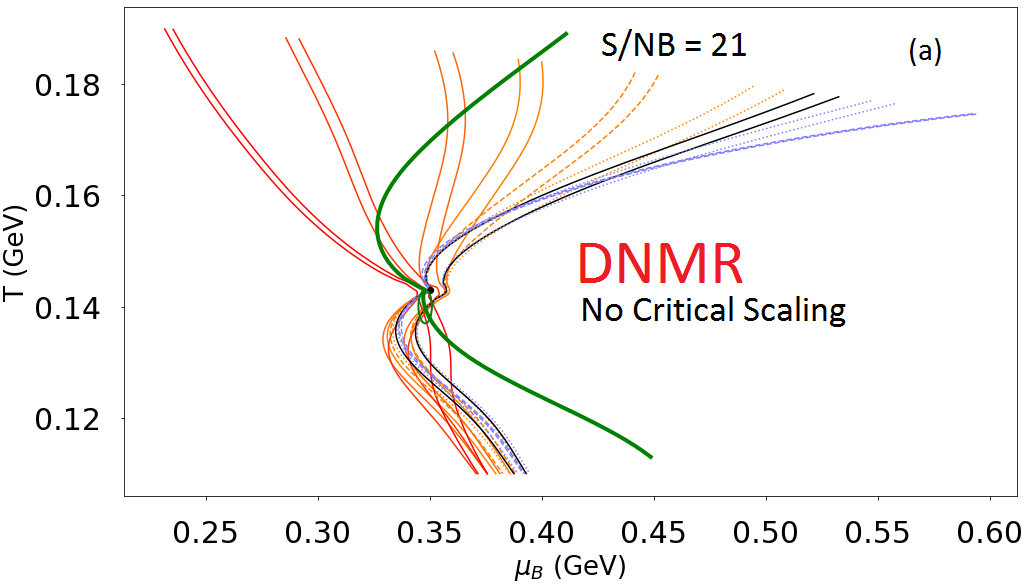}\\
    \includegraphics[width=\linewidth]{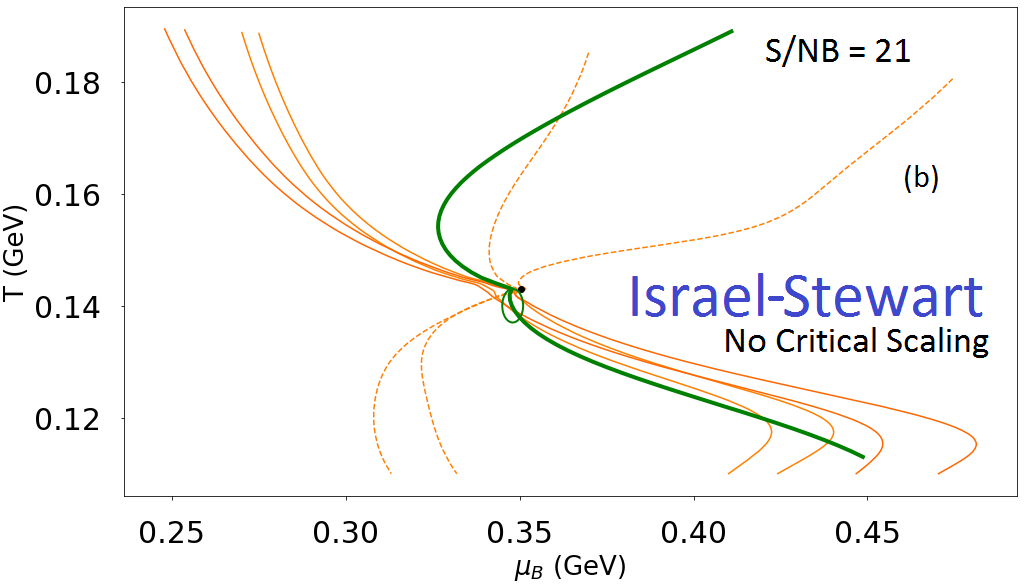}
    \end{tabular}
    \caption{Trajectories in the QCD phase diagram for different hydrodynamic equations of motion. For these trajectories, the bulk viscosity does not include critical scaling. }
    \label{fig:CS}
\end{figure}

In Fig.\ \ref{fig:CS} we observe the $\left\{T,\mu_B\right\}$ trajectories across the critical point when $\zeta T/w$ does not have critical scaling. When comparing these trajectories to the critically scaled ones in Fig.\ \ref{fig:Runs} we find that there are not very large differences. However, for Israel-Stewart equations of motion at low temperatures both scenarios seem like they might run along the first order phase transition line for a bit before the system turns into hadrons.  The biggest difference with and without critical scaling is that the critically scaled $\zeta T/w$ then jumps up to the left of the phase diagram (towards higher temperatures) within the hadron gas phase, whereas the regular $\zeta T/w $ scenario exhibits a more regular trajectory and always progresses downwards (towards low temperatures) in the phase diagram.

\section{Entropy production and trajectory Constraints}
\label{sec:entropy}
It has so far been demonstrated that, due to the existence of a potential attractor for the time evolution of $\chi$ and $\Omega$, there may be a degeneracy in the final freeze-out state of the system. That is, many different trajectories in the phase diagram that are initially very different come extremely close to each other at late times. We have put an emphasis on entropy production as a conceptual basis for understanding the deviations from isentropes. However, we are currently unaware of any rigorous calculation of entropy production for DNMR (or Israel-Stewart when derived from kinetic theory).

This sort of calculation would be extremely useful in allowing for quantitative cuts on what kinds of initial conditions and trajectories are possible, via the second law of thermodynamics. We note that it is clear that due to the deviation of our results from the isentropes, there must be a large effect on the entropy production due to our choice in initial conditions and transport coefficients.  We point out that our chosen transport coefficients are reasonable and not unrealistic since current relativistic viscous hydrodynamic models used within heavy-ion collisions are based on the DNMR formalism \cite{Ryu:2015vwa}.

One can also put some constraints on the choices of initial viscous conditions by taking a similar approach as was done in \cite{Janik:2005zt}. In that paper, the weak energy condition is used to put physical bounds on possible values for the shear-stress throughout the evolution, in a system that undergoes Bjorken flow. The weak energy condition is the condition that

\begin{eqnarray}
    T^{\mu\nu}t_\mu t_\nu \geq 0
\end{eqnarray}
were $t_\mu$ is any time-like vector. This condition has the simple interpretation that  the energy density of the fluid should be non-negative for any observer. Using this constraint, one can extend further the work done in \cite{Janik:2005zt} to put constraints on a non-conformal system. Then, instead of a constraint on just $\chi$, the constraints involve both $\chi$ and $\Omega$. Doing the derivation, one finds the following
\begin{eqnarray}
    \frac{\chi}{2} - \Omega \geq -1\\
    \Omega - \chi \geq -1
    \label{scndConst}
\end{eqnarray}
which must be satisfied simultaneously throughout the evolution. Notice that none of our choices of initial $\chi$ and $\Omega$ violate these bounds, but the choice $\{\chi,\Omega\} = \{0.5,-0.5\}$ does hit the bound in Eq.\ \eqref{scndConst}.

\section{Conclusions}\label{sec:conclusions}

In this paper we analyzed how far-from-equilibrium initial conditions of heavy-ion collisions could affect the search for the QCD critical point.  For a single freeze-out point there exists a multitude of potential trajectories that could have lead to that point because of the entropy that is produced when one considers realistic transport coefficients. Each trajectory is defined by its initial conditions that not only includes the initial energy density and baryon density but also its initial shear stress tensor and bulk pressure. These trajectories diverge far from isentropes, which are calculated along lines of constant $S/N_B$, and depend strongly on the sign of the initial $\Pi$ and $\pi^\eta_\eta$. The non-uniqueness of a freeze-out point with respect to a given initial condition presents an interesting problem in both determining the initial state given the final state freeze-out conditions, as well as in determining the possible late time properties of the fluid (e.g. the possibility of only certain events passing through the critical point for a fixed beam energy). 

We studied both DNMR and Israel-Stewart equations of motion. Perhaps, unsurprisingly, we find that DNMR is better equipped to handle larger initial inverse Reynolds numbers and we did not find any trajectories that led to runaway trajectories, which we interpret as a consequence of DNMR having a more well-controlled expansion \cite{Denicol:2012cn,Chattopadhyay:2019jqj}. Within DNMR the potential attractors appeared to be relatively flat in $\chi$ and $\Omega$ even at the critical point.  In contrast, the Israel-Stewart equation of motion also appear as if they will eventually reach an attractor. However, at the critical point a large, negative spike in $\Omega$ was seen, well outside the range of applicability for hydrodynamics.  Despite this spike the solutions still returned to a potential attractor by freeze-out (we emphasize {\it potential} because due to the finite lifetime of hydrodynamics this would occur beyond our freeze-out point).  We note, however, that the potential attractor line appears significantly different in Israel-Stewart and is no longer flat but rather looks like a hill at the critical point. For phenomenological purposes, this work indicates that codes that solve Israel-Stewart vs. DNMR equations of motion should expect different results when exploring the QCD phase diagram at large baryon densities. Therefore, since the main difference between DNMR and Israel-Stewart lies only in how they treat far from equilibrium transient effects (since they have the same Navier-Stokes limit), our results indicate that the out-of-equilibrium properties of the hot and baryon rich QGP must be taken into account in experimentally-driven attempts to locate the QCD critical point using heavy-ion collisions.

On an event-by-event basis each event may pass through the QCD phase diagram in radically different ways, even if hydrodynamics is only initialized at very low temperatures, as was shown here. Additionally, it was previously pointed out \cite{Feng:2018anl} that viscosity affects the time scale of the phase transition (across a first order line).   Instead, we suggest that one should think of observables that could tag individual events (or groups of events) by similar trajectories through the phase diagram in order to better understand the QCD equation of state at large baryon densities.

The next step in our future studies involves going beyond Bjorken flow, taking into account a more realistic spacetime evolution of the medium. This would then allow us to incorporate the effects of baryon diffusion, which would lead to further entropy production and likely cause an even larger divergence from isentropes. Additionally it has been shown that $\mu_B$ can vary with rapidity \cite{Brewer:2018abr,Li:2018ini} even at LHC collisions, so this would provide a new knob to turn in this type of analysis.  Further obvious extensions of this work would be to include multiple conserved charges, which has already been shown to shift the path of the isentropes even for ideal hydrodynamics \cite{Noronha-Hostler:2019ayj,Monnai:2019hkn}, and also critical fluctuations (although we believe that no consensus has yet been reached on the proper way to include them in state-of-the-art numerical relativistic viscous fluids).

This work also presents a direct challenge for the extraction of the QCD equation of state from relativistic heavy-ion collisions at large baryon densities. In fact, far-from-equilibrium effects are likely even larger at low beam energies (regardless if the degrees of freedom are hadrons or quarks/gluons), which makes previous claims of an EOS extracted from heavy-ion collisions probably unrealistic \cite{Danielewicz:2002pu} (especially considering this previous work assumed $T=0$ whereas these beam energies have now experimental evidence of temperatures greater than $T>70$ MeV \cite{Adamczewski-Musch:2019byl}).  Thus, heavy-ion collision constraints on the EOS can, at best, be applicable only to the neutron star mergers themselves (as was discussed extensively in \cite{Most:2018eaw,Most:2019onn,Adamczewski-Musch:2019byl}). This means that one should not use a heavy-ion extracted EOS, which includes temperature effects even at low center of mass collision energies, when placing constraints on the EOS of cold neutron stars (i.e. $T\sim 0$ MeV) \cite{Horowitz:2020evx} (for a detailed understanding of the EOS relevant for neutron stars see, for instance, \cite{Baym:2017whm}). 

While the original intention of our work was to focus on low energy heavy-ion collision energies relevant to the RHIC Beam Energy Scan, HADES, FAIR, and NICA, a similar study to show the connection between viscosity and the EOS may also be relevant to explore in neutron star mergers. Our results do indicate that the inclusion of viscosity can dramatically change the trajectories through the QCD phase diagram and it would be very interesting to see similar studies of this nature in neutron star mergers.  We emphasize here, however, that the bulk viscosity in neutron star mergers arises from weak interactions \cite{Kolomeitsev:2014gfa,Alford:2017rxf,Ofengeim:2019fjy,Alford:2019kdw,Alford:2019qtm} so their values and consequences are not the same as in the present study.  Additionally, it is not clear how the presence of general relativity would affect the overall evolution of the viscous fluid \cite{Bemfica:2019cop,Bemfica:2019knx} or the existence of an attractor, nor are we aware of estimates of the magnitude of the initial shear stress tensor or bulk pressure in realistic neutron star merger conditions, which could affect their trajectories across the low temperature, high baryon density region of the QCD phase diagram.

\section*{Acknowledgements}
Thus authors would like to thank Jorge Noronha, Claudia Ratti, Paolo Parotto, Chun Shen, and Michael Strickland for discussions and their insights into this work.
J.N.H. acknowledges support from the US-DOE Nuclear Science Grant No. DE-SC0019175 and the Alfred P. Sloan Foundation. E.M. was been supported by the
National Science Foundation via grant PHY-1560077. The authors also acknowledge support from the Illinois Campus Cluster, a computing resource that is operated by the Illinois Campus Cluster Program (ICCP) in conjunction with the National Center for Supercomputing Applications (NCSA), and which is supported by funds from the University of Illinois at Urbana-Champaign.

\appendix
\section{Israel-Stewart and $\dot{\beta}$ terms}\label{sec:dbetas}

\begin{figure}
    \centering
    \includegraphics[width=\linewidth]{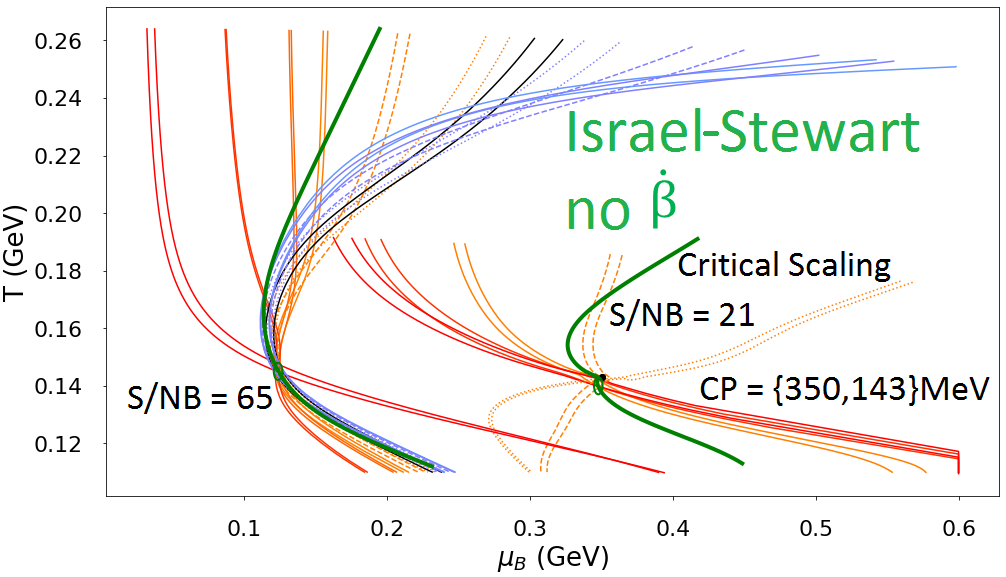}
    \caption{$\left\{T,\mu_B\right\}$ trajectories far from the critical point and at the critical point using Israel-Stewart equations of motion without the $\dot\beta$ terms.}
    \label{fig:nobeta}
\end{figure}

In the following section we will study the influence of the $\dot \beta$ terms in the Israel-Stewart equations of motion. We generally find that Israel-Stewart equations of motion without $\dot\beta$ terms leads to an extremely wide spread across $\mu_B$ for the initial conditions that freeze-out far from the critical point.  The initial conditions that start at large $\mu_B$ trajectories are in fact ones that would not be particularly atypical for heavy-ion collisions (they start with an initial bulk pressure $\Pi\leq 0$ and a positive contribution to $\pi^{\eta}_{\eta}\geq 0$). This demonstrates that even for high beam energies that initial conditions that begin at large $\mu_B$ may be needed. At the critical point, we are limited to only initial conditions that have a positive initial $\Pi$ and a negative initial $\pi^{\eta}_{\eta}$ because all other initial conditions would start at too large of $\mu_B$ for our EOS to handle.

\begin{figure*}
    \centering
    \begin{tabular}{c c}
    \includegraphics[width=0.5\linewidth]{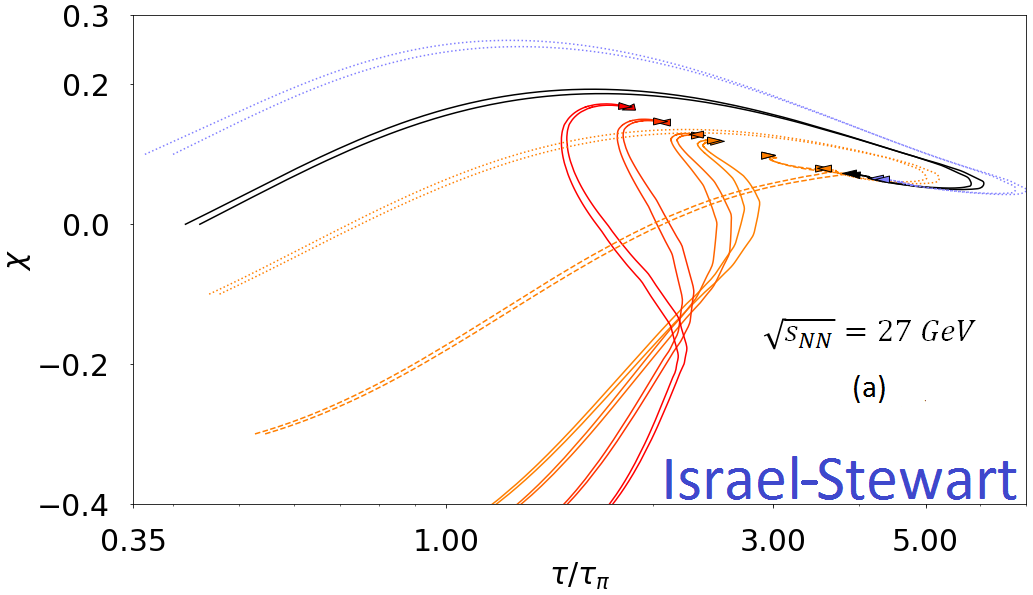} &  \includegraphics[width=0.5\linewidth]{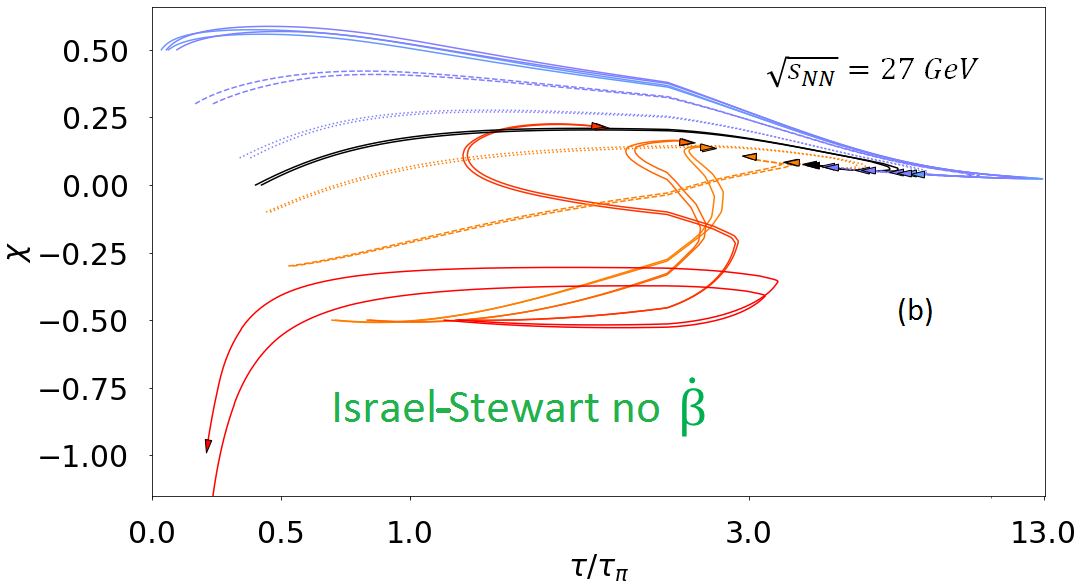}
    \end{tabular}
    \caption{Shear inverse Reynolds number $\chi=\pi^\eta_\eta/w$ trajectories far from the critical point using Israel-Stewart equations of motion with $\dot{\beta}$ terms in (a) and without $\dot{\beta}$ terms in (b).  The solid black lines assume that the initial $Re^{-1}=0$ for both shear and bulk (the band demonstrates the width of our range of the freeze-out $\left\{T,\mu_B\right\}$. }
    \label{fig:chi_far_dB}
\end{figure*}

\begin{figure*}
    \centering
    \begin{tabular}{c c}
    \includegraphics[width=0.5\linewidth]{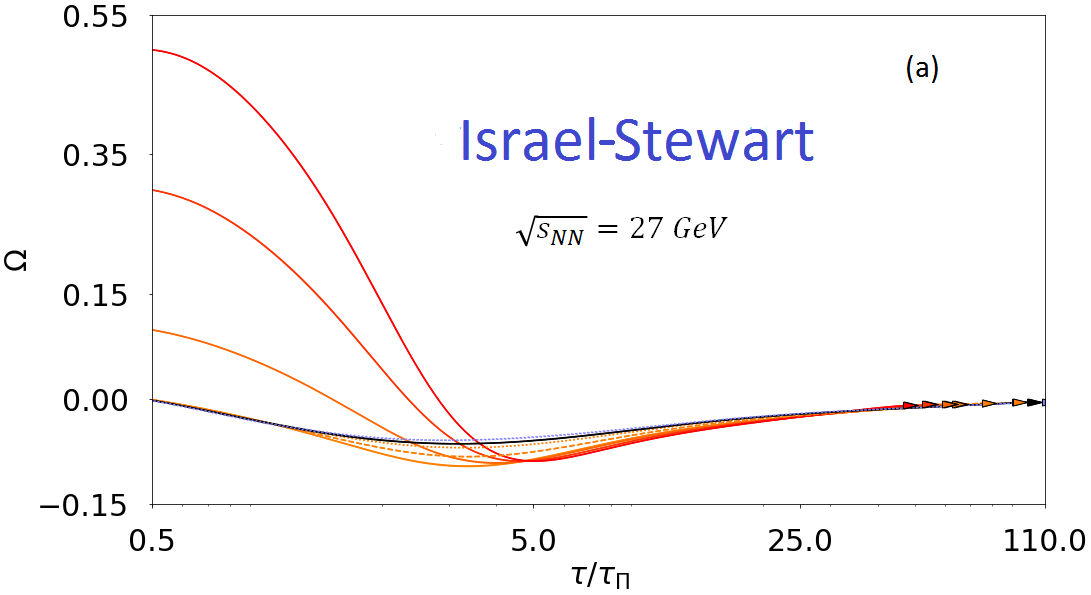} &  \includegraphics[width=0.5\linewidth]{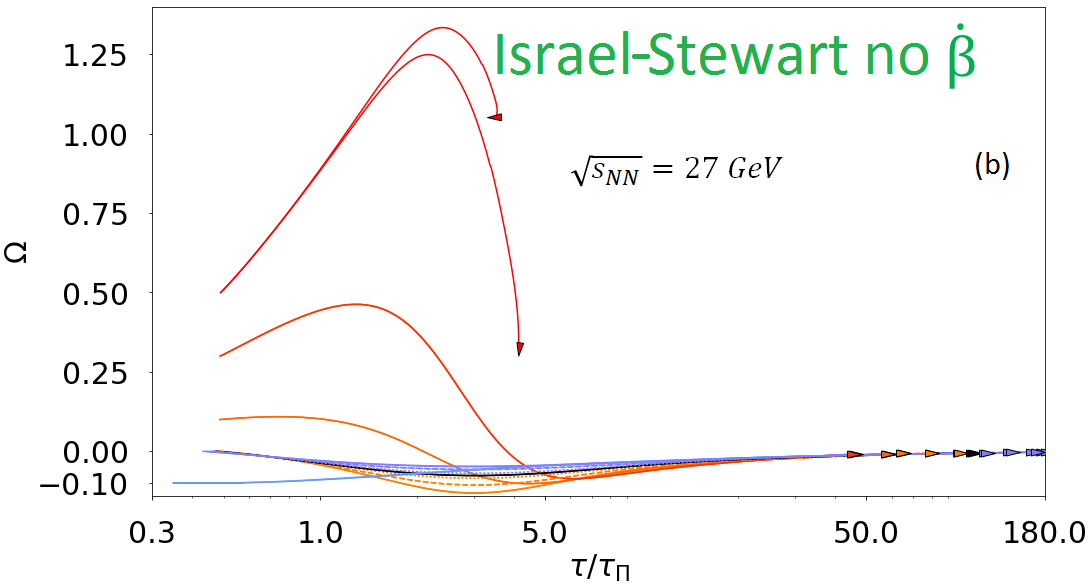}
    \end{tabular}
    \caption{Bulk inverse Reynolds number $\Omega=\Pi/w$ trajectories far from the critical point using Israel-Stewart equations of motion with $\dot{\beta}$ terms in (a) and without $\dot{\beta}$ terms in (b). The solid black lines assume that the initial $Re^{-1}=0$ for both shear and bulk (the band demonstrates the width of our range of the freeze-out $\left\{T,\mu_B\right\}$. }
    \label{fig:om_far_dB}
\end{figure*}

In Sec.\ \ref{sec:eqs} we explained that in the original Israel-Stewart paper \cite{Israel:1979wp} they neglected terms that incorporated the gradients of the temperature.  Below we study the effect of these terms and generally find that the inclusion of the $\dot{\beta}$ terms lead to better (and more well-behaved) inverse Reynolds numbers for both shear and bulk viscosity.

First, we explore the $Re^{-1}$ of shear and bulk viscosity far from the critical point (close to $\mu_B\rightarrow 0$).  The shear $Re^{-1}$ is shown in Fig.\ \ref{fig:chi_far_dB} with and without the $\dot{\beta}$ terms. In both cases we scale the time evolution by the shear relaxation time.  One can quickly see that the inclusion of $\dot{\beta}$ leads to a smaller range of $Re^{-1}$ numbers and that those $Re^{-1}$ appear to converge to a line on a relatively short time scale. The arrows at the end of the lines point in the direction of the derivative, which implies that given a long enough hydrodynamic expansion that they would eventually converge to a singular point.  We caution, though, that we stop our hydrodynamic expansion once the trajectories reach our freeze-out temperature and, therefore, it appears due to the limited run times of hydrodynamics in heavy-ion collisions at the beam energy scan that the time scales are not long enough to converge to a single point in $\chi$. 

In contrast, the $Re^{-1}$ of shear for Israel-Stewart without the $\dot{\beta}$ terms produces a much large $Re^{-1}$ and even has trajectories that appear to diverge in $\chi$ (the solid lines) becoming ever more negative with time. These trajectories are initialized to have a large, negative $\chi$ and a large positive $\Omega$.  

The bulk $Re^{-1}$, as shown in Fig.\ \ref{fig:om_far_dB}, does not appear to be as sensitive to the inclusion of $\dot{\beta}$ terms, which is likely because the $\zeta T/w$ is relatively small at initial times such that $\Omega$ quickly drops to a potential attractor.  However, even in the case of bulk viscosity, we find that the same extreme initial conditions (solid red line) that was problematic in Fig.\ \ref{fig:chi_far_dB} also produces a very large $Re^{-1}>1$ for the bulk viscosity.  Additionally, the lifetime of hydrodynamics is shorter than the other runs such that there is not enough time for $\Omega$ to reach the attractor.

Overall, we find that even far from the critical point, Israel-Stewart codes that neglect the $\dot{\beta}$ terms may run into problems for initial conditions that begin far from equilibrium and especially may see a shear stress tensor that has runaway behavior. This is bound to lead to causality problems \cite{Bemfica:2020xym}. Thus, any exploration of the QCD phase diagram using Israel-Stewart theory in the far from equilibrium regime should, at the bare minimum, include the $\dot{\beta}$ terms.

\begin{figure*}
    \centering
    \begin{tabular}{c c}
    \includegraphics[width=0.5\linewidth]{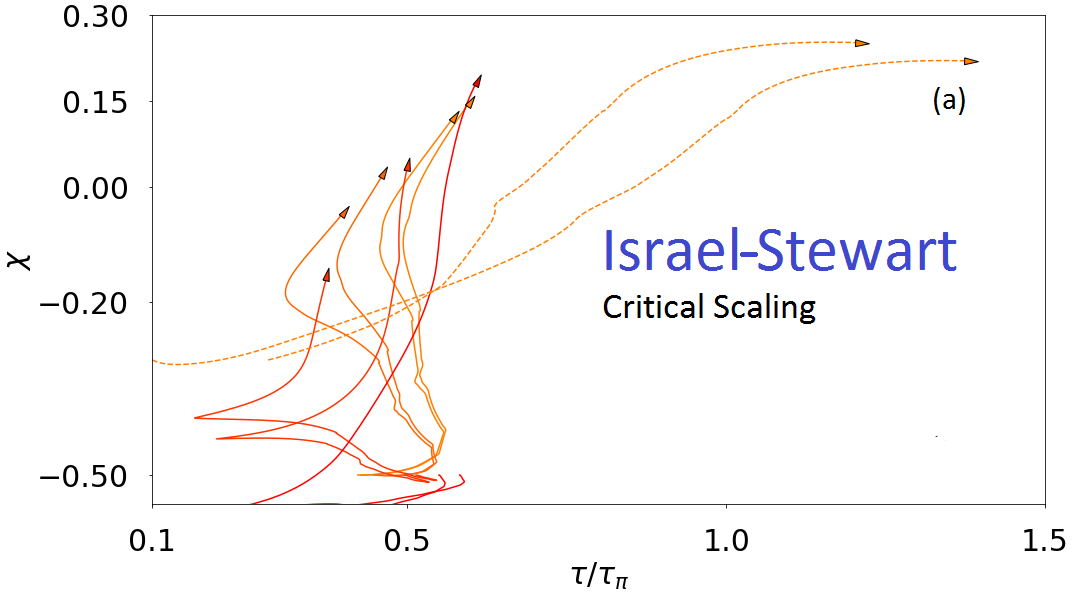} &  \includegraphics[width=0.5\linewidth]{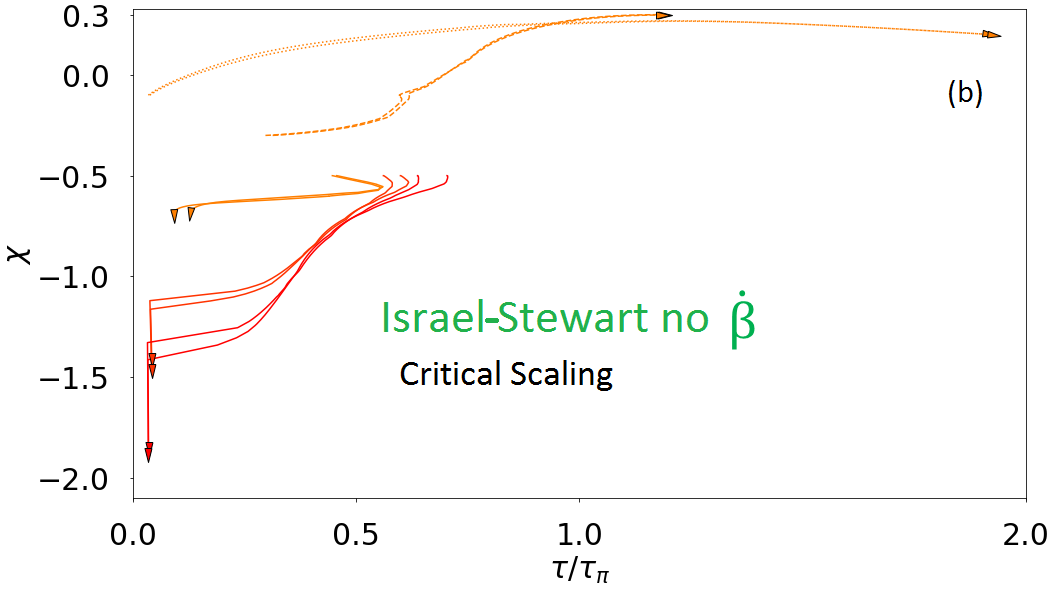}
    \end{tabular}
    \caption{Shear inverse Reynolds number $\chi=\pi^\eta_\eta/w$ trajectories far from the critical point using Israel-Stewart equations of motion with $\dot{\beta}$ terms in (a) and without $\dot{\beta}$ terms in (b).  Here only the critically scaled $\zeta T/w$ is considered. }
    \label{fig:chi_at_dB}
\end{figure*}

Next, we explore the influence of the inclusion of the $\dot{\beta}$ terms when the trajectories pass through (or very close) to the critical point. First we consider the $Re^{-1}$ for shear viscosity in Fig.\ \ref{fig:chi_at_dB}. When we include $\dot{\beta}$ terms, we see that $\chi$ appears to have some sort of universal line that all the trajectories are pointing towards.  We do find that the time scales are too short for the curves to truly converge but this hints that with the $\dot{\beta}$ terms one could reach an attractor if the time scales were long enough. Unlike in Fig.\ \ref{fig:chi_far_dB} where we found only extreme trajectories that diverged in $\chi$ as hydrodynamics evolved in time when $\dot{\beta}$ terms are excluded, at the critical point we find that {\it most} trajectories diverge at the critical point for $\chi$ if we neglect $\dot{\beta}$ terms.  This demonstrates the importance of using the full equations of motion for Israel-Stewart if one wants to study the QCD phase diagram, especially close to a phase transition.

\begin{figure*}
    \centering
    \begin{tabular}{c c}
    \includegraphics[width=0.5\linewidth]{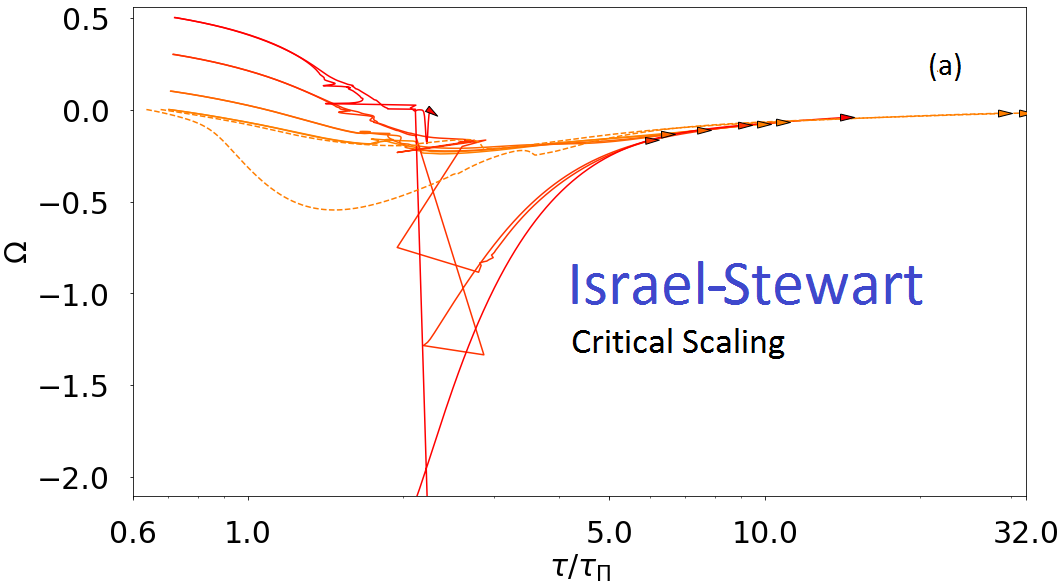} & 
    \includegraphics[width=0.5\linewidth]{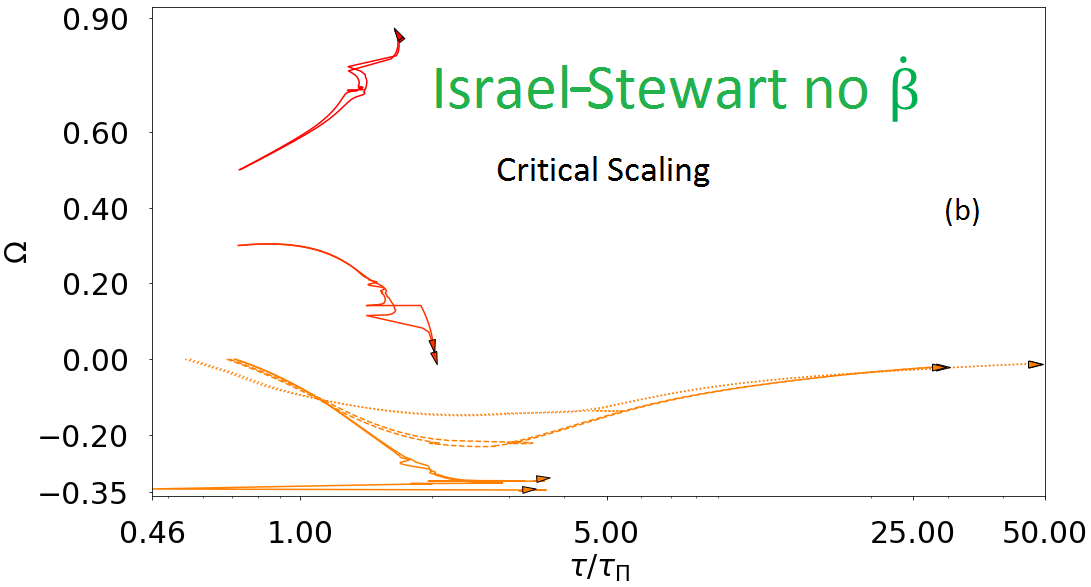}
    \end{tabular}
    \caption{Bulk inverse Reynolds number $\Omega=\Pi/w$ trajectories at the critical point using Israel-Stewart equations of motion with $\dot{\beta}$ terms in (a) and without $\dot{\beta}$ terms in (b).  Here only the critically scaled $\zeta T/w$ is considered. }
    \label{fig:om_CP_dB}
\end{figure*}

In this section, we only consider the critically scaled $\zeta T/w$ because we wanted to test the limits of Israel-Stewart with and without the $\dot{\beta}$ terms.  In Fig.\ \ref{fig:om_CP_dB} we plot the $Re^{-1}$ for the bulk viscosity with and without the $\dot{\beta}$ terms.  In both cases we can obtain very large values of $\Omega$ (in fact, much larger than DNMR) but it is clear from Fig.\ \ref{fig:om_CP_dB}  that while $\Omega$ briefly diverges as one crosses the critical point (due to the large value of $\zeta T/w$) with the inclusion of $\dot{\beta}$ terms, it quickly recovers and is able to return to the potential attractor very quickly.  In contrast, Israel-Stewart without $\dot{\beta}$ terms diverges in a multitude of directions and it is not clear if an attractor is obtained even for the few trajectories that do not diverge. Thus, we argue that Israel-Stewart without $\dot{\beta}$ terms should definitely not be used near a critical point, nor even when the system is far from equilibrium because it can lead to diverging solutions. 

\begin{figure*}
    \centering
    \begin{tabular}{c c}
    \includegraphics[width=0.5\linewidth]{tmutraj_Israel_Stewart_Dbetas_NoCS_Energy0_15_FO_0348,014.png} &  \includegraphics[width=0.5\linewidth]{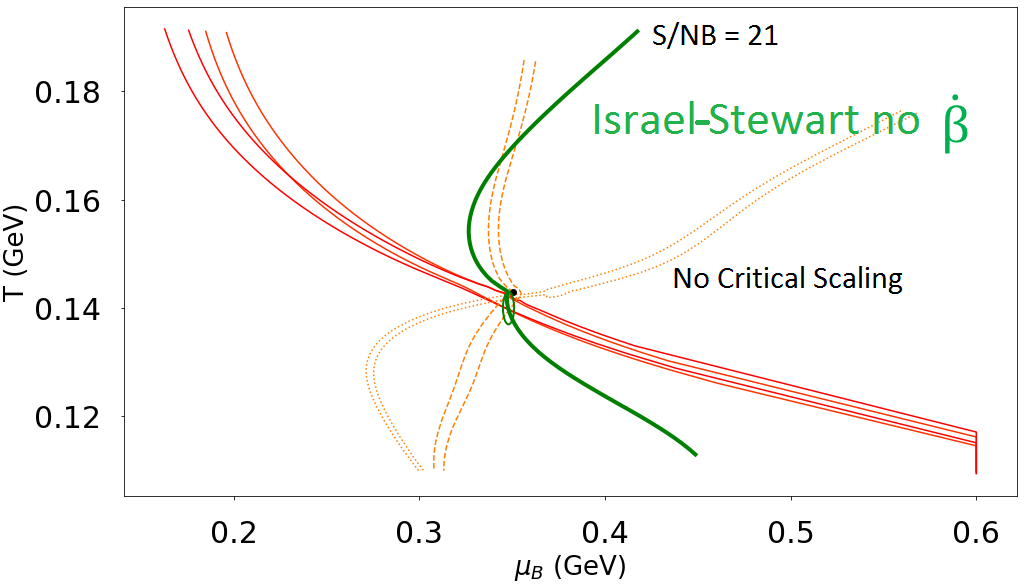}
    \end{tabular}
    \caption{$\left\{T,\mu_B\right\}$ trajectories at the critical point without critically scaled $\zeta T/w$ using Israel-Stewart equations of motion with $\dot{\beta}$ terms in (a) and without $\dot{\beta}$ terms in (b) }
    \label{fig:Tmu_dB}
\end{figure*}

Finally, in Fig.\ \ref{fig:Tmu_dB} we compare the trajectories of Israel-Stewart with and without the $\dot{\beta}$ terms at the critical point. Surprisingly enough, if one were only to look at the $\left\{T,\mu_B\right\}$ trajectories, it would appear that the equations of motions are well behaved at the critical point.  There are no particular red flags here, nor do they appear drastically different than the DNMR shown in Fig.\ \ref{fig:CS}.  Therefore, it is important to investigate the $Re^{-1}$ as one studies the QCD phase diagram to ensure that hydrodynamic is still applicable.  This is especially important when close to the critical point because transport coefficients from the dynamic universality class H diverge at the critical point \cite{Son:2004iv}.  Thus, this is a serious issue that realistic hydrodynamic models will need to contend with. 

One interesting consequence of the $\dot{\beta}$ terms is that they produce trajectories that appear to cross the critical point and pass into large regions of $\mu_B$ that may potentially remain in the deconfined phase.  It is difficult to know precisely what is happening in these low temperature regions because this is precisely the part of the phase diagram (large $\mu_B$ along the first order phase transition) where our EOS begins to break down.  However, this does raise the possibility that there may be trajectories that cross the critical point, but enter into deconfined matter and then cross over the first order phase transition line at lower temperatures/higher $\mu_B$.  The consequence of such a trajectory we leave for a future paper when we have an improved EOS where we can extend to large $\mu_B$'s to explore this part of the QCD phase diagram in more detail. 

\bibliography{BIG}

\begin{thebibliography}{175}%
\makeatletter
\providecommand \@ifxundefined [1]{%
 \@ifx{#1\undefined}
}%
\providecommand \@ifnum [1]{%
 \ifnum #1\expandafter \@firstoftwo
 \else \expandafter \@secondoftwo
 \fi
}%
\providecommand \@ifx [1]{%
 \ifx #1\expandafter \@firstoftwo
 \else \expandafter \@secondoftwo
 \fi
}%
\providecommand \natexlab [1]{#1}%
\providecommand \enquote  [1]{``#1''}%
\providecommand \bibnamefont  [1]{#1}%
\providecommand \bibfnamefont [1]{#1}%
\providecommand \citenamefont [1]{#1}%
\providecommand \href@noop [0]{\@secondoftwo}%
\providecommand \href [0]{\begingroup \@sanitize@url \@href}%
\providecommand \@href[1]{\@@startlink{#1}\@@href}%
\providecommand \@@href[1]{\endgroup#1\@@endlink}%
\providecommand \@sanitize@url [0]{\catcode `\\12\catcode `\$12\catcode
  `\&12\catcode `\#12\catcode `\^12\catcode `\_12\catcode `\%12\relax}%
\providecommand \@@startlink[1]{}%
\providecommand \@@endlink[0]{}%
\providecommand \url  [0]{\begingroup\@sanitize@url \@url }%
\providecommand \@url [1]{\endgroup\@href {#1}{\urlprefix }}%
\providecommand \urlprefix  [0]{URL }%
\providecommand \Eprint [0]{\href }%
\providecommand \doibase [0]{http://dx.doi.org/}%
\providecommand \selectlanguage [0]{\@gobble}%
\providecommand \bibinfo  [0]{\@secondoftwo}%
\providecommand \bibfield  [0]{\@secondoftwo}%
\providecommand \translation [1]{[#1]}%
\providecommand \BibitemOpen [0]{}%
\providecommand \bibitemStop [0]{}%
\providecommand \bibitemNoStop [0]{.\EOS\space}%
\providecommand \EOS [0]{\spacefactor3000\relax}%
\providecommand \BibitemShut  [1]{\csname bibitem#1\endcsname}%
\let\auto@bib@innerbib\@empty
\bibitem [{\citenamefont {Aoki}\ \emph {et~al.}(2006)\citenamefont {Aoki},
  \citenamefont {Endrodi}, \citenamefont {Fodor}, \citenamefont {Katz},\ and\
  \citenamefont {Szabo}}]{Aoki:2006we}%
  \BibitemOpen
  \bibfield  {author} {\bibinfo {author} {\bibfnamefont {Y.}~\bibnamefont
  {Aoki}}, \bibinfo {author} {\bibfnamefont {G.}~\bibnamefont {Endrodi}},
  \bibinfo {author} {\bibfnamefont {Z.}~\bibnamefont {Fodor}}, \bibinfo
  {author} {\bibfnamefont {S.~D.}\ \bibnamefont {Katz}}, \ and\ \bibinfo
  {author} {\bibfnamefont {K.~K.}\ \bibnamefont {Szabo}},\ }\href {\doibase
  10.1038/nature05120} {\bibfield  {journal} {\bibinfo  {journal} {Nature}\
  }\textbf {\bibinfo {volume} {443}},\ \bibinfo {pages} {675} (\bibinfo {year}
  {2006})},\ \Eprint {http://arxiv.org/abs/hep-lat/0611014}
  {arXiv:hep-lat/0611014 [hep-lat]} \BibitemShut {NoStop}%
\bibitem [{\citenamefont {Borsanyi}\ \emph {et~al.}(2010)\citenamefont
  {Borsanyi}, \citenamefont {Fodor}, \citenamefont {Hoelbling}, \citenamefont
  {Katz}, \citenamefont {Krieg}, \citenamefont {Ratti},\ and\ \citenamefont
  {Szabo}}]{Borsanyi:2010bp}%
  \BibitemOpen
  \bibfield  {author} {\bibinfo {author} {\bibfnamefont {S.}~\bibnamefont
  {Borsanyi}}, \bibinfo {author} {\bibfnamefont {Z.}~\bibnamefont {Fodor}},
  \bibinfo {author} {\bibfnamefont {C.}~\bibnamefont {Hoelbling}}, \bibinfo
  {author} {\bibfnamefont {S.~D.}\ \bibnamefont {Katz}}, \bibinfo {author}
  {\bibfnamefont {S.}~\bibnamefont {Krieg}}, \bibinfo {author} {\bibfnamefont
  {C.}~\bibnamefont {Ratti}}, \ and\ \bibinfo {author} {\bibfnamefont {K.~K.}\
  \bibnamefont {Szabo}} (\bibinfo {collaboration} {Wuppertal-Budapest}),\
  }\href {\doibase 10.1007/JHEP09(2010)073} {\bibfield  {journal} {\bibinfo
  {journal} {JHEP}\ }\textbf {\bibinfo {volume} {09}},\ \bibinfo {pages} {073}
  (\bibinfo {year} {2010})},\ \Eprint {http://arxiv.org/abs/1005.3508}
  {arXiv:1005.3508 [hep-lat]} \BibitemShut {NoStop}%
\bibitem [{\citenamefont {Bazavov}\ \emph {et~al.}(2012)\citenamefont {Bazavov}
  \emph {et~al.}}]{Bazavov:2011nk}%
  \BibitemOpen
  \bibfield  {author} {\bibinfo {author} {\bibfnamefont {A.}~\bibnamefont
  {Bazavov}} \emph {et~al.},\ }\href {\doibase 10.1103/PhysRevD.85.054503}
  {\bibfield  {journal} {\bibinfo  {journal} {Phys. Rev.}\ }\textbf {\bibinfo
  {volume} {D85}},\ \bibinfo {pages} {054503} (\bibinfo {year} {2012})},\
  \Eprint {http://arxiv.org/abs/1111.1710} {arXiv:1111.1710 [hep-lat]}
  \BibitemShut {NoStop}%
\bibitem [{\citenamefont {Halasz}\ \emph {et~al.}(1998)\citenamefont {Halasz},
  \citenamefont {Jackson}, \citenamefont {Shrock}, \citenamefont {Stephanov},\
  and\ \citenamefont {Verbaarschot}}]{Halasz:1998qr}%
  \BibitemOpen
  \bibfield  {author} {\bibinfo {author} {\bibfnamefont {A.~M.}\ \bibnamefont
  {Halasz}}, \bibinfo {author} {\bibfnamefont {A.}~\bibnamefont {Jackson}},
  \bibinfo {author} {\bibfnamefont {R.}~\bibnamefont {Shrock}}, \bibinfo
  {author} {\bibfnamefont {M.~A.}\ \bibnamefont {Stephanov}}, \ and\ \bibinfo
  {author} {\bibfnamefont {J.}~\bibnamefont {Verbaarschot}},\ }\href {\doibase
  10.1103/PhysRevD.58.096007} {\bibfield  {journal} {\bibinfo  {journal} {Phys.
  Rev. D}\ }\textbf {\bibinfo {volume} {58}},\ \bibinfo {pages} {096007}
  (\bibinfo {year} {1998})},\ \Eprint {http://arxiv.org/abs/hep-ph/9804290}
  {arXiv:hep-ph/9804290} \BibitemShut {NoStop}%
\bibitem [{\citenamefont {Stephanov}\ \emph {et~al.}(1998)\citenamefont
  {Stephanov}, \citenamefont {Rajagopal},\ and\ \citenamefont
  {Shuryak}}]{Stephanov:1998dy}%
  \BibitemOpen
  \bibfield  {author} {\bibinfo {author} {\bibfnamefont {M.~A.}\ \bibnamefont
  {Stephanov}}, \bibinfo {author} {\bibfnamefont {K.}~\bibnamefont
  {Rajagopal}}, \ and\ \bibinfo {author} {\bibfnamefont {E.~V.}\ \bibnamefont
  {Shuryak}},\ }\href {\doibase 10.1103/PhysRevLett.81.4816} {\bibfield
  {journal} {\bibinfo  {journal} {Phys. Rev. Lett.}\ }\textbf {\bibinfo
  {volume} {81}},\ \bibinfo {pages} {4816} (\bibinfo {year} {1998})},\ \Eprint
  {http://arxiv.org/abs/hep-ph/9806219} {arXiv:hep-ph/9806219} \BibitemShut
  {NoStop}%
\bibitem [{\citenamefont {Stephanov}\ \emph {et~al.}(1999)\citenamefont
  {Stephanov}, \citenamefont {Rajagopal},\ and\ \citenamefont
  {Shuryak}}]{Stephanov:1999zu}%
  \BibitemOpen
  \bibfield  {author} {\bibinfo {author} {\bibfnamefont {M.~A.}\ \bibnamefont
  {Stephanov}}, \bibinfo {author} {\bibfnamefont {K.}~\bibnamefont
  {Rajagopal}}, \ and\ \bibinfo {author} {\bibfnamefont {E.~V.}\ \bibnamefont
  {Shuryak}},\ }\href {\doibase 10.1103/PhysRevD.60.114028} {\bibfield
  {journal} {\bibinfo  {journal} {Phys. Rev. D}\ }\textbf {\bibinfo {volume}
  {60}},\ \bibinfo {pages} {114028} (\bibinfo {year} {1999})},\ \Eprint
  {http://arxiv.org/abs/hep-ph/9903292} {arXiv:hep-ph/9903292} \BibitemShut
  {NoStop}%
\bibitem [{\citenamefont {Dexheimer}\ and\ \citenamefont
  {Schramm}(2010)}]{Dexheimer:2009hi}%
  \BibitemOpen
  \bibfield  {author} {\bibinfo {author} {\bibfnamefont {V.~A.}\ \bibnamefont
  {Dexheimer}}\ and\ \bibinfo {author} {\bibfnamefont {S.}~\bibnamefont
  {Schramm}},\ }\href {\doibase 10.1103/PhysRevC.81.045201} {\bibfield
  {journal} {\bibinfo  {journal} {Phys. Rev.}\ }\textbf {\bibinfo {volume}
  {C81}},\ \bibinfo {pages} {045201} (\bibinfo {year} {2010})},\ \Eprint
  {http://arxiv.org/abs/0901.1748} {arXiv:0901.1748 [astro-ph.SR]} \BibitemShut
  {NoStop}%
\bibitem [{\citenamefont {Critelli}\ \emph {et~al.}(2017)\citenamefont
  {Critelli}, \citenamefont {Noronha}, \citenamefont {Noronha-Hostler},
  \citenamefont {Portillo}, \citenamefont {Ratti},\ and\ \citenamefont
  {Rougemont}}]{Critelli:2017oub}%
  \BibitemOpen
  \bibfield  {author} {\bibinfo {author} {\bibfnamefont {R.}~\bibnamefont
  {Critelli}}, \bibinfo {author} {\bibfnamefont {J.}~\bibnamefont {Noronha}},
  \bibinfo {author} {\bibfnamefont {J.}~\bibnamefont {Noronha-Hostler}},
  \bibinfo {author} {\bibfnamefont {I.}~\bibnamefont {Portillo}}, \bibinfo
  {author} {\bibfnamefont {C.}~\bibnamefont {Ratti}}, \ and\ \bibinfo {author}
  {\bibfnamefont {R.}~\bibnamefont {Rougemont}},\ }\href {\doibase
  10.1103/PhysRevD.96.096026} {\bibfield  {journal} {\bibinfo  {journal} {Phys.
  Rev.}\ }\textbf {\bibinfo {volume} {D96}},\ \bibinfo {pages} {096026}
  (\bibinfo {year} {2017})},\ \Eprint {http://arxiv.org/abs/1706.00455}
  {arXiv:1706.00455 [nucl-th]} \BibitemShut {NoStop}%
\bibitem [{\citenamefont {Fan}\ \emph {et~al.}(2017)\citenamefont {Fan},
  \citenamefont {Luo},\ and\ \citenamefont {Zong}}]{Fan:2016ovc}%
  \BibitemOpen
  \bibfield  {author} {\bibinfo {author} {\bibfnamefont {W.}~\bibnamefont
  {Fan}}, \bibinfo {author} {\bibfnamefont {X.}~\bibnamefont {Luo}}, \ and\
  \bibinfo {author} {\bibfnamefont {H.-S.}\ \bibnamefont {Zong}},\ }\href
  {\doibase 10.1142/S0217751X17500610} {\bibfield  {journal} {\bibinfo
  {journal} {Int. J. Mod. Phys.}\ }\textbf {\bibinfo {volume} {A32}},\ \bibinfo
  {pages} {1750061} (\bibinfo {year} {2017})},\ \Eprint
  {http://arxiv.org/abs/1608.07903} {arXiv:1608.07903 [hep-ph]} \BibitemShut
  {NoStop}%
\bibitem [{\citenamefont {Fu}\ \emph {et~al.}(2019)\citenamefont {Fu},
  \citenamefont {Pawlowski},\ and\ \citenamefont {Rennecke}}]{Fu:2019hdw}%
  \BibitemOpen
  \bibfield  {author} {\bibinfo {author} {\bibfnamefont {W.-j.}\ \bibnamefont
  {Fu}}, \bibinfo {author} {\bibfnamefont {J.~M.}\ \bibnamefont {Pawlowski}}, \
  and\ \bibinfo {author} {\bibfnamefont {F.}~\bibnamefont {Rennecke}},\
  }\href@noop {} {\  (\bibinfo {year} {2019})},\ \Eprint
  {http://arxiv.org/abs/1909.02991} {arXiv:1909.02991 [hep-ph]} \BibitemShut
  {NoStop}%
\bibitem [{\citenamefont {Motornenko}\ \emph {et~al.}(2020)\citenamefont
  {Motornenko}, \citenamefont {Steinheimer}, \citenamefont {Vovchenko},
  \citenamefont {Schramm},\ and\ \citenamefont
  {Stoecker}}]{Motornenko:2019arp}%
  \BibitemOpen
  \bibfield  {author} {\bibinfo {author} {\bibfnamefont {A.}~\bibnamefont
  {Motornenko}}, \bibinfo {author} {\bibfnamefont {J.}~\bibnamefont
  {Steinheimer}}, \bibinfo {author} {\bibfnamefont {V.}~\bibnamefont
  {Vovchenko}}, \bibinfo {author} {\bibfnamefont {S.}~\bibnamefont {Schramm}},
  \ and\ \bibinfo {author} {\bibfnamefont {H.}~\bibnamefont {Stoecker}},\
  }\href {\doibase 10.1103/PhysRevC.101.034904} {\bibfield  {journal} {\bibinfo
   {journal} {Phys. Rev.}\ }\textbf {\bibinfo {volume} {C101}},\ \bibinfo
  {pages} {034904} (\bibinfo {year} {2020})},\ \Eprint
  {http://arxiv.org/abs/1905.00866} {arXiv:1905.00866 [hep-ph]} \BibitemShut
  {NoStop}%
\bibitem [{\citenamefont {Annala}\ \emph {et~al.}(2020)\citenamefont {Annala},
  \citenamefont {Gorda}, \citenamefont {Kurkela}, \citenamefont {Nättilä},\
  and\ \citenamefont {Vuorinen}}]{Annala:2019puf}%
  \BibitemOpen
  \bibfield  {author} {\bibinfo {author} {\bibfnamefont {E.}~\bibnamefont
  {Annala}}, \bibinfo {author} {\bibfnamefont {T.}~\bibnamefont {Gorda}},
  \bibinfo {author} {\bibfnamefont {A.}~\bibnamefont {Kurkela}}, \bibinfo
  {author} {\bibfnamefont {J.}~\bibnamefont {Nättilä}}, \ and\ \bibinfo
  {author} {\bibfnamefont {A.}~\bibnamefont {Vuorinen}},\ }\href {\doibase
  10.1038/s41567-020-0914-9} {\bibfield  {journal} {\bibinfo  {journal} {Nature
  Phys.}\ } (\bibinfo {year} {2020}),\ 10.1038/s41567-020-0914-9},\ \Eprint
  {http://arxiv.org/abs/1903.09121} {arXiv:1903.09121 [astro-ph.HE]}
  \BibitemShut {NoStop}%
\bibitem [{\citenamefont {Tan}\ \emph {et~al.}(2020)\citenamefont {Tan},
  \citenamefont {Noronha-Hostler},\ and\ \citenamefont {Yunes}}]{Tan:2020ics}%
  \BibitemOpen
  \bibfield  {author} {\bibinfo {author} {\bibfnamefont {H.}~\bibnamefont
  {Tan}}, \bibinfo {author} {\bibfnamefont {J.}~\bibnamefont
  {Noronha-Hostler}}, \ and\ \bibinfo {author} {\bibfnamefont {N.}~\bibnamefont
  {Yunes}},\ }\href@noop {} {\  (\bibinfo {year} {2020})},\ \Eprint
  {http://arxiv.org/abs/2006.16296} {arXiv:2006.16296 [astro-ph.HE]}
  \BibitemShut {NoStop}%
\bibitem [{\citenamefont {Bedaque}\ and\ \citenamefont
  {Steiner}(2015)}]{Bedaque:2014sqa}%
  \BibitemOpen
  \bibfield  {author} {\bibinfo {author} {\bibfnamefont {P.}~\bibnamefont
  {Bedaque}}\ and\ \bibinfo {author} {\bibfnamefont {A.~W.}\ \bibnamefont
  {Steiner}},\ }\href {\doibase 10.1103/PhysRevLett.114.031103} {\bibfield
  {journal} {\bibinfo  {journal} {Phys. Rev. Lett.}\ }\textbf {\bibinfo
  {volume} {114}},\ \bibinfo {pages} {031103} (\bibinfo {year} {2015})},\
  \Eprint {http://arxiv.org/abs/1408.5116} {arXiv:1408.5116 [nucl-th]}
  \BibitemShut {NoStop}%
\bibitem [{\citenamefont {Alford}\ \emph {et~al.}(2013)\citenamefont {Alford},
  \citenamefont {Han},\ and\ \citenamefont {Prakash}}]{Alford:2013aca}%
  \BibitemOpen
  \bibfield  {author} {\bibinfo {author} {\bibfnamefont {M.~G.}\ \bibnamefont
  {Alford}}, \bibinfo {author} {\bibfnamefont {S.}~\bibnamefont {Han}}, \ and\
  \bibinfo {author} {\bibfnamefont {M.}~\bibnamefont {Prakash}},\ }\href
  {\doibase 10.1103/PhysRevD.88.083013} {\bibfield  {journal} {\bibinfo
  {journal} {Phys. Rev.}\ }\textbf {\bibinfo {volume} {D88}},\ \bibinfo {pages}
  {083013} (\bibinfo {year} {2013})},\ \Eprint {http://arxiv.org/abs/1302.4732}
  {arXiv:1302.4732 [astro-ph.SR]} \BibitemShut {NoStop}%
\bibitem [{\citenamefont {Dexheimer}\ \emph {et~al.}(2015)\citenamefont
  {Dexheimer}, \citenamefont {Negreiros},\ and\ \citenamefont
  {Schramm}}]{Dexheimer:2014pea}%
  \BibitemOpen
  \bibfield  {author} {\bibinfo {author} {\bibfnamefont {V.}~\bibnamefont
  {Dexheimer}}, \bibinfo {author} {\bibfnamefont {R.}~\bibnamefont
  {Negreiros}}, \ and\ \bibinfo {author} {\bibfnamefont {S.}~\bibnamefont
  {Schramm}},\ }\href {\doibase 10.1103/PhysRevC.91.055808} {\bibfield
  {journal} {\bibinfo  {journal} {Phys. Rev.}\ }\textbf {\bibinfo {volume}
  {C91}},\ \bibinfo {pages} {055808} (\bibinfo {year} {2015})},\ \Eprint
  {http://arxiv.org/abs/1411.4623} {arXiv:1411.4623 [astro-ph.HE]} \BibitemShut
  {NoStop}%
\bibitem [{\citenamefont {Benic}\ \emph {et~al.}(2015)\citenamefont {Benic},
  \citenamefont {Blaschke}, \citenamefont {Alvarez-Castillo}, \citenamefont
  {Fischer},\ and\ \citenamefont {Typel}}]{Benic:2014jia}%
  \BibitemOpen
  \bibfield  {author} {\bibinfo {author} {\bibfnamefont {S.}~\bibnamefont
  {Benic}}, \bibinfo {author} {\bibfnamefont {D.}~\bibnamefont {Blaschke}},
  \bibinfo {author} {\bibfnamefont {D.~E.}\ \bibnamefont {Alvarez-Castillo}},
  \bibinfo {author} {\bibfnamefont {T.}~\bibnamefont {Fischer}}, \ and\
  \bibinfo {author} {\bibfnamefont {S.}~\bibnamefont {Typel}},\ }\href
  {\doibase 10.1051/0004-6361/201425318} {\bibfield  {journal} {\bibinfo
  {journal} {Astron. Astrophys.}\ }\textbf {\bibinfo {volume} {577}},\ \bibinfo
  {pages} {A40} (\bibinfo {year} {2015})},\ \Eprint
  {http://arxiv.org/abs/1411.2856} {arXiv:1411.2856 [astro-ph.HE]} \BibitemShut
  {NoStop}%
\bibitem [{\citenamefont {Montana}\ \emph {et~al.}(2019)\citenamefont
  {Montana}, \citenamefont {Tolos}, \citenamefont {Hanauske},\ and\
  \citenamefont {Rezzolla}}]{Montana:2018bkb}%
  \BibitemOpen
  \bibfield  {author} {\bibinfo {author} {\bibfnamefont {G.}~\bibnamefont
  {Montana}}, \bibinfo {author} {\bibfnamefont {L.}~\bibnamefont {Tolos}},
  \bibinfo {author} {\bibfnamefont {M.}~\bibnamefont {Hanauske}}, \ and\
  \bibinfo {author} {\bibfnamefont {L.}~\bibnamefont {Rezzolla}},\ }\href
  {\doibase 10.1103/PhysRevD.99.103009} {\bibfield  {journal} {\bibinfo
  {journal} {Phys. Rev.}\ }\textbf {\bibinfo {volume} {D99}},\ \bibinfo {pages}
  {103009} (\bibinfo {year} {2019})},\ \Eprint
  {http://arxiv.org/abs/1811.10929} {arXiv:1811.10929 [astro-ph.HE]}
  \BibitemShut {NoStop}%
\bibitem [{\citenamefont {Most}\ \emph {et~al.}(2019)\citenamefont {Most},
  \citenamefont {Papenfort}, \citenamefont {Dexheimer}, \citenamefont
  {Hanauske}, \citenamefont {Schramm}, \citenamefont {Stöcker},\ and\
  \citenamefont {Rezzolla}}]{Most:2018eaw}%
  \BibitemOpen
  \bibfield  {author} {\bibinfo {author} {\bibfnamefont {E.~R.}\ \bibnamefont
  {Most}}, \bibinfo {author} {\bibfnamefont {L.~J.}\ \bibnamefont {Papenfort}},
  \bibinfo {author} {\bibfnamefont {V.}~\bibnamefont {Dexheimer}}, \bibinfo
  {author} {\bibfnamefont {M.}~\bibnamefont {Hanauske}}, \bibinfo {author}
  {\bibfnamefont {S.}~\bibnamefont {Schramm}}, \bibinfo {author} {\bibfnamefont
  {H.}~\bibnamefont {Stöcker}}, \ and\ \bibinfo {author} {\bibfnamefont
  {L.}~\bibnamefont {Rezzolla}},\ }\href {\doibase
  10.1103/PhysRevLett.122.061101} {\bibfield  {journal} {\bibinfo  {journal}
  {Phys. Rev. Lett.}\ }\textbf {\bibinfo {volume} {122}},\ \bibinfo {pages}
  {061101} (\bibinfo {year} {2019})},\ \Eprint
  {http://arxiv.org/abs/1807.03684} {arXiv:1807.03684 [astro-ph.HE]}
  \BibitemShut {NoStop}%
\bibitem [{\citenamefont {Zha}\ \emph {et~al.}(2020)\citenamefont {Zha},
  \citenamefont {O'Connor}, \citenamefont {Chu}, \citenamefont {Lin},\ and\
  \citenamefont {Couch}}]{Zha:2020gjw}%
  \BibitemOpen
  \bibfield  {author} {\bibinfo {author} {\bibfnamefont {S.}~\bibnamefont
  {Zha}}, \bibinfo {author} {\bibfnamefont {E.~P.}\ \bibnamefont {O'Connor}},
  \bibinfo {author} {\bibfnamefont {M.-c.}\ \bibnamefont {Chu}}, \bibinfo
  {author} {\bibfnamefont {L.-M.}\ \bibnamefont {Lin}}, \ and\ \bibinfo
  {author} {\bibfnamefont {S.~M.}\ \bibnamefont {Couch}},\ }\href@noop {} {\
  (\bibinfo {year} {2020})},\ \Eprint {http://arxiv.org/abs/2007.04716}
  {arXiv:2007.04716 [astro-ph.HE]} \BibitemShut {NoStop}%
\bibitem [{\citenamefont {Adam}\ \emph {et~al.}(2020)\citenamefont {Adam} \emph
  {et~al.}}]{Adam:2020unf}%
  \BibitemOpen
  \bibfield  {author} {\bibinfo {author} {\bibfnamefont {J.}~\bibnamefont
  {Adam}} \emph {et~al.} (\bibinfo {collaboration} {STAR}),\ }\href@noop {} {\
  (\bibinfo {year} {2020})},\ \Eprint {http://arxiv.org/abs/2001.02852}
  {arXiv:2001.02852 [nucl-ex]} \BibitemShut {NoStop}%
\bibitem [{\citenamefont {Adamczewski-Musch}\ \emph {et~al.}(2020)\citenamefont
  {Adamczewski-Musch} \emph {et~al.}}]{Adamczewski-Musch:2020slf}%
  \BibitemOpen
  \bibfield  {author} {\bibinfo {author} {\bibfnamefont {J.}~\bibnamefont
  {Adamczewski-Musch}} \emph {et~al.} (\bibinfo {collaboration} {HADES}),\
  }\href@noop {} {\  (\bibinfo {year} {2020})},\ \Eprint
  {http://arxiv.org/abs/2002.08701} {arXiv:2002.08701 [nucl-ex]} \BibitemShut
  {NoStop}%
\bibitem [{\citenamefont {Stephanov}(2011)}]{Stephanov:2011pb}%
  \BibitemOpen
  \bibfield  {author} {\bibinfo {author} {\bibfnamefont {M.~A.}\ \bibnamefont
  {Stephanov}},\ }\href {\doibase 10.1103/PhysRevLett.107.052301} {\bibfield
  {journal} {\bibinfo  {journal} {Phys. Rev. Lett.}\ }\textbf {\bibinfo
  {volume} {107}},\ \bibinfo {pages} {052301} (\bibinfo {year} {2011})},\
  \Eprint {http://arxiv.org/abs/1104.1627} {arXiv:1104.1627 [hep-ph]}
  \BibitemShut {NoStop}%
\bibitem [{\citenamefont {Bzdak}\ and\ \citenamefont
  {Koch}(2012)}]{Bzdak:2012ab}%
  \BibitemOpen
  \bibfield  {author} {\bibinfo {author} {\bibfnamefont {A.}~\bibnamefont
  {Bzdak}}\ and\ \bibinfo {author} {\bibfnamefont {V.}~\bibnamefont {Koch}},\
  }\href {\doibase 10.1103/PhysRevC.86.044904} {\bibfield  {journal} {\bibinfo
  {journal} {Phys. Rev.}\ }\textbf {\bibinfo {volume} {C86}},\ \bibinfo {pages}
  {044904} (\bibinfo {year} {2012})},\ \Eprint {http://arxiv.org/abs/1206.4286}
  {arXiv:1206.4286 [nucl-th]} \BibitemShut {NoStop}%
\bibitem [{\citenamefont {Bzdak}\ \emph {et~al.}(2013)\citenamefont {Bzdak},
  \citenamefont {Koch},\ and\ \citenamefont {Skokov}}]{Bzdak:2012an}%
  \BibitemOpen
  \bibfield  {author} {\bibinfo {author} {\bibfnamefont {A.}~\bibnamefont
  {Bzdak}}, \bibinfo {author} {\bibfnamefont {V.}~\bibnamefont {Koch}}, \ and\
  \bibinfo {author} {\bibfnamefont {V.}~\bibnamefont {Skokov}},\ }\href
  {\doibase 10.1103/PhysRevC.87.014901} {\bibfield  {journal} {\bibinfo
  {journal} {Phys. Rev.}\ }\textbf {\bibinfo {volume} {C87}},\ \bibinfo {pages}
  {014901} (\bibinfo {year} {2013})},\ \Eprint {http://arxiv.org/abs/1203.4529}
  {arXiv:1203.4529 [hep-ph]} \BibitemShut {NoStop}%
\bibitem [{\citenamefont {Bzdak}\ and\ \citenamefont
  {Koch}(2015)}]{Bzdak:2013pha}%
  \BibitemOpen
  \bibfield  {author} {\bibinfo {author} {\bibfnamefont {A.}~\bibnamefont
  {Bzdak}}\ and\ \bibinfo {author} {\bibfnamefont {V.}~\bibnamefont {Koch}},\
  }\href {\doibase 10.1103/PhysRevC.91.027901} {\bibfield  {journal} {\bibinfo
  {journal} {Phys. Rev.}\ }\textbf {\bibinfo {volume} {C91}},\ \bibinfo {pages}
  {027901} (\bibinfo {year} {2015})},\ \Eprint {http://arxiv.org/abs/1312.4574}
  {arXiv:1312.4574 [nucl-th]} \BibitemShut {NoStop}%
\bibitem [{\citenamefont {Hippert}\ \emph {et~al.}(2016)\citenamefont
  {Hippert}, \citenamefont {Fraga},\ and\ \citenamefont
  {Santos}}]{Hippert:2015rwa}%
  \BibitemOpen
  \bibfield  {author} {\bibinfo {author} {\bibfnamefont {M.}~\bibnamefont
  {Hippert}}, \bibinfo {author} {\bibfnamefont {E.~S.}\ \bibnamefont {Fraga}},
  \ and\ \bibinfo {author} {\bibfnamefont {E.~M.}\ \bibnamefont {Santos}},\
  }\href {\doibase 10.1103/PhysRevD.93.014029} {\bibfield  {journal} {\bibinfo
  {journal} {Phys. Rev. D}\ }\textbf {\bibinfo {volume} {93}},\ \bibinfo
  {pages} {014029} (\bibinfo {year} {2016})},\ \Eprint
  {http://arxiv.org/abs/1507.04764} {arXiv:1507.04764 [hep-ph]} \BibitemShut
  {NoStop}%
\bibitem [{\citenamefont {Steinheimer}\ \emph {et~al.}(2018)\citenamefont
  {Steinheimer}, \citenamefont {Vovchenko}, \citenamefont {Aichelin},
  \citenamefont {Bleicher},\ and\ \citenamefont
  {Stöcker}}]{Steinheimer:2016cir}%
  \BibitemOpen
  \bibfield  {author} {\bibinfo {author} {\bibfnamefont {J.}~\bibnamefont
  {Steinheimer}}, \bibinfo {author} {\bibfnamefont {V.}~\bibnamefont
  {Vovchenko}}, \bibinfo {author} {\bibfnamefont {J.}~\bibnamefont {Aichelin}},
  \bibinfo {author} {\bibfnamefont {M.}~\bibnamefont {Bleicher}}, \ and\
  \bibinfo {author} {\bibfnamefont {H.}~\bibnamefont {Stöcker}},\ }\href
  {\doibase 10.1016/j.physletb.2017.11.012} {\bibfield  {journal} {\bibinfo
  {journal} {Phys. Lett. B}\ }\textbf {\bibinfo {volume} {776}},\ \bibinfo
  {pages} {32} (\bibinfo {year} {2018})},\ \Eprint
  {http://arxiv.org/abs/1608.03737} {arXiv:1608.03737 [nucl-th]} \BibitemShut
  {NoStop}%
\bibitem [{\citenamefont {Bluhm}\ \emph {et~al.}(2017)\citenamefont {Bluhm},
  \citenamefont {Nahrgang}, \citenamefont {Bass},\ and\ \citenamefont
  {Schaefer}}]{Bluhm:2016byc}%
  \BibitemOpen
  \bibfield  {author} {\bibinfo {author} {\bibfnamefont {M.}~\bibnamefont
  {Bluhm}}, \bibinfo {author} {\bibfnamefont {M.}~\bibnamefont {Nahrgang}},
  \bibinfo {author} {\bibfnamefont {S.~A.}\ \bibnamefont {Bass}}, \ and\
  \bibinfo {author} {\bibfnamefont {T.}~\bibnamefont {Schaefer}},\ }\href
  {\doibase 10.1140/epjc/s10052-017-4771-3} {\bibfield  {journal} {\bibinfo
  {journal} {Eur. Phys. J. C}\ }\textbf {\bibinfo {volume} {77}},\ \bibinfo
  {pages} {210} (\bibinfo {year} {2017})},\ \Eprint
  {http://arxiv.org/abs/1612.03889} {arXiv:1612.03889 [nucl-th]} \BibitemShut
  {NoStop}%
\bibitem [{\citenamefont {Sombun}\ \emph {et~al.}(2018)\citenamefont {Sombun},
  \citenamefont {Steinheimer}, \citenamefont {Herold}, \citenamefont
  {Limphirat}, \citenamefont {Yan},\ and\ \citenamefont
  {Bleicher}}]{Sombun:2017bxi}%
  \BibitemOpen
  \bibfield  {author} {\bibinfo {author} {\bibfnamefont {S.}~\bibnamefont
  {Sombun}}, \bibinfo {author} {\bibfnamefont {J.}~\bibnamefont {Steinheimer}},
  \bibinfo {author} {\bibfnamefont {C.}~\bibnamefont {Herold}}, \bibinfo
  {author} {\bibfnamefont {A.}~\bibnamefont {Limphirat}}, \bibinfo {author}
  {\bibfnamefont {Y.}~\bibnamefont {Yan}}, \ and\ \bibinfo {author}
  {\bibfnamefont {M.}~\bibnamefont {Bleicher}},\ }\href {\doibase
  10.1088/1361-6471/aa9c6c} {\bibfield  {journal} {\bibinfo  {journal} {J.
  Phys.}\ }\textbf {\bibinfo {volume} {G45}},\ \bibinfo {pages} {025101}
  (\bibinfo {year} {2018})},\ \Eprint {http://arxiv.org/abs/1709.00879}
  {arXiv:1709.00879 [nucl-th]} \BibitemShut {NoStop}%
\bibitem [{\citenamefont {Hippert}\ and\ \citenamefont
  {Fraga}(2017)}]{Hippert:2017xoj}%
  \BibitemOpen
  \bibfield  {author} {\bibinfo {author} {\bibfnamefont {M.}~\bibnamefont
  {Hippert}}\ and\ \bibinfo {author} {\bibfnamefont {E.~S.}\ \bibnamefont
  {Fraga}},\ }\href {\doibase 10.1103/PhysRevD.96.034011} {\bibfield  {journal}
  {\bibinfo  {journal} {Phys. Rev. D}\ }\textbf {\bibinfo {volume} {96}},\
  \bibinfo {pages} {034011} (\bibinfo {year} {2017})},\ \Eprint
  {http://arxiv.org/abs/1702.02028} {arXiv:1702.02028 [hep-ph]} \BibitemShut
  {NoStop}%
\bibitem [{\citenamefont {Agah~Nouhou}\ \emph {et~al.}(2019)\citenamefont
  {Agah~Nouhou}, \citenamefont {Bluhm}, \citenamefont {Borer}, \citenamefont
  {Nahrgang}, \citenamefont {Sami},\ and\ \citenamefont
  {Touroux}}]{Nouhou:2019nhe}%
  \BibitemOpen
  \bibfield  {author} {\bibinfo {author} {\bibfnamefont {M.}~\bibnamefont
  {Agah~Nouhou}}, \bibinfo {author} {\bibfnamefont {M.}~\bibnamefont {Bluhm}},
  \bibinfo {author} {\bibfnamefont {A.}~\bibnamefont {Borer}}, \bibinfo
  {author} {\bibfnamefont {M.}~\bibnamefont {Nahrgang}}, \bibinfo {author}
  {\bibfnamefont {T.}~\bibnamefont {Sami}}, \ and\ \bibinfo {author}
  {\bibfnamefont {N.}~\bibnamefont {Touroux}},\ }\bibfield  {booktitle} {\emph
  {\bibinfo {booktitle} {{Proceedings, 18th Hellenic School and Workshops on
  Elementary Particle Physics and Gravity (CORFU2018): Corfu, Corfu,
  Greece}}},\ }\href {\doibase 10.22323/1.347.0179} {\bibfield  {journal}
  {\bibinfo  {journal} {PoS}\ }\textbf {\bibinfo {volume} {CORFU2018}},\
  \bibinfo {pages} {179} (\bibinfo {year} {2019})},\ \Eprint
  {http://arxiv.org/abs/1906.02647} {arXiv:1906.02647 [nucl-th]} \BibitemShut
  {NoStop}%
\bibitem [{\citenamefont {Auvinen}\ and\ \citenamefont
  {Petersen}(2013)}]{Auvinen:2013sba}%
  \BibitemOpen
  \bibfield  {author} {\bibinfo {author} {\bibfnamefont {J.}~\bibnamefont
  {Auvinen}}\ and\ \bibinfo {author} {\bibfnamefont {H.}~\bibnamefont
  {Petersen}},\ }\href {\doibase 10.1103/PhysRevC.88.064908} {\bibfield
  {journal} {\bibinfo  {journal} {Phys. Rev.}\ }\textbf {\bibinfo {volume}
  {C88}},\ \bibinfo {pages} {064908} (\bibinfo {year} {2013})},\ \Eprint
  {http://arxiv.org/abs/1310.1764} {arXiv:1310.1764 [nucl-th]} \BibitemShut
  {NoStop}%
\bibitem [{\citenamefont {Steinheimer}\ \emph {et~al.}(2014)\citenamefont
  {Steinheimer}, \citenamefont {Auvinen}, \citenamefont {Petersen},
  \citenamefont {Bleicher},\ and\ \citenamefont
  {Stöcker}}]{Steinheimer:2014pfa}%
  \BibitemOpen
  \bibfield  {author} {\bibinfo {author} {\bibfnamefont {J.}~\bibnamefont
  {Steinheimer}}, \bibinfo {author} {\bibfnamefont {J.}~\bibnamefont
  {Auvinen}}, \bibinfo {author} {\bibfnamefont {H.}~\bibnamefont {Petersen}},
  \bibinfo {author} {\bibfnamefont {M.}~\bibnamefont {Bleicher}}, \ and\
  \bibinfo {author} {\bibfnamefont {H.}~\bibnamefont {Stöcker}},\ }\href
  {\doibase 10.1103/PhysRevC.89.054913} {\bibfield  {journal} {\bibinfo
  {journal} {Phys. Rev.}\ }\textbf {\bibinfo {volume} {C89}},\ \bibinfo {pages}
  {054913} (\bibinfo {year} {2014})},\ \Eprint {http://arxiv.org/abs/1402.7236}
  {arXiv:1402.7236 [nucl-th]} \BibitemShut {NoStop}%
\bibitem [{\citenamefont {Monnai}\ \emph {et~al.}(2017)\citenamefont {Monnai},
  \citenamefont {Mukherjee},\ and\ \citenamefont {Yin}}]{Monnai:2016kud}%
  \BibitemOpen
  \bibfield  {author} {\bibinfo {author} {\bibfnamefont {A.}~\bibnamefont
  {Monnai}}, \bibinfo {author} {\bibfnamefont {S.}~\bibnamefont {Mukherjee}}, \
  and\ \bibinfo {author} {\bibfnamefont {Y.}~\bibnamefont {Yin}},\ }\href
  {\doibase 10.1103/PhysRevC.95.034902} {\bibfield  {journal} {\bibinfo
  {journal} {Phys. Rev.}\ }\textbf {\bibinfo {volume} {C95}},\ \bibinfo {pages}
  {034902} (\bibinfo {year} {2017})},\ \Eprint
  {http://arxiv.org/abs/1606.00771} {arXiv:1606.00771 [nucl-th]} \BibitemShut
  {NoStop}%
\bibitem [{\citenamefont {Auvinen}\ \emph
  {et~al.}(2018{\natexlab{a}})\citenamefont {Auvinen}, \citenamefont
  {Bernhard}, \citenamefont {Bass},\ and\ \citenamefont
  {Karpenko}}]{Auvinen:2017fjw}%
  \BibitemOpen
  \bibfield  {author} {\bibinfo {author} {\bibfnamefont {J.}~\bibnamefont
  {Auvinen}}, \bibinfo {author} {\bibfnamefont {J.~E.}\ \bibnamefont
  {Bernhard}}, \bibinfo {author} {\bibfnamefont {S.~A.}\ \bibnamefont {Bass}},
  \ and\ \bibinfo {author} {\bibfnamefont {I.}~\bibnamefont {Karpenko}},\
  }\href {\doibase 10.1103/PhysRevC.97.044905} {\bibfield  {journal} {\bibinfo
  {journal} {Phys. Rev.}\ }\textbf {\bibinfo {volume} {C97}},\ \bibinfo {pages}
  {044905} (\bibinfo {year} {2018}{\natexlab{a}})},\ \Eprint
  {http://arxiv.org/abs/1706.03666} {arXiv:1706.03666 [hep-ph]} \BibitemShut
  {NoStop}%
\bibitem [{\citenamefont {Stephanov}\ and\ \citenamefont
  {Yin}(2018)}]{Stephanov:2017ghc}%
  \BibitemOpen
  \bibfield  {author} {\bibinfo {author} {\bibfnamefont {M.}~\bibnamefont
  {Stephanov}}\ and\ \bibinfo {author} {\bibfnamefont {Y.}~\bibnamefont
  {Yin}},\ }\href {\doibase 10.1103/PhysRevD.98.036006} {\bibfield  {journal}
  {\bibinfo  {journal} {Phys. Rev.}\ }\textbf {\bibinfo {volume} {D98}},\
  \bibinfo {pages} {036006} (\bibinfo {year} {2018})},\ \Eprint
  {http://arxiv.org/abs/1712.10305} {arXiv:1712.10305 [nucl-th]} \BibitemShut
  {NoStop}%
\bibitem [{\citenamefont {Shen}\ and\ \citenamefont
  {Schenke}(2018)}]{Shen:2017bsr}%
  \BibitemOpen
  \bibfield  {author} {\bibinfo {author} {\bibfnamefont {C.}~\bibnamefont
  {Shen}}\ and\ \bibinfo {author} {\bibfnamefont {B.}~\bibnamefont {Schenke}},\
  }\href {\doibase 10.1103/PhysRevC.97.024907} {\bibfield  {journal} {\bibinfo
  {journal} {Phys. Rev.}\ }\textbf {\bibinfo {volume} {C97}},\ \bibinfo {pages}
  {024907} (\bibinfo {year} {2018})},\ \Eprint
  {http://arxiv.org/abs/1710.00881} {arXiv:1710.00881 [nucl-th]} \BibitemShut
  {NoStop}%
\bibitem [{\citenamefont {Nahrgang}\ \emph {et~al.}(2019)\citenamefont
  {Nahrgang}, \citenamefont {Bluhm}, \citenamefont {Schaefer},\ and\
  \citenamefont {Bass}}]{Nahrgang:2018afz}%
  \BibitemOpen
  \bibfield  {author} {\bibinfo {author} {\bibfnamefont {M.}~\bibnamefont
  {Nahrgang}}, \bibinfo {author} {\bibfnamefont {M.}~\bibnamefont {Bluhm}},
  \bibinfo {author} {\bibfnamefont {T.}~\bibnamefont {Schaefer}}, \ and\
  \bibinfo {author} {\bibfnamefont {S.~A.}\ \bibnamefont {Bass}},\ }\href
  {\doibase 10.1103/PhysRevD.99.116015} {\bibfield  {journal} {\bibinfo
  {journal} {Phys. Rev.}\ }\textbf {\bibinfo {volume} {D99}},\ \bibinfo {pages}
  {116015} (\bibinfo {year} {2019})},\ \Eprint
  {http://arxiv.org/abs/1804.05728} {arXiv:1804.05728 [nucl-th]} \BibitemShut
  {NoStop}%
\bibitem [{\citenamefont {Akamatsu}\ \emph {et~al.}(2018)\citenamefont
  {Akamatsu}, \citenamefont {Asakawa}, \citenamefont {Hirano}, \citenamefont
  {Kitazawa}, \citenamefont {Morita}, \citenamefont {Murase}, \citenamefont
  {Nara}, \citenamefont {Nonaka},\ and\ \citenamefont
  {Ohnishi}}]{Akamatsu:2018olk}%
  \BibitemOpen
  \bibfield  {author} {\bibinfo {author} {\bibfnamefont {Y.}~\bibnamefont
  {Akamatsu}}, \bibinfo {author} {\bibfnamefont {M.}~\bibnamefont {Asakawa}},
  \bibinfo {author} {\bibfnamefont {T.}~\bibnamefont {Hirano}}, \bibinfo
  {author} {\bibfnamefont {M.}~\bibnamefont {Kitazawa}}, \bibinfo {author}
  {\bibfnamefont {K.}~\bibnamefont {Morita}}, \bibinfo {author} {\bibfnamefont
  {K.}~\bibnamefont {Murase}}, \bibinfo {author} {\bibfnamefont
  {Y.}~\bibnamefont {Nara}}, \bibinfo {author} {\bibfnamefont {C.}~\bibnamefont
  {Nonaka}}, \ and\ \bibinfo {author} {\bibfnamefont {A.}~\bibnamefont
  {Ohnishi}},\ }\href {\doibase 10.1103/PhysRevC.98.024909} {\bibfield
  {journal} {\bibinfo  {journal} {Phys. Rev. C}\ }\textbf {\bibinfo {volume}
  {98}},\ \bibinfo {pages} {024909} (\bibinfo {year} {2018})},\ \Eprint
  {http://arxiv.org/abs/1805.09024} {arXiv:1805.09024 [nucl-th]} \BibitemShut
  {NoStop}%
\bibitem [{\citenamefont {Kanakubo}\ \emph {et~al.}(2020)\citenamefont
  {Kanakubo}, \citenamefont {Tachibana},\ and\ \citenamefont
  {Hirano}}]{Kanakubo:2019ogh}%
  \BibitemOpen
  \bibfield  {author} {\bibinfo {author} {\bibfnamefont {Y.}~\bibnamefont
  {Kanakubo}}, \bibinfo {author} {\bibfnamefont {Y.}~\bibnamefont {Tachibana}},
  \ and\ \bibinfo {author} {\bibfnamefont {T.}~\bibnamefont {Hirano}},\ }\href
  {\doibase 10.1103/PhysRevC.101.024912} {\bibfield  {journal} {\bibinfo
  {journal} {Phys. Rev. C}\ }\textbf {\bibinfo {volume} {101}},\ \bibinfo
  {pages} {024912} (\bibinfo {year} {2020})},\ \Eprint
  {http://arxiv.org/abs/1910.10556} {arXiv:1910.10556 [nucl-th]} \BibitemShut
  {NoStop}%
\bibitem [{\citenamefont {Du}\ and\ \citenamefont {Heinz}(2019)}]{Du:2019obx}%
  \BibitemOpen
  \bibfield  {author} {\bibinfo {author} {\bibfnamefont {L.}~\bibnamefont
  {Du}}\ and\ \bibinfo {author} {\bibfnamefont {U.}~\bibnamefont {Heinz}},\
  }\href {\doibase 10.1016/j.cpc.2019.107090} {\  (\bibinfo {year} {2019}),\
  10.1016/j.cpc.2019.107090},\ \Eprint {http://arxiv.org/abs/1906.11181}
  {arXiv:1906.11181 [nucl-th]} \BibitemShut {NoStop}%
\bibitem [{\citenamefont {Fotakis}\ \emph {et~al.}(2019)\citenamefont
  {Fotakis}, \citenamefont {Greif}, \citenamefont {Denicol}, \citenamefont
  {Niemi},\ and\ \citenamefont {Greiner}}]{Fotakis:2019nbq}%
  \BibitemOpen
  \bibfield  {author} {\bibinfo {author} {\bibfnamefont {J.~A.}\ \bibnamefont
  {Fotakis}}, \bibinfo {author} {\bibfnamefont {M.}~\bibnamefont {Greif}},
  \bibinfo {author} {\bibfnamefont {G.}~\bibnamefont {Denicol}}, \bibinfo
  {author} {\bibfnamefont {H.}~\bibnamefont {Niemi}}, \ and\ \bibinfo {author}
  {\bibfnamefont {C.}~\bibnamefont {Greiner}},\ }\href@noop {} {\  (\bibinfo
  {year} {2019})},\ \Eprint {http://arxiv.org/abs/1912.09103} {arXiv:1912.09103
  [hep-ph]} \BibitemShut {NoStop}%
\bibitem [{\citenamefont {Martinez}\ \emph
  {et~al.}(2019{\natexlab{a}})\citenamefont {Martinez}, \citenamefont
  {Sievert}, \citenamefont {Wertepny},\ and\ \citenamefont
  {Noronha-Hostler}}]{Martinez:2019rlp}%
  \BibitemOpen
  \bibfield  {author} {\bibinfo {author} {\bibfnamefont {M.}~\bibnamefont
  {Martinez}}, \bibinfo {author} {\bibfnamefont {M.~D.}\ \bibnamefont
  {Sievert}}, \bibinfo {author} {\bibfnamefont {D.~E.}\ \bibnamefont
  {Wertepny}}, \ and\ \bibinfo {author} {\bibfnamefont {J.}~\bibnamefont
  {Noronha-Hostler}},\ }\href@noop {} {\  (\bibinfo {year}
  {2019}{\natexlab{a}})},\ \Eprint {http://arxiv.org/abs/1911.12454}
  {arXiv:1911.12454 [nucl-th]} \BibitemShut {NoStop}%
\bibitem [{\citenamefont {Martinez}\ \emph
  {et~al.}(2019{\natexlab{b}})\citenamefont {Martinez}, \citenamefont
  {Sievert}, \citenamefont {Wertepny},\ and\ \citenamefont
  {Noronha-Hostler}}]{Martinez:2019jbu}%
  \BibitemOpen
  \bibfield  {author} {\bibinfo {author} {\bibfnamefont {M.}~\bibnamefont
  {Martinez}}, \bibinfo {author} {\bibfnamefont {M.~D.}\ \bibnamefont
  {Sievert}}, \bibinfo {author} {\bibfnamefont {D.~E.}\ \bibnamefont
  {Wertepny}}, \ and\ \bibinfo {author} {\bibfnamefont {J.}~\bibnamefont
  {Noronha-Hostler}},\ }\href@noop {} {\  (\bibinfo {year}
  {2019}{\natexlab{b}})},\ \Eprint {http://arxiv.org/abs/1911.10272}
  {arXiv:1911.10272 [nucl-th]} \BibitemShut {NoStop}%
\bibitem [{\citenamefont {Moreau}\ \emph {et~al.}(2019)\citenamefont {Moreau},
  \citenamefont {Soloveva}, \citenamefont {Oliva}, \citenamefont {Song},
  \citenamefont {Cassing},\ and\ \citenamefont
  {Bratkovskaya}}]{Moreau:2019vhw}%
  \BibitemOpen
  \bibfield  {author} {\bibinfo {author} {\bibfnamefont {P.}~\bibnamefont
  {Moreau}}, \bibinfo {author} {\bibfnamefont {O.}~\bibnamefont {Soloveva}},
  \bibinfo {author} {\bibfnamefont {L.}~\bibnamefont {Oliva}}, \bibinfo
  {author} {\bibfnamefont {T.}~\bibnamefont {Song}}, \bibinfo {author}
  {\bibfnamefont {W.}~\bibnamefont {Cassing}}, \ and\ \bibinfo {author}
  {\bibfnamefont {E.}~\bibnamefont {Bratkovskaya}},\ }\href {\doibase
  10.1103/PhysRevC.100.014911} {\bibfield  {journal} {\bibinfo  {journal}
  {Phys. Rev. C}\ }\textbf {\bibinfo {volume} {100}},\ \bibinfo {pages}
  {014911} (\bibinfo {year} {2019})},\ \Eprint
  {http://arxiv.org/abs/1903.10257} {arXiv:1903.10257 [nucl-th]} \BibitemShut
  {NoStop}%
\bibitem [{\citenamefont {Soloveva}\ \emph {et~al.}(2020)\citenamefont
  {Soloveva}, \citenamefont {Moreau},\ and\ \citenamefont
  {Bratkovskaya}}]{Soloveva:2019xph}%
  \BibitemOpen
  \bibfield  {author} {\bibinfo {author} {\bibfnamefont {O.}~\bibnamefont
  {Soloveva}}, \bibinfo {author} {\bibfnamefont {P.}~\bibnamefont {Moreau}}, \
  and\ \bibinfo {author} {\bibfnamefont {E.}~\bibnamefont {Bratkovskaya}},\
  }\href {\doibase 10.1103/PhysRevC.101.045203} {\bibfield  {journal} {\bibinfo
   {journal} {Phys. Rev. C}\ }\textbf {\bibinfo {volume} {101}},\ \bibinfo
  {pages} {045203} (\bibinfo {year} {2020})},\ \Eprint
  {http://arxiv.org/abs/1911.08547} {arXiv:1911.08547 [nucl-th]} \BibitemShut
  {NoStop}%
\bibitem [{\citenamefont {An}\ \emph {et~al.}(2019)\citenamefont {An},
  \citenamefont {Başar}, \citenamefont {Stephanov},\ and\ \citenamefont
  {Yee}}]{An:2019csj}%
  \BibitemOpen
  \bibfield  {author} {\bibinfo {author} {\bibfnamefont {X.}~\bibnamefont
  {An}}, \bibinfo {author} {\bibfnamefont {G.}~\bibnamefont {Başar}}, \bibinfo
  {author} {\bibfnamefont {M.}~\bibnamefont {Stephanov}}, \ and\ \bibinfo
  {author} {\bibfnamefont {H.-U.}\ \bibnamefont {Yee}},\ }\href@noop {} {\
  (\bibinfo {year} {2019})},\ \Eprint {http://arxiv.org/abs/1912.13456}
  {arXiv:1912.13456 [hep-th]} \BibitemShut {NoStop}%
\bibitem [{\citenamefont {Bluhm}\ \emph {et~al.}(2020)\citenamefont {Bluhm}
  \emph {et~al.}}]{Bluhm:2020mpc}%
  \BibitemOpen
  \bibfield  {author} {\bibinfo {author} {\bibfnamefont {M.}~\bibnamefont
  {Bluhm}} \emph {et~al.},\ }\href@noop {} {\  (\bibinfo {year} {2020})},\
  \Eprint {http://arxiv.org/abs/2001.08831} {arXiv:2001.08831 [nucl-th]}
  \BibitemShut {NoStop}%
\bibitem [{\citenamefont {Rao}\ \emph {et~al.}(2019)\citenamefont {Rao},
  \citenamefont {Sievert},\ and\ \citenamefont
  {Noronha-Hostler}}]{Rao:2019vgy}%
  \BibitemOpen
  \bibfield  {author} {\bibinfo {author} {\bibfnamefont {S.}~\bibnamefont
  {Rao}}, \bibinfo {author} {\bibfnamefont {M.}~\bibnamefont {Sievert}}, \ and\
  \bibinfo {author} {\bibfnamefont {J.}~\bibnamefont {Noronha-Hostler}},\
  }\href@noop {} {\  (\bibinfo {year} {2019})},\ \Eprint
  {http://arxiv.org/abs/1910.03677} {arXiv:1910.03677 [nucl-th]} \BibitemShut
  {NoStop}%
\bibitem [{\citenamefont {Giacalone}\ \emph {et~al.}(2017)\citenamefont
  {Giacalone}, \citenamefont {Noronha-Hostler},\ and\ \citenamefont
  {Ollitrault}}]{Giacalone:2017uqx}%
  \BibitemOpen
  \bibfield  {author} {\bibinfo {author} {\bibfnamefont {G.}~\bibnamefont
  {Giacalone}}, \bibinfo {author} {\bibfnamefont {J.}~\bibnamefont
  {Noronha-Hostler}}, \ and\ \bibinfo {author} {\bibfnamefont {J.-Y.}\
  \bibnamefont {Ollitrault}},\ }\href {\doibase 10.1103/PhysRevC.95.054910}
  {\bibfield  {journal} {\bibinfo  {journal} {Phys. Rev.}\ }\textbf {\bibinfo
  {volume} {C95}},\ \bibinfo {pages} {054910} (\bibinfo {year} {2017})},\
  \Eprint {http://arxiv.org/abs/1702.01730} {arXiv:1702.01730 [nucl-th]}
  \BibitemShut {NoStop}%
\bibitem [{\citenamefont {Pratt}\ \emph {et~al.}(2015)\citenamefont {Pratt},
  \citenamefont {Sangaline}, \citenamefont {Sorensen},\ and\ \citenamefont
  {Wang}}]{Pratt:2015zsa}%
  \BibitemOpen
  \bibfield  {author} {\bibinfo {author} {\bibfnamefont {S.}~\bibnamefont
  {Pratt}}, \bibinfo {author} {\bibfnamefont {E.}~\bibnamefont {Sangaline}},
  \bibinfo {author} {\bibfnamefont {P.}~\bibnamefont {Sorensen}}, \ and\
  \bibinfo {author} {\bibfnamefont {H.}~\bibnamefont {Wang}},\ }\href {\doibase
  10.1103/PhysRevLett.114.202301} {\bibfield  {journal} {\bibinfo  {journal}
  {Phys. Rev. Lett.}\ }\textbf {\bibinfo {volume} {114}},\ \bibinfo {pages}
  {202301} (\bibinfo {year} {2015})},\ \Eprint
  {http://arxiv.org/abs/1501.04042} {arXiv:1501.04042 [nucl-th]} \BibitemShut
  {NoStop}%
\bibitem [{\citenamefont {Moreland}\ and\ \citenamefont
  {Soltz}(2016)}]{Moreland:2015dvc}%
  \BibitemOpen
  \bibfield  {author} {\bibinfo {author} {\bibfnamefont {J.~S.}\ \bibnamefont
  {Moreland}}\ and\ \bibinfo {author} {\bibfnamefont {R.~A.}\ \bibnamefont
  {Soltz}},\ }\href {\doibase 10.1103/PhysRevC.93.044913} {\bibfield  {journal}
  {\bibinfo  {journal} {Phys. Rev.}\ }\textbf {\bibinfo {volume} {C93}},\
  \bibinfo {pages} {044913} (\bibinfo {year} {2016})},\ \Eprint
  {http://arxiv.org/abs/1512.02189} {arXiv:1512.02189 [nucl-th]} \BibitemShut
  {NoStop}%
\bibitem [{\citenamefont {Alba}\ \emph {et~al.}(2018)\citenamefont {Alba},
  \citenamefont {Mantovani~Sarti}, \citenamefont {Noronha}, \citenamefont
  {Noronha-Hostler}, \citenamefont {Parotto}, \citenamefont
  {Portillo~Vazquez},\ and\ \citenamefont {Ratti}}]{Alba:2017hhe}%
  \BibitemOpen
  \bibfield  {author} {\bibinfo {author} {\bibfnamefont {P.}~\bibnamefont
  {Alba}}, \bibinfo {author} {\bibfnamefont {V.}~\bibnamefont
  {Mantovani~Sarti}}, \bibinfo {author} {\bibfnamefont {J.}~\bibnamefont
  {Noronha}}, \bibinfo {author} {\bibfnamefont {J.}~\bibnamefont
  {Noronha-Hostler}}, \bibinfo {author} {\bibfnamefont {P.}~\bibnamefont
  {Parotto}}, \bibinfo {author} {\bibfnamefont {I.}~\bibnamefont
  {Portillo~Vazquez}}, \ and\ \bibinfo {author} {\bibfnamefont
  {C.}~\bibnamefont {Ratti}},\ }\href {\doibase 10.1103/PhysRevC.98.034909}
  {\bibfield  {journal} {\bibinfo  {journal} {Phys. Rev.}\ }\textbf {\bibinfo
  {volume} {C98}},\ \bibinfo {pages} {034909} (\bibinfo {year} {2018})},\
  \Eprint {http://arxiv.org/abs/1711.05207} {arXiv:1711.05207 [nucl-th]}
  \BibitemShut {NoStop}%
\bibitem [{\citenamefont {Auvinen}\ \emph
  {et~al.}(2018{\natexlab{b}})\citenamefont {Auvinen}, \citenamefont {Eskola},
  \citenamefont {Huovinen}, \citenamefont {Niemi}, \citenamefont
  {Paatelainen},\ and\ \citenamefont {Petreczky}}]{Auvinen:2018uej}%
  \BibitemOpen
  \bibfield  {author} {\bibinfo {author} {\bibfnamefont {J.}~\bibnamefont
  {Auvinen}}, \bibinfo {author} {\bibfnamefont {K.~J.}\ \bibnamefont {Eskola}},
  \bibinfo {author} {\bibfnamefont {P.}~\bibnamefont {Huovinen}}, \bibinfo
  {author} {\bibfnamefont {H.}~\bibnamefont {Niemi}}, \bibinfo {author}
  {\bibfnamefont {R.}~\bibnamefont {Paatelainen}}, \ and\ \bibinfo {author}
  {\bibfnamefont {P.}~\bibnamefont {Petreczky}},\ }\bibfield  {booktitle}
  {\emph {\bibinfo {booktitle} {{Hadronic contributions to the muon anomalous
  magnetic moment Workshop. $(g-2)\_{\mu}$ Workshop. Mini proceedings}}},\
  }\href {\doibase 10.22323/1.336.0135} {\bibfield  {journal} {\bibinfo
  {journal} {PoS}\ }\textbf {\bibinfo {volume} {Confinement2018}},\ \bibinfo
  {pages} {135} (\bibinfo {year} {2018}{\natexlab{b}})},\ \Eprint
  {http://arxiv.org/abs/1811.01792} {arXiv:1811.01792 [hep-ph]} \BibitemShut
  {NoStop}%
\bibitem [{\citenamefont {Noronha-Hostler}(2015)}]{Noronha-Hostler:2015qmd}%
  \BibitemOpen
  \bibfield  {author} {\bibinfo {author} {\bibfnamefont {J.}~\bibnamefont
  {Noronha-Hostler}},\ }in\ \href@noop {} {\emph {\bibinfo {booktitle}
  {{Proceedings, 12th Conference on the Intersections of Particle and Nuclear
  Physics (CIPANP 2015): Vail, Colorado, USA, May 19-24, 2015}}}}\ (\bibinfo
  {year} {2015})\ \Eprint {http://arxiv.org/abs/1512.06315} {arXiv:1512.06315
  [nucl-th]} \BibitemShut {NoStop}%
\bibitem [{\citenamefont {Niemi}\ \emph {et~al.}(2016)\citenamefont {Niemi},
  \citenamefont {Eskola},\ and\ \citenamefont {Paatelainen}}]{Niemi:2015qia}%
  \BibitemOpen
  \bibfield  {author} {\bibinfo {author} {\bibfnamefont {H.}~\bibnamefont
  {Niemi}}, \bibinfo {author} {\bibfnamefont {K.~J.}\ \bibnamefont {Eskola}}, \
  and\ \bibinfo {author} {\bibfnamefont {R.}~\bibnamefont {Paatelainen}},\
  }\href {\doibase 10.1103/PhysRevC.93.024907} {\bibfield  {journal} {\bibinfo
  {journal} {Phys. Rev.}\ }\textbf {\bibinfo {volume} {C93}},\ \bibinfo {pages}
  {024907} (\bibinfo {year} {2016})},\ \Eprint
  {http://arxiv.org/abs/1505.02677} {arXiv:1505.02677 [hep-ph]} \BibitemShut
  {NoStop}%
\bibitem [{\citenamefont {Bernhard}\ \emph {et~al.}(2019)\citenamefont
  {Bernhard}, \citenamefont {Moreland},\ and\ \citenamefont
  {Bass}}]{Bernhard:2019bmu}%
  \BibitemOpen
  \bibfield  {author} {\bibinfo {author} {\bibfnamefont {J.~E.}\ \bibnamefont
  {Bernhard}}, \bibinfo {author} {\bibfnamefont {J.~S.}\ \bibnamefont
  {Moreland}}, \ and\ \bibinfo {author} {\bibfnamefont {S.~A.}\ \bibnamefont
  {Bass}},\ }\href {\doibase 10.1038/s41567-019-0611-8} {\bibfield  {journal}
  {\bibinfo  {journal} {Nature Phys.}\ }\textbf {\bibinfo {volume} {15}},\
  \bibinfo {pages} {1113} (\bibinfo {year} {2019})}\BibitemShut {NoStop}%
\bibitem [{\citenamefont {Heller}\ and\ \citenamefont
  {Spalinski}(2015)}]{Heller:2015dha}%
  \BibitemOpen
  \bibfield  {author} {\bibinfo {author} {\bibfnamefont {M.~P.}\ \bibnamefont
  {Heller}}\ and\ \bibinfo {author} {\bibfnamefont {M.}~\bibnamefont
  {Spalinski}},\ }\href {\doibase 10.1103/PhysRevLett.115.072501} {\bibfield
  {journal} {\bibinfo  {journal} {Phys. Rev. Lett.}\ }\textbf {\bibinfo
  {volume} {115}},\ \bibinfo {pages} {072501} (\bibinfo {year} {2015})},\
  \Eprint {http://arxiv.org/abs/1503.07514} {arXiv:1503.07514 [hep-th]}
  \BibitemShut {NoStop}%
\bibitem [{\citenamefont {Buchel}\ \emph {et~al.}(2016)\citenamefont {Buchel},
  \citenamefont {Heller},\ and\ \citenamefont {Noronha}}]{Buchel:2016cbj}%
  \BibitemOpen
  \bibfield  {author} {\bibinfo {author} {\bibfnamefont {A.}~\bibnamefont
  {Buchel}}, \bibinfo {author} {\bibfnamefont {M.~P.}\ \bibnamefont {Heller}},
  \ and\ \bibinfo {author} {\bibfnamefont {J.}~\bibnamefont {Noronha}},\ }\href
  {\doibase 10.1103/PhysRevD.94.106011} {\bibfield  {journal} {\bibinfo
  {journal} {Phys. Rev.}\ }\textbf {\bibinfo {volume} {D94}},\ \bibinfo {pages}
  {106011} (\bibinfo {year} {2016})},\ \Eprint
  {http://arxiv.org/abs/1603.05344} {arXiv:1603.05344 [hep-th]} \BibitemShut
  {NoStop}%
\bibitem [{\citenamefont {Heller}\ \emph {et~al.}(2018)\citenamefont {Heller},
  \citenamefont {Kurkela}, \citenamefont {Spaliński},\ and\ \citenamefont
  {Svensson}}]{Heller:2016rtz}%
  \BibitemOpen
  \bibfield  {author} {\bibinfo {author} {\bibfnamefont {M.~P.}\ \bibnamefont
  {Heller}}, \bibinfo {author} {\bibfnamefont {A.}~\bibnamefont {Kurkela}},
  \bibinfo {author} {\bibfnamefont {M.}~\bibnamefont {Spaliński}}, \ and\
  \bibinfo {author} {\bibfnamefont {V.}~\bibnamefont {Svensson}},\ }\href
  {\doibase 10.1103/PhysRevD.97.091503} {\bibfield  {journal} {\bibinfo
  {journal} {Phys. Rev.}\ }\textbf {\bibinfo {volume} {D97}},\ \bibinfo {pages}
  {091503} (\bibinfo {year} {2018})},\ \Eprint
  {http://arxiv.org/abs/1609.04803} {arXiv:1609.04803 [nucl-th]} \BibitemShut
  {NoStop}%
\bibitem [{\citenamefont {Spaliński}(2018)}]{Spalinski:2017mel}%
  \BibitemOpen
  \bibfield  {author} {\bibinfo {author} {\bibfnamefont {M.}~\bibnamefont
  {Spaliński}},\ }\href {\doibase 10.1016/j.physletb.2017.11.059} {\bibfield
  {journal} {\bibinfo  {journal} {Phys. Lett.}\ }\textbf {\bibinfo {volume}
  {B776}},\ \bibinfo {pages} {468} (\bibinfo {year} {2018})},\ \Eprint
  {http://arxiv.org/abs/1708.01921} {arXiv:1708.01921 [hep-th]} \BibitemShut
  {NoStop}%
\bibitem [{\citenamefont
  {Romatschke}(2017{\natexlab{a}})}]{Romatschke:2017acs}%
  \BibitemOpen
  \bibfield  {author} {\bibinfo {author} {\bibfnamefont {P.}~\bibnamefont
  {Romatschke}},\ }\href {\doibase 10.1007/JHEP12(2017)079} {\bibfield
  {journal} {\bibinfo  {journal} {JHEP}\ }\textbf {\bibinfo {volume} {12}},\
  \bibinfo {pages} {079} (\bibinfo {year} {2017}{\natexlab{a}})},\ \Eprint
  {http://arxiv.org/abs/1710.03234} {arXiv:1710.03234 [hep-th]} \BibitemShut
  {NoStop}%
\bibitem [{\citenamefont {Romatschke}(2018)}]{Romatschke:2017vte}%
  \BibitemOpen
  \bibfield  {author} {\bibinfo {author} {\bibfnamefont {P.}~\bibnamefont
  {Romatschke}},\ }\href {\doibase 10.1103/PhysRevLett.120.012301} {\bibfield
  {journal} {\bibinfo  {journal} {Phys. Rev. Lett.}\ }\textbf {\bibinfo
  {volume} {120}},\ \bibinfo {pages} {012301} (\bibinfo {year} {2018})},\
  \Eprint {http://arxiv.org/abs/1704.08699} {arXiv:1704.08699 [hep-th]}
  \BibitemShut {NoStop}%
\bibitem [{\citenamefont {Behtash}\ \emph {et~al.}(2018)\citenamefont
  {Behtash}, \citenamefont {Cruz-Camacho},\ and\ \citenamefont
  {Martinez}}]{Behtash:2017wqg}%
  \BibitemOpen
  \bibfield  {author} {\bibinfo {author} {\bibfnamefont {A.}~\bibnamefont
  {Behtash}}, \bibinfo {author} {\bibfnamefont {C.~N.}\ \bibnamefont
  {Cruz-Camacho}}, \ and\ \bibinfo {author} {\bibfnamefont {M.}~\bibnamefont
  {Martinez}},\ }\href {\doibase 10.1103/PhysRevD.97.044041} {\bibfield
  {journal} {\bibinfo  {journal} {Phys. Rev.}\ }\textbf {\bibinfo {volume}
  {D97}},\ \bibinfo {pages} {044041} (\bibinfo {year} {2018})},\ \Eprint
  {http://arxiv.org/abs/1711.01745} {arXiv:1711.01745 [hep-th]} \BibitemShut
  {NoStop}%
\bibitem [{\citenamefont {Strickland}\ \emph {et~al.}(2018)\citenamefont
  {Strickland}, \citenamefont {Noronha},\ and\ \citenamefont
  {Denicol}}]{Strickland:2017kux}%
  \BibitemOpen
  \bibfield  {author} {\bibinfo {author} {\bibfnamefont {M.}~\bibnamefont
  {Strickland}}, \bibinfo {author} {\bibfnamefont {J.}~\bibnamefont {Noronha}},
  \ and\ \bibinfo {author} {\bibfnamefont {G.}~\bibnamefont {Denicol}},\ }\href
  {\doibase 10.1103/PhysRevD.97.036020} {\bibfield  {journal} {\bibinfo
  {journal} {Phys. Rev.}\ }\textbf {\bibinfo {volume} {D97}},\ \bibinfo {pages}
  {036020} (\bibinfo {year} {2018})},\ \Eprint
  {http://arxiv.org/abs/1709.06644} {arXiv:1709.06644 [nucl-th]} \BibitemShut
  {NoStop}%
\bibitem [{\citenamefont {Denicol}\ and\ \citenamefont
  {Noronha}(2018)}]{Denicol:2017lxn}%
  \BibitemOpen
  \bibfield  {author} {\bibinfo {author} {\bibfnamefont {G.~S.}\ \bibnamefont
  {Denicol}}\ and\ \bibinfo {author} {\bibfnamefont {J.}~\bibnamefont
  {Noronha}},\ }\href {\doibase 10.1103/PhysRevD.97.056021} {\bibfield
  {journal} {\bibinfo  {journal} {Phys. Rev.}\ }\textbf {\bibinfo {volume}
  {D97}},\ \bibinfo {pages} {056021} (\bibinfo {year} {2018})},\ \Eprint
  {http://arxiv.org/abs/1711.01657} {arXiv:1711.01657 [nucl-th]} \BibitemShut
  {NoStop}%
\bibitem [{\citenamefont {Blaizot}\ and\ \citenamefont
  {Yan}(2018)}]{Blaizot:2017ucy}%
  \BibitemOpen
  \bibfield  {author} {\bibinfo {author} {\bibfnamefont {J.-P.}\ \bibnamefont
  {Blaizot}}\ and\ \bibinfo {author} {\bibfnamefont {L.}~\bibnamefont {Yan}},\
  }\href {\doibase 10.1016/j.physletb.2018.02.058} {\bibfield  {journal}
  {\bibinfo  {journal} {Phys. Lett.}\ }\textbf {\bibinfo {volume} {B780}},\
  \bibinfo {pages} {283} (\bibinfo {year} {2018})},\ \Eprint
  {http://arxiv.org/abs/1712.03856} {arXiv:1712.03856 [nucl-th]} \BibitemShut
  {NoStop}%
\bibitem [{\citenamefont {Casalderrey-Solana}\ \emph
  {et~al.}(2018)\citenamefont {Casalderrey-Solana}, \citenamefont {Gushterov},\
  and\ \citenamefont {Meiring}}]{Casalderrey-Solana:2017zyh}%
  \BibitemOpen
  \bibfield  {author} {\bibinfo {author} {\bibfnamefont {J.}~\bibnamefont
  {Casalderrey-Solana}}, \bibinfo {author} {\bibfnamefont {N.~I.}\ \bibnamefont
  {Gushterov}}, \ and\ \bibinfo {author} {\bibfnamefont {B.}~\bibnamefont
  {Meiring}},\ }\href {\doibase 10.1007/JHEP04(2018)042} {\bibfield  {journal}
  {\bibinfo  {journal} {JHEP}\ }\textbf {\bibinfo {volume} {04}},\ \bibinfo
  {pages} {042} (\bibinfo {year} {2018})},\ \Eprint
  {http://arxiv.org/abs/1712.02772} {arXiv:1712.02772 [hep-th]} \BibitemShut
  {NoStop}%
\bibitem [{\citenamefont {Florkowski}\ \emph {et~al.}(2018)\citenamefont
  {Florkowski}, \citenamefont {Heller},\ and\ \citenamefont
  {Spalinski}}]{Florkowski:2017olj}%
  \BibitemOpen
  \bibfield  {author} {\bibinfo {author} {\bibfnamefont {W.}~\bibnamefont
  {Florkowski}}, \bibinfo {author} {\bibfnamefont {M.~P.}\ \bibnamefont
  {Heller}}, \ and\ \bibinfo {author} {\bibfnamefont {M.}~\bibnamefont
  {Spalinski}},\ }\href {\doibase 10.1088/1361-6633/aaa091} {\bibfield
  {journal} {\bibinfo  {journal} {Rept. Prog. Phys.}\ }\textbf {\bibinfo
  {volume} {81}},\ \bibinfo {pages} {046001} (\bibinfo {year} {2018})},\
  \Eprint {http://arxiv.org/abs/1707.02282} {arXiv:1707.02282 [hep-ph]}
  \BibitemShut {NoStop}%
\bibitem [{\citenamefont {Heller}\ and\ \citenamefont
  {Svensson}(2018)}]{Heller:2018qvh}%
  \BibitemOpen
  \bibfield  {author} {\bibinfo {author} {\bibfnamefont {M.~P.}\ \bibnamefont
  {Heller}}\ and\ \bibinfo {author} {\bibfnamefont {V.}~\bibnamefont
  {Svensson}},\ }\href {\doibase 10.1103/PhysRevD.98.054016} {\bibfield
  {journal} {\bibinfo  {journal} {Phys. Rev.}\ }\textbf {\bibinfo {volume}
  {D98}},\ \bibinfo {pages} {054016} (\bibinfo {year} {2018})},\ \Eprint
  {http://arxiv.org/abs/1802.08225} {arXiv:1802.08225 [nucl-th]} \BibitemShut
  {NoStop}%
\bibitem [{\citenamefont {Rougemont}\ \emph {et~al.}(2018)\citenamefont
  {Rougemont}, \citenamefont {Critelli},\ and\ \citenamefont
  {Noronha}}]{Rougemont:2018ivt}%
  \BibitemOpen
  \bibfield  {author} {\bibinfo {author} {\bibfnamefont {R.}~\bibnamefont
  {Rougemont}}, \bibinfo {author} {\bibfnamefont {R.}~\bibnamefont {Critelli}},
  \ and\ \bibinfo {author} {\bibfnamefont {J.}~\bibnamefont {Noronha}},\ }\href
  {\doibase 10.1103/PhysRevD.98.034028} {\bibfield  {journal} {\bibinfo
  {journal} {Phys. Rev.}\ }\textbf {\bibinfo {volume} {D98}},\ \bibinfo {pages}
  {034028} (\bibinfo {year} {2018})},\ \Eprint
  {http://arxiv.org/abs/1804.00189} {arXiv:1804.00189 [hep-ph]} \BibitemShut
  {NoStop}%
\bibitem [{\citenamefont {Denicol}\ and\ \citenamefont
  {Noronha}(2019{\natexlab{a}})}]{Denicol:2018pak}%
  \BibitemOpen
  \bibfield  {author} {\bibinfo {author} {\bibfnamefont {G.~S.}\ \bibnamefont
  {Denicol}}\ and\ \bibinfo {author} {\bibfnamefont {J.}~\bibnamefont
  {Noronha}},\ }\href {\doibase 10.1103/PhysRevD.99.116004} {\bibfield
  {journal} {\bibinfo  {journal} {Phys. Rev.}\ }\textbf {\bibinfo {volume}
  {D99}},\ \bibinfo {pages} {116004} (\bibinfo {year} {2019}{\natexlab{a}})},\
  \Eprint {http://arxiv.org/abs/1804.04771} {arXiv:1804.04771 [nucl-th]}
  \BibitemShut {NoStop}%
\bibitem [{\citenamefont {Almaalol}\ and\ \citenamefont
  {Strickland}(2018)}]{Almaalol:2018ynx}%
  \BibitemOpen
  \bibfield  {author} {\bibinfo {author} {\bibfnamefont {D.}~\bibnamefont
  {Almaalol}}\ and\ \bibinfo {author} {\bibfnamefont {M.}~\bibnamefont
  {Strickland}},\ }\href {\doibase 10.1103/PhysRevC.97.044911} {\bibfield
  {journal} {\bibinfo  {journal} {Phys. Rev.}\ }\textbf {\bibinfo {volume}
  {C97}},\ \bibinfo {pages} {044911} (\bibinfo {year} {2018})},\ \Eprint
  {http://arxiv.org/abs/1801.10173} {arXiv:1801.10173 [hep-ph]} \BibitemShut
  {NoStop}%
\bibitem [{\citenamefont {Casalderrey-Solana}\ \emph
  {et~al.}(2019)\citenamefont {Casalderrey-Solana}, \citenamefont {Herzog},\
  and\ \citenamefont {Meiring}}]{Casalderrey-Solana:2018uag}%
  \BibitemOpen
  \bibfield  {author} {\bibinfo {author} {\bibfnamefont {J.}~\bibnamefont
  {Casalderrey-Solana}}, \bibinfo {author} {\bibfnamefont {C.~P.}\ \bibnamefont
  {Herzog}}, \ and\ \bibinfo {author} {\bibfnamefont {B.}~\bibnamefont
  {Meiring}},\ }\href {\doibase 10.1007/JHEP01(2019)181} {\bibfield  {journal}
  {\bibinfo  {journal} {JHEP}\ }\textbf {\bibinfo {volume} {01}},\ \bibinfo
  {pages} {181} (\bibinfo {year} {2019})},\ \Eprint
  {http://arxiv.org/abs/1810.02314} {arXiv:1810.02314 [hep-th]} \BibitemShut
  {NoStop}%
\bibitem [{\citenamefont {Behtash}\ \emph
  {et~al.}(2019{\natexlab{a}})\citenamefont {Behtash}, \citenamefont
  {Cruz-Camacho}, \citenamefont {Kamata},\ and\ \citenamefont
  {Martinez}}]{Behtash:2018moe}%
  \BibitemOpen
  \bibfield  {author} {\bibinfo {author} {\bibfnamefont {A.}~\bibnamefont
  {Behtash}}, \bibinfo {author} {\bibfnamefont {C.~N.}\ \bibnamefont
  {Cruz-Camacho}}, \bibinfo {author} {\bibfnamefont {S.}~\bibnamefont
  {Kamata}}, \ and\ \bibinfo {author} {\bibfnamefont {M.}~\bibnamefont
  {Martinez}},\ }\href {\doibase 10.1016/j.physletb.2019.134914} {\bibfield
  {journal} {\bibinfo  {journal} {Phys. Lett.}\ }\textbf {\bibinfo {volume}
  {B797}},\ \bibinfo {pages} {134914} (\bibinfo {year} {2019}{\natexlab{a}})},\
  \Eprint {http://arxiv.org/abs/1805.07881} {arXiv:1805.07881 [hep-th]}
  \BibitemShut {NoStop}%
\bibitem [{\citenamefont {Behtash}\ \emph
  {et~al.}(2019{\natexlab{b}})\citenamefont {Behtash}, \citenamefont {Kamata},
  \citenamefont {Martinez},\ and\ \citenamefont {Shi}}]{Behtash:2019txb}%
  \BibitemOpen
  \bibfield  {author} {\bibinfo {author} {\bibfnamefont {A.}~\bibnamefont
  {Behtash}}, \bibinfo {author} {\bibfnamefont {S.}~\bibnamefont {Kamata}},
  \bibinfo {author} {\bibfnamefont {M.}~\bibnamefont {Martinez}}, \ and\
  \bibinfo {author} {\bibfnamefont {H.}~\bibnamefont {Shi}},\ }\href {\doibase
  10.1103/PhysRevD.99.116012} {\bibfield  {journal} {\bibinfo  {journal} {Phys.
  Rev.}\ }\textbf {\bibinfo {volume} {D99}},\ \bibinfo {pages} {116012}
  (\bibinfo {year} {2019}{\natexlab{b}})},\ \Eprint
  {http://arxiv.org/abs/1901.08632} {arXiv:1901.08632 [hep-th]} \BibitemShut
  {NoStop}%
\bibitem [{\citenamefont {Strickland}(2018)}]{Strickland:2018ayk}%
  \BibitemOpen
  \bibfield  {author} {\bibinfo {author} {\bibfnamefont {M.}~\bibnamefont
  {Strickland}},\ }\href {\doibase 10.1007/JHEP12(2018)128} {\bibfield
  {journal} {\bibinfo  {journal} {JHEP}\ }\textbf {\bibinfo {volume} {12}},\
  \bibinfo {pages} {128} (\bibinfo {year} {2018})},\ \Eprint
  {http://arxiv.org/abs/1809.01200} {arXiv:1809.01200 [nucl-th]} \BibitemShut
  {NoStop}%
\bibitem [{\citenamefont {Kurkela}\ \emph
  {et~al.}(2019{\natexlab{a}})\citenamefont {Kurkela}, \citenamefont
  {Mazeliauskas}, \citenamefont {Paquet}, \citenamefont {Schlichting},\ and\
  \citenamefont {Teaney}}]{Kurkela:2018wud}%
  \BibitemOpen
  \bibfield  {author} {\bibinfo {author} {\bibfnamefont {A.}~\bibnamefont
  {Kurkela}}, \bibinfo {author} {\bibfnamefont {A.}~\bibnamefont
  {Mazeliauskas}}, \bibinfo {author} {\bibfnamefont {J.-F.}\ \bibnamefont
  {Paquet}}, \bibinfo {author} {\bibfnamefont {S.}~\bibnamefont {Schlichting}},
  \ and\ \bibinfo {author} {\bibfnamefont {D.}~\bibnamefont {Teaney}},\ }\href
  {\doibase 10.1103/PhysRevLett.122.122302} {\bibfield  {journal} {\bibinfo
  {journal} {Phys. Rev. Lett.}\ }\textbf {\bibinfo {volume} {122}},\ \bibinfo
  {pages} {122302} (\bibinfo {year} {2019}{\natexlab{a}})},\ \Eprint
  {http://arxiv.org/abs/1805.01604} {arXiv:1805.01604 [hep-ph]} \BibitemShut
  {NoStop}%
\bibitem [{\citenamefont {Strickland}\ and\ \citenamefont
  {Tantary}(2019)}]{Strickland:2019hff}%
  \BibitemOpen
  \bibfield  {author} {\bibinfo {author} {\bibfnamefont {M.}~\bibnamefont
  {Strickland}}\ and\ \bibinfo {author} {\bibfnamefont {U.}~\bibnamefont
  {Tantary}},\ }\href {\doibase 10.1007/JHEP10(2019)069} {\bibfield  {journal}
  {\bibinfo  {journal} {JHEP}\ }\textbf {\bibinfo {volume} {10}},\ \bibinfo
  {pages} {069} (\bibinfo {year} {2019})},\ \Eprint
  {http://arxiv.org/abs/1903.03145} {arXiv:1903.03145 [hep-ph]} \BibitemShut
  {NoStop}%
\bibitem [{\citenamefont {Kurkela}\ \emph
  {et~al.}(2019{\natexlab{b}})\citenamefont {Kurkela}, \citenamefont {van~der
  Schee}, \citenamefont {Wiedemann},\ and\ \citenamefont
  {Wu}}]{Kurkela:2019set}%
  \BibitemOpen
  \bibfield  {author} {\bibinfo {author} {\bibfnamefont {A.}~\bibnamefont
  {Kurkela}}, \bibinfo {author} {\bibfnamefont {W.}~\bibnamefont {van~der
  Schee}}, \bibinfo {author} {\bibfnamefont {U.~A.}\ \bibnamefont {Wiedemann}},
  \ and\ \bibinfo {author} {\bibfnamefont {B.}~\bibnamefont {Wu}},\ }\href@noop
  {} {\  (\bibinfo {year} {2019}{\natexlab{b}})},\ \Eprint
  {http://arxiv.org/abs/1907.08101} {arXiv:1907.08101 [hep-ph]} \BibitemShut
  {NoStop}%
\bibitem [{\citenamefont {Jaiswal}\ \emph {et~al.}(2019)\citenamefont
  {Jaiswal}, \citenamefont {Chattopadhyay}, \citenamefont {Jaiswal},
  \citenamefont {Pal},\ and\ \citenamefont {Heinz}}]{Jaiswal:2019cju}%
  \BibitemOpen
  \bibfield  {author} {\bibinfo {author} {\bibfnamefont {S.}~\bibnamefont
  {Jaiswal}}, \bibinfo {author} {\bibfnamefont {C.}~\bibnamefont
  {Chattopadhyay}}, \bibinfo {author} {\bibfnamefont {A.}~\bibnamefont
  {Jaiswal}}, \bibinfo {author} {\bibfnamefont {S.}~\bibnamefont {Pal}}, \ and\
  \bibinfo {author} {\bibfnamefont {U.}~\bibnamefont {Heinz}},\ }\href
  {\doibase 10.1103/PhysRevC.100.034901} {\bibfield  {journal} {\bibinfo
  {journal} {Phys. Rev.}\ }\textbf {\bibinfo {volume} {C100}},\ \bibinfo
  {pages} {034901} (\bibinfo {year} {2019})},\ \Eprint
  {http://arxiv.org/abs/1907.07965} {arXiv:1907.07965 [nucl-th]} \BibitemShut
  {NoStop}%
\bibitem [{\citenamefont {Denicol}\ and\ \citenamefont
  {Noronha}(2019{\natexlab{b}})}]{Denicol:2019lio}%
  \BibitemOpen
  \bibfield  {author} {\bibinfo {author} {\bibfnamefont {G.~S.}\ \bibnamefont
  {Denicol}}\ and\ \bibinfo {author} {\bibfnamefont {J.}~\bibnamefont
  {Noronha}},\ }\href@noop {} {\  (\bibinfo {year} {2019}{\natexlab{b}})},\
  \Eprint {http://arxiv.org/abs/1908.09957} {arXiv:1908.09957 [nucl-th]}
  \BibitemShut {NoStop}%
\bibitem [{\citenamefont {Brewer}\ \emph {et~al.}(2019)\citenamefont {Brewer},
  \citenamefont {Yan},\ and\ \citenamefont {Yin}}]{Brewer:2019oha}%
  \BibitemOpen
  \bibfield  {author} {\bibinfo {author} {\bibfnamefont {J.}~\bibnamefont
  {Brewer}}, \bibinfo {author} {\bibfnamefont {L.}~\bibnamefont {Yan}}, \ and\
  \bibinfo {author} {\bibfnamefont {Y.}~\bibnamefont {Yin}},\ }\href@noop {} {\
   (\bibinfo {year} {2019})},\ \Eprint {http://arxiv.org/abs/1910.00021}
  {arXiv:1910.00021 [nucl-th]} \BibitemShut {NoStop}%
\bibitem [{\citenamefont {Almaalol}\ \emph {et~al.}(2020)\citenamefont
  {Almaalol}, \citenamefont {Kurkela},\ and\ \citenamefont
  {Strickland}}]{Almaalol:2020rnu}%
  \BibitemOpen
  \bibfield  {author} {\bibinfo {author} {\bibfnamefont {D.}~\bibnamefont
  {Almaalol}}, \bibinfo {author} {\bibfnamefont {A.}~\bibnamefont {Kurkela}}, \
  and\ \bibinfo {author} {\bibfnamefont {M.}~\bibnamefont {Strickland}},\
  }\href@noop {} {\  (\bibinfo {year} {2020})},\ \Eprint
  {http://arxiv.org/abs/2004.05195} {arXiv:2004.05195 [hep-ph]} \BibitemShut
  {NoStop}%
\bibitem [{\citenamefont {Berges}\ \emph {et~al.}(2020)\citenamefont {Berges},
  \citenamefont {Heller}, \citenamefont {Mazeliauskas},\ and\ \citenamefont
  {Venugopalan}}]{Berges:2020fwq}%
  \BibitemOpen
  \bibfield  {author} {\bibinfo {author} {\bibfnamefont {J.}~\bibnamefont
  {Berges}}, \bibinfo {author} {\bibfnamefont {M.~P.}\ \bibnamefont {Heller}},
  \bibinfo {author} {\bibfnamefont {A.}~\bibnamefont {Mazeliauskas}}, \ and\
  \bibinfo {author} {\bibfnamefont {R.}~\bibnamefont {Venugopalan}},\
  }\href@noop {} {\  (\bibinfo {year} {2020})},\ \Eprint
  {http://arxiv.org/abs/2005.12299} {arXiv:2005.12299 [hep-th]} \BibitemShut
  {NoStop}%
\bibitem [{\citenamefont {Bemfica}\ \emph {et~al.}(2020)\citenamefont
  {Bemfica}, \citenamefont {Disconzi}, \citenamefont {Hoang}, \citenamefont
  {Noronha},\ and\ \citenamefont {Radosz}}]{Bemfica:2020xym}%
  \BibitemOpen
  \bibfield  {author} {\bibinfo {author} {\bibfnamefont {F.~S.}\ \bibnamefont
  {Bemfica}}, \bibinfo {author} {\bibfnamefont {M.~M.}\ \bibnamefont
  {Disconzi}}, \bibinfo {author} {\bibfnamefont {V.}~\bibnamefont {Hoang}},
  \bibinfo {author} {\bibfnamefont {J.}~\bibnamefont {Noronha}}, \ and\
  \bibinfo {author} {\bibfnamefont {M.}~\bibnamefont {Radosz}},\ }\href@noop {}
  {\  (\bibinfo {year} {2020})},\ \Eprint {http://arxiv.org/abs/2005.11632}
  {arXiv:2005.11632 [hep-th]} \BibitemShut {NoStop}%
\bibitem [{\citenamefont {Nunes~da Silva}\ \emph {et~al.}(2020)\citenamefont
  {Nunes~da Silva}, \citenamefont {Chinellato}, \citenamefont {Hippert},
  \citenamefont {Serenone}, \citenamefont {Takahashi}, \citenamefont {Denicol},
  \citenamefont {Luzum},\ and\ \citenamefont {Noronha}}]{NunesdaSilva:2020bfs}%
  \BibitemOpen
  \bibfield  {author} {\bibinfo {author} {\bibfnamefont {T.}~\bibnamefont
  {Nunes~da Silva}}, \bibinfo {author} {\bibfnamefont {D.}~\bibnamefont
  {Chinellato}}, \bibinfo {author} {\bibfnamefont {M.}~\bibnamefont {Hippert}},
  \bibinfo {author} {\bibfnamefont {W.}~\bibnamefont {Serenone}}, \bibinfo
  {author} {\bibfnamefont {J.}~\bibnamefont {Takahashi}}, \bibinfo {author}
  {\bibfnamefont {G.~S.}\ \bibnamefont {Denicol}}, \bibinfo {author}
  {\bibfnamefont {M.}~\bibnamefont {Luzum}}, \ and\ \bibinfo {author}
  {\bibfnamefont {J.}~\bibnamefont {Noronha}},\ }\href@noop {} {\  (\bibinfo
  {year} {2020})},\ \Eprint {http://arxiv.org/abs/2006.02324} {arXiv:2006.02324
  [nucl-th]} \BibitemShut {NoStop}%
\bibitem [{\citenamefont {Aguiar}\ \emph {et~al.}(2007)\citenamefont {Aguiar},
  \citenamefont {Kodama}, \citenamefont {Koide},\ and\ \citenamefont
  {Hama}}]{Aguiar:2007zz}%
  \BibitemOpen
  \bibfield  {author} {\bibinfo {author} {\bibfnamefont {C.~E.}\ \bibnamefont
  {Aguiar}}, \bibinfo {author} {\bibfnamefont {T.}~\bibnamefont {Kodama}},
  \bibinfo {author} {\bibfnamefont {T.}~\bibnamefont {Koide}}, \ and\ \bibinfo
  {author} {\bibfnamefont {Y.}~\bibnamefont {Hama}},\ }\bibfield  {booktitle}
  {\emph {\bibinfo {booktitle} {{Hadron interactions. Proceedings, 18th
  Workshop, RETINHA 18, Sao Paulo, Brazil, May 22-24, 2006}}},\ }\href
  {\doibase 10.1590/S0103-97332007000100028} {\bibfield  {journal} {\bibinfo
  {journal} {Braz. J. Phys.}\ }\textbf {\bibinfo {volume} {37}},\ \bibinfo
  {pages} {95} (\bibinfo {year} {2007})}\BibitemShut {NoStop}%
\bibitem [{\citenamefont {Petersen}\ \emph {et~al.}(2008)\citenamefont
  {Petersen}, \citenamefont {Steinheimer}, \citenamefont {Burau}, \citenamefont
  {Bleicher},\ and\ \citenamefont {Stocker}}]{Petersen:2008dd}%
  \BibitemOpen
  \bibfield  {author} {\bibinfo {author} {\bibfnamefont {H.}~\bibnamefont
  {Petersen}}, \bibinfo {author} {\bibfnamefont {J.}~\bibnamefont
  {Steinheimer}}, \bibinfo {author} {\bibfnamefont {G.}~\bibnamefont {Burau}},
  \bibinfo {author} {\bibfnamefont {M.}~\bibnamefont {Bleicher}}, \ and\
  \bibinfo {author} {\bibfnamefont {H.}~\bibnamefont {Stocker}},\ }\href
  {\doibase 10.1103/PhysRevC.78.044901} {\bibfield  {journal} {\bibinfo
  {journal} {Phys. Rev.}\ }\textbf {\bibinfo {volume} {C78}},\ \bibinfo {pages}
  {044901} (\bibinfo {year} {2008})},\ \Eprint {http://arxiv.org/abs/0806.1695}
  {arXiv:0806.1695 [nucl-th]} \BibitemShut {NoStop}%
\bibitem [{\citenamefont {Steinheimer}\ \emph {et~al.}(2010)\citenamefont
  {Steinheimer}, \citenamefont {Dexheimer}, \citenamefont {Petersen},
  \citenamefont {Bleicher}, \citenamefont {Schramm},\ and\ \citenamefont
  {Stoecker}}]{Steinheimer:2009nn}%
  \BibitemOpen
  \bibfield  {author} {\bibinfo {author} {\bibfnamefont {J.}~\bibnamefont
  {Steinheimer}}, \bibinfo {author} {\bibfnamefont {V.}~\bibnamefont
  {Dexheimer}}, \bibinfo {author} {\bibfnamefont {H.}~\bibnamefont {Petersen}},
  \bibinfo {author} {\bibfnamefont {M.}~\bibnamefont {Bleicher}}, \bibinfo
  {author} {\bibfnamefont {S.}~\bibnamefont {Schramm}}, \ and\ \bibinfo
  {author} {\bibfnamefont {H.}~\bibnamefont {Stoecker}},\ }\href {\doibase
  10.1103/PhysRevC.81.044913} {\bibfield  {journal} {\bibinfo  {journal} {Phys.
  Rev.}\ }\textbf {\bibinfo {volume} {C81}},\ \bibinfo {pages} {044913}
  (\bibinfo {year} {2010})},\ \Eprint {http://arxiv.org/abs/0905.3099}
  {arXiv:0905.3099 [hep-ph]} \BibitemShut {NoStop}%
\bibitem [{\citenamefont {Steinheimer}\ \emph {et~al.}(2011)\citenamefont
  {Steinheimer}, \citenamefont {Schramm},\ and\ \citenamefont
  {Stocker}}]{Steinheimer:2011ea}%
  \BibitemOpen
  \bibfield  {author} {\bibinfo {author} {\bibfnamefont {J.}~\bibnamefont
  {Steinheimer}}, \bibinfo {author} {\bibfnamefont {S.}~\bibnamefont
  {Schramm}}, \ and\ \bibinfo {author} {\bibfnamefont {H.}~\bibnamefont
  {Stocker}},\ }\href {\doibase 10.1103/PhysRevC.84.045208} {\bibfield
  {journal} {\bibinfo  {journal} {Phys. Rev. C}\ }\textbf {\bibinfo {volume}
  {84}},\ \bibinfo {pages} {045208} (\bibinfo {year} {2011})},\ \Eprint
  {http://arxiv.org/abs/1108.2596} {arXiv:1108.2596 [hep-ph]} \BibitemShut
  {NoStop}%
\bibitem [{\citenamefont {Karpenko}\ \emph {et~al.}(2014)\citenamefont
  {Karpenko}, \citenamefont {Huovinen},\ and\ \citenamefont
  {Bleicher}}]{Karpenko:2013wva}%
  \BibitemOpen
  \bibfield  {author} {\bibinfo {author} {\bibfnamefont {I.}~\bibnamefont
  {Karpenko}}, \bibinfo {author} {\bibfnamefont {P.}~\bibnamefont {Huovinen}},
  \ and\ \bibinfo {author} {\bibfnamefont {M.}~\bibnamefont {Bleicher}},\
  }\href {\doibase 10.1016/j.cpc.2014.07.010} {\bibfield  {journal} {\bibinfo
  {journal} {Comput. Phys. Commun.}\ }\textbf {\bibinfo {volume} {185}},\
  \bibinfo {pages} {3016} (\bibinfo {year} {2014})},\ \Eprint
  {http://arxiv.org/abs/1312.4160} {arXiv:1312.4160 [nucl-th]} \BibitemShut
  {NoStop}%
\bibitem [{\citenamefont {Rougemont}\ \emph {et~al.}(2015)\citenamefont
  {Rougemont}, \citenamefont {Noronha},\ and\ \citenamefont
  {Noronha-Hostler}}]{Rougemont:2015ona}%
  \BibitemOpen
  \bibfield  {author} {\bibinfo {author} {\bibfnamefont {R.}~\bibnamefont
  {Rougemont}}, \bibinfo {author} {\bibfnamefont {J.}~\bibnamefont {Noronha}},
  \ and\ \bibinfo {author} {\bibfnamefont {J.}~\bibnamefont
  {Noronha-Hostler}},\ }\href {\doibase 10.1103/PhysRevLett.115.202301}
  {\bibfield  {journal} {\bibinfo  {journal} {Phys. Rev. Lett.}\ }\textbf
  {\bibinfo {volume} {115}},\ \bibinfo {pages} {202301} (\bibinfo {year}
  {2015})},\ \Eprint {http://arxiv.org/abs/1507.06972} {arXiv:1507.06972
  [hep-ph]} \BibitemShut {NoStop}%
\bibitem [{\citenamefont {Feng}\ \emph {et~al.}(2018)\citenamefont {Feng},
  \citenamefont {Greiner}, \citenamefont {Shi},\ and\ \citenamefont
  {Xu}}]{Feng:2018anl}%
  \BibitemOpen
  \bibfield  {author} {\bibinfo {author} {\bibfnamefont {B.}~\bibnamefont
  {Feng}}, \bibinfo {author} {\bibfnamefont {C.}~\bibnamefont {Greiner}},
  \bibinfo {author} {\bibfnamefont {S.}~\bibnamefont {Shi}}, \ and\ \bibinfo
  {author} {\bibfnamefont {Z.}~\bibnamefont {Xu}},\ }\href {\doibase
  10.1016/j.physletb.2018.05.030} {\bibfield  {journal} {\bibinfo  {journal}
  {Phys. Lett. B}\ }\textbf {\bibinfo {volume} {782}},\ \bibinfo {pages} {262}
  (\bibinfo {year} {2018})},\ \Eprint {http://arxiv.org/abs/1802.02494}
  {arXiv:1802.02494 [hep-ph]} \BibitemShut {NoStop}%
\bibitem [{\citenamefont {Denicol}\ \emph {et~al.}(2018)\citenamefont
  {Denicol}, \citenamefont {Gale}, \citenamefont {Jeon}, \citenamefont
  {Monnai}, \citenamefont {Schenke},\ and\ \citenamefont
  {Shen}}]{Denicol:2018wdp}%
  \BibitemOpen
  \bibfield  {author} {\bibinfo {author} {\bibfnamefont {G.~S.}\ \bibnamefont
  {Denicol}}, \bibinfo {author} {\bibfnamefont {C.}~\bibnamefont {Gale}},
  \bibinfo {author} {\bibfnamefont {S.}~\bibnamefont {Jeon}}, \bibinfo {author}
  {\bibfnamefont {A.}~\bibnamefont {Monnai}}, \bibinfo {author} {\bibfnamefont
  {B.}~\bibnamefont {Schenke}}, \ and\ \bibinfo {author} {\bibfnamefont
  {C.}~\bibnamefont {Shen}},\ }\href {\doibase 10.1103/PhysRevC.98.034916}
  {\bibfield  {journal} {\bibinfo  {journal} {Phys. Rev.}\ }\textbf {\bibinfo
  {volume} {C98}},\ \bibinfo {pages} {034916} (\bibinfo {year} {2018})},\
  \Eprint {http://arxiv.org/abs/1804.10557} {arXiv:1804.10557 [nucl-th]}
  \BibitemShut {NoStop}%
\bibitem [{\citenamefont {Batyuk}\ \emph {et~al.}(2018)\citenamefont {Batyuk},
  \citenamefont {Blaschke}, \citenamefont {Bleicher}, \citenamefont {Ivanov},
  \citenamefont {Karpenko}, \citenamefont {Malinina}, \citenamefont {Merts},
  \citenamefont {Nahrgang}, \citenamefont {Petersen},\ and\ \citenamefont
  {Rogachevsky}}]{Batyuk:2017sku}%
  \BibitemOpen
  \bibfield  {author} {\bibinfo {author} {\bibfnamefont {P.}~\bibnamefont
  {Batyuk}}, \bibinfo {author} {\bibfnamefont {D.}~\bibnamefont {Blaschke}},
  \bibinfo {author} {\bibfnamefont {M.}~\bibnamefont {Bleicher}}, \bibinfo
  {author} {\bibfnamefont {{\relax Yu}.~B.}\ \bibnamefont {Ivanov}}, \bibinfo
  {author} {\bibfnamefont {I.}~\bibnamefont {Karpenko}}, \bibinfo {author}
  {\bibfnamefont {L.}~\bibnamefont {Malinina}}, \bibinfo {author}
  {\bibfnamefont {S.}~\bibnamefont {Merts}}, \bibinfo {author} {\bibfnamefont
  {M.}~\bibnamefont {Nahrgang}}, \bibinfo {author} {\bibfnamefont
  {H.}~\bibnamefont {Petersen}}, \ and\ \bibinfo {author} {\bibfnamefont
  {O.}~\bibnamefont {Rogachevsky}},\ }\bibfield  {booktitle} {\emph {\bibinfo
  {booktitle} {{Proceedings, 6th International Conference on New Frontiers in
  Physics (ICNFP 2017): Crete, Greece, August 17-29, 2017}}},\ }\href {\doibase
  10.1051/epjconf/201818202056} {\bibfield  {journal} {\bibinfo  {journal} {EPJ
  Web Conf.}\ }\textbf {\bibinfo {volume} {182}},\ \bibinfo {pages} {02056}
  (\bibinfo {year} {2018})},\ \Eprint {http://arxiv.org/abs/1711.07959}
  {arXiv:1711.07959 [nucl-th]} \BibitemShut {NoStop}%
\bibitem [{\citenamefont {Weil}\ \emph {et~al.}(2016)\citenamefont {Weil} \emph
  {et~al.}}]{Weil:2016zrk}%
  \BibitemOpen
  \bibfield  {author} {\bibinfo {author} {\bibfnamefont {J.}~\bibnamefont
  {Weil}} \emph {et~al.},\ }\href {\doibase 10.1103/PhysRevC.94.054905}
  {\bibfield  {journal} {\bibinfo  {journal} {Phys. Rev. C}\ }\textbf {\bibinfo
  {volume} {94}},\ \bibinfo {pages} {054905} (\bibinfo {year} {2016})},\
  \Eprint {http://arxiv.org/abs/1606.06642} {arXiv:1606.06642 [nucl-th]}
  \BibitemShut {NoStop}%
\bibitem [{\citenamefont {Steinheimer}\ \emph {et~al.}(2008)\citenamefont
  {Steinheimer}, \citenamefont {Bleicher}, \citenamefont {Petersen},
  \citenamefont {Schramm}, \citenamefont {Stocker},\ and\ \citenamefont
  {Zschiesche}}]{Steinheimer:2007iy}%
  \BibitemOpen
  \bibfield  {author} {\bibinfo {author} {\bibfnamefont {J.}~\bibnamefont
  {Steinheimer}}, \bibinfo {author} {\bibfnamefont {M.}~\bibnamefont
  {Bleicher}}, \bibinfo {author} {\bibfnamefont {H.}~\bibnamefont {Petersen}},
  \bibinfo {author} {\bibfnamefont {S.}~\bibnamefont {Schramm}}, \bibinfo
  {author} {\bibfnamefont {H.}~\bibnamefont {Stocker}}, \ and\ \bibinfo
  {author} {\bibfnamefont {D.}~\bibnamefont {Zschiesche}},\ }\href {\doibase
  10.1103/PhysRevC.77.034901} {\bibfield  {journal} {\bibinfo  {journal} {Phys.
  Rev.}\ }\textbf {\bibinfo {volume} {C77}},\ \bibinfo {pages} {034901}
  (\bibinfo {year} {2008})},\ \Eprint {http://arxiv.org/abs/0710.0332}
  {arXiv:0710.0332 [nucl-th]} \BibitemShut {NoStop}%
\bibitem [{\citenamefont {Werner}(1993)}]{Werner:1993uh}%
  \BibitemOpen
  \bibfield  {author} {\bibinfo {author} {\bibfnamefont {K.}~\bibnamefont
  {Werner}},\ }\emph {\bibinfo {title} {{Strings, pomerons, and the venus model
  of hadronic interactions at ultrarelativistic energies}}},\ \href {\doibase
  10.1016/0370-1573(93)90078-R} {\bibinfo {type} {Other thesis}} (\bibinfo
  {year} {1993})\BibitemShut {NoStop}%
\bibitem [{\citenamefont
  {Romatschke}(2017{\natexlab{b}})}]{Romatschke:2016hle}%
  \BibitemOpen
  \bibfield  {author} {\bibinfo {author} {\bibfnamefont {P.}~\bibnamefont
  {Romatschke}},\ }\href {\doibase 10.1140/epjc/s10052-016-4567-x} {\bibfield
  {journal} {\bibinfo  {journal} {Eur. Phys. J.}\ }\textbf {\bibinfo {volume}
  {C77}},\ \bibinfo {pages} {21} (\bibinfo {year} {2017}{\natexlab{b}})},\
  \Eprint {http://arxiv.org/abs/1609.02820} {arXiv:1609.02820 [nucl-th]}
  \BibitemShut {NoStop}%
\bibitem [{\citenamefont {Most}\ \emph {et~al.}(2020)\citenamefont {Most},
  \citenamefont {Jens~Papenfort}, \citenamefont {Dexheimer}, \citenamefont
  {Hanauske}, \citenamefont {Stoecker},\ and\ \citenamefont
  {Rezzolla}}]{Most:2019onn}%
  \BibitemOpen
  \bibfield  {author} {\bibinfo {author} {\bibfnamefont {E.~R.}\ \bibnamefont
  {Most}}, \bibinfo {author} {\bibfnamefont {L.}~\bibnamefont
  {Jens~Papenfort}}, \bibinfo {author} {\bibfnamefont {V.}~\bibnamefont
  {Dexheimer}}, \bibinfo {author} {\bibfnamefont {M.}~\bibnamefont {Hanauske}},
  \bibinfo {author} {\bibfnamefont {H.}~\bibnamefont {Stoecker}}, \ and\
  \bibinfo {author} {\bibfnamefont {L.}~\bibnamefont {Rezzolla}},\ }\href
  {\doibase 10.1140/epja/s10050-020-00073-4} {\bibfield  {journal} {\bibinfo
  {journal} {Eur. Phys. J. A}\ }\textbf {\bibinfo {volume} {56}},\ \bibinfo
  {pages} {59} (\bibinfo {year} {2020})},\ \Eprint
  {http://arxiv.org/abs/1910.13893} {arXiv:1910.13893 [astro-ph.HE]}
  \BibitemShut {NoStop}%
\bibitem [{\citenamefont {Haensel}\ \emph {et~al.}(2000)\citenamefont
  {Haensel}, \citenamefont {Levenfish},\ and\ \citenamefont
  {Yakovlev}}]{Haensel:2000vz}%
  \BibitemOpen
  \bibfield  {author} {\bibinfo {author} {\bibfnamefont {P.}~\bibnamefont
  {Haensel}}, \bibinfo {author} {\bibfnamefont {K.}~\bibnamefont {Levenfish}},
  \ and\ \bibinfo {author} {\bibfnamefont {D.}~\bibnamefont {Yakovlev}},\
  }\href@noop {} {\bibfield  {journal} {\bibinfo  {journal} {Astron.
  Astrophys.}\ }\textbf {\bibinfo {volume} {357}},\ \bibinfo {pages} {1157}
  (\bibinfo {year} {2000})},\ \Eprint {http://arxiv.org/abs/astro-ph/0004183}
  {arXiv:astro-ph/0004183} \BibitemShut {NoStop}%
\bibitem [{\citenamefont {Alford}\ \emph {et~al.}(2010)\citenamefont {Alford},
  \citenamefont {Mahmoodifar},\ and\ \citenamefont
  {Schwenzer}}]{Alford:2010gw}%
  \BibitemOpen
  \bibfield  {author} {\bibinfo {author} {\bibfnamefont {M.~G.}\ \bibnamefont
  {Alford}}, \bibinfo {author} {\bibfnamefont {S.}~\bibnamefont {Mahmoodifar}},
  \ and\ \bibinfo {author} {\bibfnamefont {K.}~\bibnamefont {Schwenzer}},\
  }\href {\doibase 10.1088/0954-3899/37/12/125202} {\bibfield  {journal}
  {\bibinfo  {journal} {J. Phys. G}\ }\textbf {\bibinfo {volume} {37}},\
  \bibinfo {pages} {125202} (\bibinfo {year} {2010})},\ \Eprint
  {http://arxiv.org/abs/1005.3769} {arXiv:1005.3769 [nucl-th]} \BibitemShut
  {NoStop}%
\bibitem [{\citenamefont {Alford}\ \emph {et~al.}(2018)\citenamefont {Alford},
  \citenamefont {Bovard}, \citenamefont {Hanauske}, \citenamefont {Rezzolla},\
  and\ \citenamefont {Schwenzer}}]{Alford:2017rxf}%
  \BibitemOpen
  \bibfield  {author} {\bibinfo {author} {\bibfnamefont {M.~G.}\ \bibnamefont
  {Alford}}, \bibinfo {author} {\bibfnamefont {L.}~\bibnamefont {Bovard}},
  \bibinfo {author} {\bibfnamefont {M.}~\bibnamefont {Hanauske}}, \bibinfo
  {author} {\bibfnamefont {L.}~\bibnamefont {Rezzolla}}, \ and\ \bibinfo
  {author} {\bibfnamefont {K.}~\bibnamefont {Schwenzer}},\ }\href {\doibase
  10.1103/PhysRevLett.120.041101} {\bibfield  {journal} {\bibinfo  {journal}
  {Phys. Rev. Lett.}\ }\textbf {\bibinfo {volume} {120}},\ \bibinfo {pages}
  {041101} (\bibinfo {year} {2018})},\ \Eprint
  {http://arxiv.org/abs/1707.09475} {arXiv:1707.09475 [gr-qc]} \BibitemShut
  {NoStop}%
\bibitem [{\citenamefont {Alford}\ and\ \citenamefont
  {Harris}(2019)}]{Alford:2019qtm}%
  \BibitemOpen
  \bibfield  {author} {\bibinfo {author} {\bibfnamefont {M.~G.}\ \bibnamefont
  {Alford}}\ and\ \bibinfo {author} {\bibfnamefont {S.~P.}\ \bibnamefont
  {Harris}},\ }\href {\doibase 10.1103/PhysRevC.100.035803} {\bibfield
  {journal} {\bibinfo  {journal} {Phys. Rev. C}\ }\textbf {\bibinfo {volume}
  {100}},\ \bibinfo {pages} {035803} (\bibinfo {year} {2019})},\ \Eprint
  {http://arxiv.org/abs/1907.03795} {arXiv:1907.03795 [nucl-th]} \BibitemShut
  {NoStop}%
\bibitem [{\citenamefont {Bemfica}\ \emph
  {et~al.}(2019{\natexlab{a}})\citenamefont {Bemfica}, \citenamefont
  {Disconzi},\ and\ \citenamefont {Noronha}}]{Bemfica:2019cop}%
  \BibitemOpen
  \bibfield  {author} {\bibinfo {author} {\bibfnamefont {F.~S.}\ \bibnamefont
  {Bemfica}}, \bibinfo {author} {\bibfnamefont {M.~M.}\ \bibnamefont
  {Disconzi}}, \ and\ \bibinfo {author} {\bibfnamefont {J.}~\bibnamefont
  {Noronha}},\ }\href {\doibase 10.1103/PhysRevLett.122.221602} {\bibfield
  {journal} {\bibinfo  {journal} {Phys. Rev. Lett.}\ }\textbf {\bibinfo
  {volume} {122}},\ \bibinfo {pages} {221602} (\bibinfo {year}
  {2019}{\natexlab{a}})},\ \Eprint {http://arxiv.org/abs/1901.06701}
  {arXiv:1901.06701 [gr-qc]} \BibitemShut {NoStop}%
\bibitem [{\citenamefont {Bemfica}\ \emph {et~al.}(2018)\citenamefont
  {Bemfica}, \citenamefont {Disconzi},\ and\ \citenamefont
  {Noronha}}]{Bemfica:2017wps}%
  \BibitemOpen
  \bibfield  {author} {\bibinfo {author} {\bibfnamefont {F.~S.}\ \bibnamefont
  {Bemfica}}, \bibinfo {author} {\bibfnamefont {M.~M.}\ \bibnamefont
  {Disconzi}}, \ and\ \bibinfo {author} {\bibfnamefont {J.}~\bibnamefont
  {Noronha}},\ }\href {\doibase 10.1103/PhysRevD.98.104064} {\bibfield
  {journal} {\bibinfo  {journal} {Phys. Rev. D}\ }\textbf {\bibinfo {volume}
  {98}},\ \bibinfo {pages} {104064} (\bibinfo {year} {2018})},\ \Eprint
  {http://arxiv.org/abs/1708.06255} {arXiv:1708.06255 [gr-qc]} \BibitemShut
  {NoStop}%
\bibitem [{\citenamefont {Bebie}\ \emph {et~al.}(1992)\citenamefont {Bebie},
  \citenamefont {Gerber}, \citenamefont {Goity},\ and\ \citenamefont
  {Leutwyler}}]{Bebie:1991ij}%
  \BibitemOpen
  \bibfield  {author} {\bibinfo {author} {\bibfnamefont {H.}~\bibnamefont
  {Bebie}}, \bibinfo {author} {\bibfnamefont {P.}~\bibnamefont {Gerber}},
  \bibinfo {author} {\bibfnamefont {J.}~\bibnamefont {Goity}}, \ and\ \bibinfo
  {author} {\bibfnamefont {H.}~\bibnamefont {Leutwyler}},\ }\href {\doibase
  10.1016/0550-3213(92)90005-V} {\bibfield  {journal} {\bibinfo  {journal}
  {Nucl. Phys. B}\ }\textbf {\bibinfo {volume} {378}},\ \bibinfo {pages} {95}
  (\bibinfo {year} {1992})}\BibitemShut {NoStop}%
\bibitem [{\citenamefont {Günther}\ \emph {et~al.}(2017)\citenamefont
  {Günther}, \citenamefont {Bellwied}, \citenamefont {Borsanyi}, \citenamefont
  {Fodor}, \citenamefont {Katz}, \citenamefont {Pasztor},\ and\ \citenamefont
  {Ratti}}]{Gunther:2017sxn}%
  \BibitemOpen
  \bibfield  {author} {\bibinfo {author} {\bibfnamefont {J.}~\bibnamefont
  {Günther}}, \bibinfo {author} {\bibfnamefont {R.}~\bibnamefont {Bellwied}},
  \bibinfo {author} {\bibfnamefont {S.}~\bibnamefont {Borsanyi}}, \bibinfo
  {author} {\bibfnamefont {Z.}~\bibnamefont {Fodor}}, \bibinfo {author}
  {\bibfnamefont {S.~D.}\ \bibnamefont {Katz}}, \bibinfo {author}
  {\bibfnamefont {A.}~\bibnamefont {Pasztor}}, \ and\ \bibinfo {author}
  {\bibfnamefont {C.}~\bibnamefont {Ratti}},\ }\bibfield  {booktitle} {\emph
  {\bibinfo {booktitle} {{Proceedings, 12th Conference on Quark Confinement and
  the Hadron Spectrum (Confinement XII): Thessaloniki, Greece}}},\ }\href
  {\doibase 10.1051/epjconf/201713707008} {\bibfield  {journal} {\bibinfo
  {journal} {EPJ Web Conf.}\ }\textbf {\bibinfo {volume} {137}},\ \bibinfo
  {pages} {07008} (\bibinfo {year} {2017})}\BibitemShut {NoStop}%
\bibitem [{\citenamefont {Bellwied}\ \emph {et~al.}(2019)\citenamefont
  {Bellwied}, \citenamefont {Noronha-Hostler}, \citenamefont {Parotto},
  \citenamefont {Portillo~Vazquez}, \citenamefont {Ratti},\ and\ \citenamefont
  {Stafford}}]{Bellwied:2018tkc}%
  \BibitemOpen
  \bibfield  {author} {\bibinfo {author} {\bibfnamefont {R.}~\bibnamefont
  {Bellwied}}, \bibinfo {author} {\bibfnamefont {J.}~\bibnamefont
  {Noronha-Hostler}}, \bibinfo {author} {\bibfnamefont {P.}~\bibnamefont
  {Parotto}}, \bibinfo {author} {\bibfnamefont {I.}~\bibnamefont
  {Portillo~Vazquez}}, \bibinfo {author} {\bibfnamefont {C.}~\bibnamefont
  {Ratti}}, \ and\ \bibinfo {author} {\bibfnamefont {J.~M.}\ \bibnamefont
  {Stafford}},\ }\href {\doibase 10.1103/PhysRevC.99.034912} {\bibfield
  {journal} {\bibinfo  {journal} {Phys. Rev.}\ }\textbf {\bibinfo {volume}
  {C99}},\ \bibinfo {pages} {034912} (\bibinfo {year} {2019})},\ \Eprint
  {http://arxiv.org/abs/1805.00088} {arXiv:1805.00088 [hep-ph]} \BibitemShut
  {NoStop}%
\bibitem [{\citenamefont {Parotto}\ \emph {et~al.}(2018)\citenamefont
  {Parotto}, \citenamefont {Bluhm}, \citenamefont {Mroczek}, \citenamefont
  {Nahrgang}, \citenamefont {Noronha-Hostler}, \citenamefont {Rajagopal},
  \citenamefont {Ratti}, \citenamefont {Schäfer},\ and\ \citenamefont
  {Stephanov}}]{Parotto:2018pwx}%
  \BibitemOpen
  \bibfield  {author} {\bibinfo {author} {\bibfnamefont {P.}~\bibnamefont
  {Parotto}}, \bibinfo {author} {\bibfnamefont {M.}~\bibnamefont {Bluhm}},
  \bibinfo {author} {\bibfnamefont {D.}~\bibnamefont {Mroczek}}, \bibinfo
  {author} {\bibfnamefont {M.}~\bibnamefont {Nahrgang}}, \bibinfo {author}
  {\bibfnamefont {J.}~\bibnamefont {Noronha-Hostler}}, \bibinfo {author}
  {\bibfnamefont {K.}~\bibnamefont {Rajagopal}}, \bibinfo {author}
  {\bibfnamefont {C.}~\bibnamefont {Ratti}}, \bibinfo {author} {\bibfnamefont
  {T.}~\bibnamefont {Schäfer}}, \ and\ \bibinfo {author} {\bibfnamefont
  {M.}~\bibnamefont {Stephanov}},\ }\href@noop {} {\  (\bibinfo {year}
  {2018})},\ \Eprint {http://arxiv.org/abs/1805.05249} {arXiv:1805.05249
  [hep-ph]} \BibitemShut {NoStop}%
\bibitem [{\citenamefont {Noronha-Hostler}\ \emph {et~al.}(2019)\citenamefont
  {Noronha-Hostler}, \citenamefont {Parotto}, \citenamefont {Ratti},\ and\
  \citenamefont {Stafford}}]{Noronha-Hostler:2019ayj}%
  \BibitemOpen
  \bibfield  {author} {\bibinfo {author} {\bibfnamefont {J.}~\bibnamefont
  {Noronha-Hostler}}, \bibinfo {author} {\bibfnamefont {P.}~\bibnamefont
  {Parotto}}, \bibinfo {author} {\bibfnamefont {C.}~\bibnamefont {Ratti}}, \
  and\ \bibinfo {author} {\bibfnamefont {J.~M.}\ \bibnamefont {Stafford}},\
  }\href {\doibase 10.1103/PhysRevC.100.064910} {\bibfield  {journal} {\bibinfo
   {journal} {Phys. Rev.}\ }\textbf {\bibinfo {volume} {C100}},\ \bibinfo
  {pages} {064910} (\bibinfo {year} {2019})},\ \Eprint
  {http://arxiv.org/abs/1902.06723} {arXiv:1902.06723 [hep-ph]} \BibitemShut
  {NoStop}%
\bibitem [{\citenamefont {Monnai}\ \emph {et~al.}(2019)\citenamefont {Monnai},
  \citenamefont {Schenke},\ and\ \citenamefont {Shen}}]{Monnai:2019hkn}%
  \BibitemOpen
  \bibfield  {author} {\bibinfo {author} {\bibfnamefont {A.}~\bibnamefont
  {Monnai}}, \bibinfo {author} {\bibfnamefont {B.}~\bibnamefont {Schenke}}, \
  and\ \bibinfo {author} {\bibfnamefont {C.}~\bibnamefont {Shen}},\ }\href
  {\doibase 10.1103/PhysRevC.100.024907} {\bibfield  {journal} {\bibinfo
  {journal} {Phys. Rev.}\ }\textbf {\bibinfo {volume} {C100}},\ \bibinfo
  {pages} {024907} (\bibinfo {year} {2019})},\ \Eprint
  {http://arxiv.org/abs/1902.05095} {arXiv:1902.05095 [nucl-th]} \BibitemShut
  {NoStop}%
\bibitem [{\citenamefont {Stafford}\ \emph {et~al.}(2019)\citenamefont
  {Stafford}, \citenamefont {Alba}, \citenamefont {Bellwied}, \citenamefont
  {Mantovani-Sarti}, \citenamefont {Noronha-Hostler}, \citenamefont {Parotto},
  \citenamefont {Portillo-Vazquez},\ and\ \citenamefont
  {Ratti}}]{Stafford:2019yuy}%
  \BibitemOpen
  \bibfield  {author} {\bibinfo {author} {\bibfnamefont {J.~M.}\ \bibnamefont
  {Stafford}}, \bibinfo {author} {\bibfnamefont {P.}~\bibnamefont {Alba}},
  \bibinfo {author} {\bibfnamefont {R.}~\bibnamefont {Bellwied}}, \bibinfo
  {author} {\bibfnamefont {V.}~\bibnamefont {Mantovani-Sarti}}, \bibinfo
  {author} {\bibfnamefont {J.}~\bibnamefont {Noronha-Hostler}}, \bibinfo
  {author} {\bibfnamefont {P.}~\bibnamefont {Parotto}}, \bibinfo {author}
  {\bibfnamefont {I.}~\bibnamefont {Portillo-Vazquez}}, \ and\ \bibinfo
  {author} {\bibfnamefont {C.}~\bibnamefont {Ratti}},\ }in\ \href@noop {}
  {\emph {\bibinfo {booktitle} {{18th International Conference on Strangeness
  in Quark Matter (SQM 2019) Bari, Italy, June 10-15, 2019}}}}\ (\bibinfo
  {year} {2019})\ \Eprint {http://arxiv.org/abs/1912.12968} {arXiv:1912.12968
  [hep-ph]} \BibitemShut {NoStop}%
\bibitem [{\citenamefont {Demir}\ and\ \citenamefont
  {Bass}(2009)}]{Demir:2008tr}%
  \BibitemOpen
  \bibfield  {author} {\bibinfo {author} {\bibfnamefont {N.}~\bibnamefont
  {Demir}}\ and\ \bibinfo {author} {\bibfnamefont {S.~A.}\ \bibnamefont
  {Bass}},\ }\href {\doibase 10.1103/PhysRevLett.102.172302} {\bibfield
  {journal} {\bibinfo  {journal} {Phys. Rev. Lett.}\ }\textbf {\bibinfo
  {volume} {102}},\ \bibinfo {pages} {172302} (\bibinfo {year} {2009})},\
  \Eprint {http://arxiv.org/abs/0812.2422} {arXiv:0812.2422 [nucl-th]}
  \BibitemShut {NoStop}%
\bibitem [{\citenamefont {Denicol}\ \emph {et~al.}(2013)\citenamefont
  {Denicol}, \citenamefont {Gale}, \citenamefont {Jeon},\ and\ \citenamefont
  {Noronha}}]{Denicol:2013nua}%
  \BibitemOpen
  \bibfield  {author} {\bibinfo {author} {\bibfnamefont {G.~S.}\ \bibnamefont
  {Denicol}}, \bibinfo {author} {\bibfnamefont {C.}~\bibnamefont {Gale}},
  \bibinfo {author} {\bibfnamefont {S.}~\bibnamefont {Jeon}}, \ and\ \bibinfo
  {author} {\bibfnamefont {J.}~\bibnamefont {Noronha}},\ }\href {\doibase
  10.1103/PhysRevC.88.064901} {\bibfield  {journal} {\bibinfo  {journal} {Phys.
  Rev.}\ }\textbf {\bibinfo {volume} {C88}},\ \bibinfo {pages} {064901}
  (\bibinfo {year} {2013})},\ \Eprint {http://arxiv.org/abs/1308.1923}
  {arXiv:1308.1923 [nucl-th]} \BibitemShut {NoStop}%
\bibitem [{\citenamefont {Kadam}\ and\ \citenamefont
  {Mishra}(2014)}]{Kadam:2014cua}%
  \BibitemOpen
  \bibfield  {author} {\bibinfo {author} {\bibfnamefont {G.~P.}\ \bibnamefont
  {Kadam}}\ and\ \bibinfo {author} {\bibfnamefont {H.}~\bibnamefont {Mishra}},\
  }\href {\doibase 10.1016/j.nuclphysa.2014.12.004} {\bibfield  {journal}
  {\bibinfo  {journal} {Nucl. Phys.}\ }\textbf {\bibinfo {volume} {A934}},\
  \bibinfo {pages} {133} (\bibinfo {year} {2014})},\ \Eprint
  {http://arxiv.org/abs/1408.6329} {arXiv:1408.6329 [hep-ph]} \BibitemShut
  {NoStop}%
\bibitem [{\citenamefont {Martinez}\ \emph
  {et~al.}(2019{\natexlab{c}})\citenamefont {Martinez}, \citenamefont
  {Schäfer},\ and\ \citenamefont {Skokov}}]{Martinez:2019bsn}%
  \BibitemOpen
  \bibfield  {author} {\bibinfo {author} {\bibfnamefont {M.}~\bibnamefont
  {Martinez}}, \bibinfo {author} {\bibfnamefont {T.}~\bibnamefont {Schäfer}},
  \ and\ \bibinfo {author} {\bibfnamefont {V.}~\bibnamefont {Skokov}},\ }\href
  {\doibase 10.1103/PhysRevD.100.074017} {\bibfield  {journal} {\bibinfo
  {journal} {Phys. Rev.}\ }\textbf {\bibinfo {volume} {D100}},\ \bibinfo
  {pages} {074017} (\bibinfo {year} {2019}{\natexlab{c}})},\ \Eprint
  {http://arxiv.org/abs/1906.11306} {arXiv:1906.11306 [hep-ph]} \BibitemShut
  {NoStop}%
\bibitem [{\citenamefont {Rougemont}\ \emph {et~al.}(2017)\citenamefont
  {Rougemont}, \citenamefont {Critelli}, \citenamefont {Noronha-Hostler},
  \citenamefont {Noronha},\ and\ \citenamefont {Ratti}}]{Rougemont:2017tlu}%
  \BibitemOpen
  \bibfield  {author} {\bibinfo {author} {\bibfnamefont {R.}~\bibnamefont
  {Rougemont}}, \bibinfo {author} {\bibfnamefont {R.}~\bibnamefont {Critelli}},
  \bibinfo {author} {\bibfnamefont {J.}~\bibnamefont {Noronha-Hostler}},
  \bibinfo {author} {\bibfnamefont {J.}~\bibnamefont {Noronha}}, \ and\
  \bibinfo {author} {\bibfnamefont {C.}~\bibnamefont {Ratti}},\ }\href
  {\doibase 10.1103/PhysRevD.96.014032} {\bibfield  {journal} {\bibinfo
  {journal} {Phys. Rev.}\ }\textbf {\bibinfo {volume} {D96}},\ \bibinfo {pages}
  {014032} (\bibinfo {year} {2017})},\ \Eprint
  {http://arxiv.org/abs/1704.05558} {arXiv:1704.05558 [hep-ph]} \BibitemShut
  {NoStop}%
\bibitem [{\citenamefont {Son}\ and\ \citenamefont
  {Stephanov}(2004)}]{Son:2004iv}%
  \BibitemOpen
  \bibfield  {author} {\bibinfo {author} {\bibfnamefont {D.}~\bibnamefont
  {Son}}\ and\ \bibinfo {author} {\bibfnamefont {M.}~\bibnamefont
  {Stephanov}},\ }\href {\doibase 10.1103/PhysRevD.70.056001} {\bibfield
  {journal} {\bibinfo  {journal} {Phys. Rev. D}\ }\textbf {\bibinfo {volume}
  {70}},\ \bibinfo {pages} {056001} (\bibinfo {year} {2004})},\ \Eprint
  {http://arxiv.org/abs/hep-ph/0401052} {arXiv:hep-ph/0401052} \BibitemShut
  {NoStop}%
\bibitem [{\citenamefont {Bjorken}(1983)}]{Bjorken:1982qr}%
  \BibitemOpen
  \bibfield  {author} {\bibinfo {author} {\bibfnamefont {J.}~\bibnamefont
  {Bjorken}},\ }\href {\doibase 10.1103/PhysRevD.27.140} {\bibfield  {journal}
  {\bibinfo  {journal} {Phys. Rev. D}\ }\textbf {\bibinfo {volume} {27}},\
  \bibinfo {pages} {140} (\bibinfo {year} {1983})}\BibitemShut {NoStop}%
\bibitem [{\citenamefont {Muronga}(2004)}]{Muronga:2003ta}%
  \BibitemOpen
  \bibfield  {author} {\bibinfo {author} {\bibfnamefont {A.}~\bibnamefont
  {Muronga}},\ }\href {\doibase 10.1103/PhysRevC.69.034903} {\bibfield
  {journal} {\bibinfo  {journal} {Phys. Rev. C}\ }\textbf {\bibinfo {volume}
  {69}},\ \bibinfo {pages} {034903} (\bibinfo {year} {2004})},\ \Eprint
  {http://arxiv.org/abs/nucl-th/0309055} {arXiv:nucl-th/0309055} \BibitemShut
  {NoStop}%
\bibitem [{\citenamefont {Denicol}\ \emph {et~al.}(2012)\citenamefont
  {Denicol}, \citenamefont {Niemi}, \citenamefont {Molnar},\ and\ \citenamefont
  {Rischke}}]{Denicol:2012cn}%
  \BibitemOpen
  \bibfield  {author} {\bibinfo {author} {\bibfnamefont {G.~S.}\ \bibnamefont
  {Denicol}}, \bibinfo {author} {\bibfnamefont {H.}~\bibnamefont {Niemi}},
  \bibinfo {author} {\bibfnamefont {E.}~\bibnamefont {Molnar}}, \ and\ \bibinfo
  {author} {\bibfnamefont {D.~H.}\ \bibnamefont {Rischke}},\ }\href {\doibase
  10.1103/PhysRevD.85.114047, 10.1103/PhysRevD.91.039902} {\bibfield  {journal}
  {\bibinfo  {journal} {Phys. Rev.}\ }\textbf {\bibinfo {volume} {D85}},\
  \bibinfo {pages} {114047} (\bibinfo {year} {2012})},\ \bibinfo {note}
  {[Erratum: Phys. Rev.D91,no.3,039902(2015)]},\ \Eprint
  {http://arxiv.org/abs/1202.4551} {arXiv:1202.4551 [nucl-th]} \BibitemShut
  {NoStop}%
\bibitem [{\citenamefont {Israel}\ and\ \citenamefont
  {Stewart}(1979)}]{Israel:1979wp}%
  \BibitemOpen
  \bibfield  {author} {\bibinfo {author} {\bibfnamefont {W.}~\bibnamefont
  {Israel}}\ and\ \bibinfo {author} {\bibfnamefont {J.~M.}\ \bibnamefont
  {Stewart}},\ }\href {\doibase 10.1016/0003-4916(79)90130-1} {\bibfield
  {journal} {\bibinfo  {journal} {Annals Phys.}\ }\textbf {\bibinfo {volume}
  {118}},\ \bibinfo {pages} {341} (\bibinfo {year} {1979})}\BibitemShut
  {NoStop}%
\bibitem [{\citenamefont {Bazow}\ \emph {et~al.}(2018)\citenamefont {Bazow},
  \citenamefont {Heinz},\ and\ \citenamefont {Strickland}}]{Bazow:2016yra}%
  \BibitemOpen
  \bibfield  {author} {\bibinfo {author} {\bibfnamefont {D.}~\bibnamefont
  {Bazow}}, \bibinfo {author} {\bibfnamefont {U.~W.}\ \bibnamefont {Heinz}}, \
  and\ \bibinfo {author} {\bibfnamefont {M.}~\bibnamefont {Strickland}},\
  }\href {\doibase 10.1016/j.cpc.2017.01.015} {\bibfield  {journal} {\bibinfo
  {journal} {Comput. Phys. Commun.}\ }\textbf {\bibinfo {volume} {225}},\
  \bibinfo {pages} {92} (\bibinfo {year} {2018})},\ \Eprint
  {http://arxiv.org/abs/1608.06577} {arXiv:1608.06577 [physics.comp-ph]}
  \BibitemShut {NoStop}%
\bibitem [{\citenamefont {Denicol}\ \emph
  {et~al.}(2014{\natexlab{a}})\citenamefont {Denicol}, \citenamefont {Jeon},\
  and\ \citenamefont {Gale}}]{Denicol:2014vaa}%
  \BibitemOpen
  \bibfield  {author} {\bibinfo {author} {\bibfnamefont {G.~S.}\ \bibnamefont
  {Denicol}}, \bibinfo {author} {\bibfnamefont {S.}~\bibnamefont {Jeon}}, \
  and\ \bibinfo {author} {\bibfnamefont {C.}~\bibnamefont {Gale}},\ }\href
  {\doibase 10.1103/PhysRevC.90.024912} {\bibfield  {journal} {\bibinfo
  {journal} {Phys. Rev.}\ }\textbf {\bibinfo {volume} {C90}},\ \bibinfo {pages}
  {024912} (\bibinfo {year} {2014}{\natexlab{a}})},\ \Eprint
  {http://arxiv.org/abs/1403.0962} {arXiv:1403.0962 [nucl-th]} \BibitemShut
  {NoStop}%
\bibitem [{\citenamefont {Denicol}\ \emph
  {et~al.}(2014{\natexlab{b}})\citenamefont {Denicol}, \citenamefont
  {Florkowski}, \citenamefont {Ryblewski},\ and\ \citenamefont
  {Strickland}}]{Denicol:2014mca}%
  \BibitemOpen
  \bibfield  {author} {\bibinfo {author} {\bibfnamefont {G.~S.}\ \bibnamefont
  {Denicol}}, \bibinfo {author} {\bibfnamefont {W.}~\bibnamefont {Florkowski}},
  \bibinfo {author} {\bibfnamefont {R.}~\bibnamefont {Ryblewski}}, \ and\
  \bibinfo {author} {\bibfnamefont {M.}~\bibnamefont {Strickland}},\ }\href
  {\doibase 10.1103/PhysRevC.90.044905} {\bibfield  {journal} {\bibinfo
  {journal} {Phys. Rev. C}\ }\textbf {\bibinfo {volume} {90}},\ \bibinfo
  {pages} {044905} (\bibinfo {year} {2014}{\natexlab{b}})},\ \Eprint
  {http://arxiv.org/abs/1407.4767} {arXiv:1407.4767 [hep-ph]} \BibitemShut
  {NoStop}%
\bibitem [{\citenamefont {Noronha-Hostler}\ \emph {et~al.}(2012)\citenamefont
  {Noronha-Hostler}, \citenamefont {Noronha},\ and\ \citenamefont
  {Greiner}}]{NoronhaHostler:2012ug}%
  \BibitemOpen
  \bibfield  {author} {\bibinfo {author} {\bibfnamefont {J.}~\bibnamefont
  {Noronha-Hostler}}, \bibinfo {author} {\bibfnamefont {J.}~\bibnamefont
  {Noronha}}, \ and\ \bibinfo {author} {\bibfnamefont {C.}~\bibnamefont
  {Greiner}},\ }\href {\doibase 10.1103/PhysRevC.86.024913} {\bibfield
  {journal} {\bibinfo  {journal} {Phys. Rev.}\ }\textbf {\bibinfo {volume}
  {C86}},\ \bibinfo {pages} {024913} (\bibinfo {year} {2012})},\ \Eprint
  {http://arxiv.org/abs/1206.5138} {arXiv:1206.5138 [nucl-th]} \BibitemShut
  {NoStop}%
\bibitem [{\citenamefont {Christiansen}\ \emph {et~al.}(2015)\citenamefont
  {Christiansen}, \citenamefont {Haas}, \citenamefont {Pawlowski},\ and\
  \citenamefont {Strodthoff}}]{Christiansen:2014ypa}%
  \BibitemOpen
  \bibfield  {author} {\bibinfo {author} {\bibfnamefont {N.}~\bibnamefont
  {Christiansen}}, \bibinfo {author} {\bibfnamefont {M.}~\bibnamefont {Haas}},
  \bibinfo {author} {\bibfnamefont {J.~M.}\ \bibnamefont {Pawlowski}}, \ and\
  \bibinfo {author} {\bibfnamefont {N.}~\bibnamefont {Strodthoff}},\ }\href
  {\doibase 10.1103/PhysRevLett.115.112002} {\bibfield  {journal} {\bibinfo
  {journal} {Phys. Rev. Lett.}\ }\textbf {\bibinfo {volume} {115}},\ \bibinfo
  {pages} {112002} (\bibinfo {year} {2015})},\ \Eprint
  {http://arxiv.org/abs/1411.7986} {arXiv:1411.7986 [hep-ph]} \BibitemShut
  {NoStop}%
\bibitem [{\citenamefont {Dubla}\ \emph {et~al.}(2018)\citenamefont {Dubla},
  \citenamefont {Masciocchi}, \citenamefont {Pawlowski}, \citenamefont
  {Schenke}, \citenamefont {Shen},\ and\ \citenamefont
  {Stachel}}]{Dubla:2018czx}%
  \BibitemOpen
  \bibfield  {author} {\bibinfo {author} {\bibfnamefont {A.}~\bibnamefont
  {Dubla}}, \bibinfo {author} {\bibfnamefont {S.}~\bibnamefont {Masciocchi}},
  \bibinfo {author} {\bibfnamefont {J.~M.}\ \bibnamefont {Pawlowski}}, \bibinfo
  {author} {\bibfnamefont {B.}~\bibnamefont {Schenke}}, \bibinfo {author}
  {\bibfnamefont {C.}~\bibnamefont {Shen}}, \ and\ \bibinfo {author}
  {\bibfnamefont {J.}~\bibnamefont {Stachel}},\ }\href {\doibase
  10.1016/j.nuclphysa.2018.09.046} {\bibfield  {journal} {\bibinfo  {journal}
  {Nucl. Phys.}\ }\textbf {\bibinfo {volume} {A979}},\ \bibinfo {pages} {251}
  (\bibinfo {year} {2018})},\ \Eprint {http://arxiv.org/abs/1805.02985}
  {arXiv:1805.02985 [nucl-th]} \BibitemShut {NoStop}%
\bibitem [{\citenamefont {Bernhard}\ \emph {et~al.}(2016)\citenamefont
  {Bernhard}, \citenamefont {Moreland}, \citenamefont {Bass}, \citenamefont
  {Liu},\ and\ \citenamefont {Heinz}}]{Bernhard:2016tnd}%
  \BibitemOpen
  \bibfield  {author} {\bibinfo {author} {\bibfnamefont {J.~E.}\ \bibnamefont
  {Bernhard}}, \bibinfo {author} {\bibfnamefont {J.~S.}\ \bibnamefont
  {Moreland}}, \bibinfo {author} {\bibfnamefont {S.~A.}\ \bibnamefont {Bass}},
  \bibinfo {author} {\bibfnamefont {J.}~\bibnamefont {Liu}}, \ and\ \bibinfo
  {author} {\bibfnamefont {U.}~\bibnamefont {Heinz}},\ }\href {\doibase
  10.1103/PhysRevC.94.024907} {\bibfield  {journal} {\bibinfo  {journal} {Phys.
  Rev.}\ }\textbf {\bibinfo {volume} {C94}},\ \bibinfo {pages} {024907}
  (\bibinfo {year} {2016})},\ \Eprint {http://arxiv.org/abs/1605.03954}
  {arXiv:1605.03954 [nucl-th]} \BibitemShut {NoStop}%
\bibitem [{\citenamefont {Finazzo}\ \emph {et~al.}(2015)\citenamefont
  {Finazzo}, \citenamefont {Rougemont}, \citenamefont {Marrochio},\ and\
  \citenamefont {Noronha}}]{Finazzo:2014cna}%
  \BibitemOpen
  \bibfield  {author} {\bibinfo {author} {\bibfnamefont {S.~I.}\ \bibnamefont
  {Finazzo}}, \bibinfo {author} {\bibfnamefont {R.}~\bibnamefont {Rougemont}},
  \bibinfo {author} {\bibfnamefont {H.}~\bibnamefont {Marrochio}}, \ and\
  \bibinfo {author} {\bibfnamefont {J.}~\bibnamefont {Noronha}},\ }\href
  {\doibase 10.1007/JHEP02(2015)051} {\bibfield  {journal} {\bibinfo  {journal}
  {JHEP}\ }\textbf {\bibinfo {volume} {02}},\ \bibinfo {pages} {051} (\bibinfo
  {year} {2015})},\ \Eprint {http://arxiv.org/abs/1412.2968} {arXiv:1412.2968
  [hep-ph]} \BibitemShut {NoStop}%
\bibitem [{\citenamefont {Alqahtani}\ \emph {et~al.}(2018)\citenamefont
  {Alqahtani}, \citenamefont {Nopoush},\ and\ \citenamefont
  {Strickland}}]{Alqahtani:2017mhy}%
  \BibitemOpen
  \bibfield  {author} {\bibinfo {author} {\bibfnamefont {M.}~\bibnamefont
  {Alqahtani}}, \bibinfo {author} {\bibfnamefont {M.}~\bibnamefont {Nopoush}},
  \ and\ \bibinfo {author} {\bibfnamefont {M.}~\bibnamefont {Strickland}},\
  }\href {\doibase 10.1016/j.ppnp.2018.05.004} {\bibfield  {journal} {\bibinfo
  {journal} {Prog. Part. Nucl. Phys.}\ }\textbf {\bibinfo {volume} {101}},\
  \bibinfo {pages} {204} (\bibinfo {year} {2018})},\ \Eprint
  {http://arxiv.org/abs/1712.03282} {arXiv:1712.03282 [nucl-th]} \BibitemShut
  {NoStop}%
\bibitem [{\citenamefont {Almaalol}\ \emph {et~al.}(2019)\citenamefont
  {Almaalol}, \citenamefont {Alqahtani},\ and\ \citenamefont
  {Strickland}}]{Almaalol:2018gjh}%
  \BibitemOpen
  \bibfield  {author} {\bibinfo {author} {\bibfnamefont {D.}~\bibnamefont
  {Almaalol}}, \bibinfo {author} {\bibfnamefont {M.}~\bibnamefont {Alqahtani}},
  \ and\ \bibinfo {author} {\bibfnamefont {M.}~\bibnamefont {Strickland}},\
  }\href {\doibase 10.1103/PhysRevC.99.044902} {\bibfield  {journal} {\bibinfo
  {journal} {Phys. Rev.}\ }\textbf {\bibinfo {volume} {C99}},\ \bibinfo {pages}
  {044902} (\bibinfo {year} {2019})},\ \Eprint
  {http://arxiv.org/abs/1807.04337} {arXiv:1807.04337 [nucl-th]} \BibitemShut
  {NoStop}%
\bibitem [{\citenamefont {Ryu}\ \emph {et~al.}(2015)\citenamefont {Ryu},
  \citenamefont {Paquet}, \citenamefont {Shen}, \citenamefont {Denicol},
  \citenamefont {Schenke}, \citenamefont {Jeon},\ and\ \citenamefont
  {Gale}}]{Ryu:2015vwa}%
  \BibitemOpen
  \bibfield  {author} {\bibinfo {author} {\bibfnamefont {S.}~\bibnamefont
  {Ryu}}, \bibinfo {author} {\bibfnamefont {J.~F.}\ \bibnamefont {Paquet}},
  \bibinfo {author} {\bibfnamefont {C.}~\bibnamefont {Shen}}, \bibinfo {author}
  {\bibfnamefont {G.~S.}\ \bibnamefont {Denicol}}, \bibinfo {author}
  {\bibfnamefont {B.}~\bibnamefont {Schenke}}, \bibinfo {author} {\bibfnamefont
  {S.}~\bibnamefont {Jeon}}, \ and\ \bibinfo {author} {\bibfnamefont
  {C.}~\bibnamefont {Gale}},\ }\href {\doibase 10.1103/PhysRevLett.115.132301}
  {\bibfield  {journal} {\bibinfo  {journal} {Phys. Rev. Lett.}\ }\textbf
  {\bibinfo {volume} {115}},\ \bibinfo {pages} {132301} (\bibinfo {year}
  {2015})},\ \Eprint {http://arxiv.org/abs/1502.01675} {arXiv:1502.01675
  [nucl-th]} \BibitemShut {NoStop}%
\bibitem [{\citenamefont {Borsanyi}\ \emph {et~al.}(2020)\citenamefont
  {Borsanyi}, \citenamefont {Fodor}, \citenamefont {Guenther}, \citenamefont
  {Kara}, \citenamefont {Katz}, \citenamefont {Parotto}, \citenamefont
  {Pasztor}, \citenamefont {Ratti},\ and\ \citenamefont
  {Szabo}}]{Borsanyi:2020fev}%
  \BibitemOpen
  \bibfield  {author} {\bibinfo {author} {\bibfnamefont {S.}~\bibnamefont
  {Borsanyi}}, \bibinfo {author} {\bibfnamefont {Z.}~\bibnamefont {Fodor}},
  \bibinfo {author} {\bibfnamefont {J.~N.}\ \bibnamefont {Guenther}}, \bibinfo
  {author} {\bibfnamefont {R.}~\bibnamefont {Kara}}, \bibinfo {author}
  {\bibfnamefont {S.~D.}\ \bibnamefont {Katz}}, \bibinfo {author}
  {\bibfnamefont {P.}~\bibnamefont {Parotto}}, \bibinfo {author} {\bibfnamefont
  {A.}~\bibnamefont {Pasztor}}, \bibinfo {author} {\bibfnamefont
  {C.}~\bibnamefont {Ratti}}, \ and\ \bibinfo {author} {\bibfnamefont {K.~K.}\
  \bibnamefont {Szabo}},\ }\href@noop {} {\  (\bibinfo {year} {2020})},\
  \Eprint {http://arxiv.org/abs/2002.02821} {arXiv:2002.02821 [hep-lat]}
  \BibitemShut {NoStop}%
\bibitem [{\citenamefont {Jiang}\ \emph {et~al.}(2016)\citenamefont {Jiang},
  \citenamefont {Li},\ and\ \citenamefont {Song}}]{Jiang:2015hri}%
  \BibitemOpen
  \bibfield  {author} {\bibinfo {author} {\bibfnamefont {L.}~\bibnamefont
  {Jiang}}, \bibinfo {author} {\bibfnamefont {P.}~\bibnamefont {Li}}, \ and\
  \bibinfo {author} {\bibfnamefont {H.}~\bibnamefont {Song}},\ }\href {\doibase
  10.1103/PhysRevC.94.024918} {\bibfield  {journal} {\bibinfo  {journal} {Phys.
  Rev.}\ }\textbf {\bibinfo {volume} {C94}},\ \bibinfo {pages} {024918}
  (\bibinfo {year} {2016})},\ \Eprint {http://arxiv.org/abs/1512.06164}
  {arXiv:1512.06164 [nucl-th]} \BibitemShut {NoStop}%
\bibitem [{\citenamefont {Mukherjee}\ \emph {et~al.}(2016)\citenamefont
  {Mukherjee}, \citenamefont {Venugopalan},\ and\ \citenamefont
  {Yin}}]{Mukherjee:2016kyu}%
  \BibitemOpen
  \bibfield  {author} {\bibinfo {author} {\bibfnamefont {S.}~\bibnamefont
  {Mukherjee}}, \bibinfo {author} {\bibfnamefont {R.}~\bibnamefont
  {Venugopalan}}, \ and\ \bibinfo {author} {\bibfnamefont {Y.}~\bibnamefont
  {Yin}},\ }\href {\doibase 10.1103/PhysRevLett.117.222301} {\bibfield
  {journal} {\bibinfo  {journal} {Phys. Rev. Lett.}\ }\textbf {\bibinfo
  {volume} {117}},\ \bibinfo {pages} {222301} (\bibinfo {year} {2016})},\
  \Eprint {http://arxiv.org/abs/1605.09341} {arXiv:1605.09341 [hep-ph]}
  \BibitemShut {NoStop}%
\bibitem [{\citenamefont {Akamatsu}\ \emph {et~al.}(2019)\citenamefont
  {Akamatsu}, \citenamefont {Teaney}, \citenamefont {Yan},\ and\ \citenamefont
  {Yin}}]{Akamatsu:2018vjr}%
  \BibitemOpen
  \bibfield  {author} {\bibinfo {author} {\bibfnamefont {Y.}~\bibnamefont
  {Akamatsu}}, \bibinfo {author} {\bibfnamefont {D.}~\bibnamefont {Teaney}},
  \bibinfo {author} {\bibfnamefont {F.}~\bibnamefont {Yan}}, \ and\ \bibinfo
  {author} {\bibfnamefont {Y.}~\bibnamefont {Yin}},\ }\href {\doibase
  10.1103/PhysRevC.100.044901} {\bibfield  {journal} {\bibinfo  {journal}
  {Phys. Rev. C}\ }\textbf {\bibinfo {volume} {100}},\ \bibinfo {pages}
  {044901} (\bibinfo {year} {2019})},\ \Eprint
  {http://arxiv.org/abs/1811.05081} {arXiv:1811.05081 [nucl-th]} \BibitemShut
  {NoStop}%
\bibitem [{\citenamefont {Ejiri}\ \emph {et~al.}(2006)\citenamefont {Ejiri},
  \citenamefont {Karsch}, \citenamefont {Laermann},\ and\ \citenamefont
  {Schmidt}}]{Ejiri:2005uv}%
  \BibitemOpen
  \bibfield  {author} {\bibinfo {author} {\bibfnamefont {S.}~\bibnamefont
  {Ejiri}}, \bibinfo {author} {\bibfnamefont {F.}~\bibnamefont {Karsch}},
  \bibinfo {author} {\bibfnamefont {E.}~\bibnamefont {Laermann}}, \ and\
  \bibinfo {author} {\bibfnamefont {C.}~\bibnamefont {Schmidt}},\ }\href
  {\doibase 10.1103/PhysRevD.73.054506} {\bibfield  {journal} {\bibinfo
  {journal} {Phys. Rev.}\ }\textbf {\bibinfo {volume} {D73}},\ \bibinfo {pages}
  {054506} (\bibinfo {year} {2006})},\ \Eprint
  {http://arxiv.org/abs/hep-lat/0512040} {arXiv:hep-lat/0512040 [hep-lat]}
  \BibitemShut {NoStop}%
\bibitem [{\citenamefont {Schmid}(2008)}]{Schmid:2008sy}%
  \BibitemOpen
  \bibfield  {author} {\bibinfo {author} {\bibfnamefont {C.}~\bibnamefont
  {Schmid}} (\bibinfo {collaboration} {COL-NOTE = RBC, HotQCD}),\ }in\
  \href@noop {} {\emph {\bibinfo {booktitle} {{Proceedings, 3rd International
  Conference on Hard and Electromagnetic Probes of High-Energy Nuclear
  Collisions (Hard Probes 2008): Illa da Toxa, Spain, June 8-14, 2008}}}}\
  (\bibinfo {year} {2008})\ \Eprint {http://arxiv.org/abs/0810.0374}
  {arXiv:0810.0374 [hep-lat]} \BibitemShut {NoStop}%
\bibitem [{\citenamefont {Bellwied}\ \emph {et~al.}(2016)\citenamefont
  {Bellwied}, \citenamefont {Borsanyi}, \citenamefont {Fodor}, \citenamefont
  {Gunther}, \citenamefont {Katz}, \citenamefont {Pasztor}, \citenamefont
  {Ratti},\ and\ \citenamefont {Szabo}}]{Bellwied:2016cpq}%
  \BibitemOpen
  \bibfield  {author} {\bibinfo {author} {\bibfnamefont {R.}~\bibnamefont
  {Bellwied}}, \bibinfo {author} {\bibfnamefont {S.}~\bibnamefont {Borsanyi}},
  \bibinfo {author} {\bibfnamefont {Z.}~\bibnamefont {Fodor}}, \bibinfo
  {author} {\bibfnamefont {J.}~\bibnamefont {Gunther}}, \bibinfo {author}
  {\bibfnamefont {S.~D.}\ \bibnamefont {Katz}}, \bibinfo {author}
  {\bibfnamefont {A.}~\bibnamefont {Pasztor}}, \bibinfo {author} {\bibfnamefont
  {C.}~\bibnamefont {Ratti}}, \ and\ \bibinfo {author} {\bibfnamefont {K.~K.}\
  \bibnamefont {Szabo}},\ }\bibfield  {booktitle} {\emph {\bibinfo {booktitle}
  {{Proceedings, 25th International Conference on Ultra-Relativistic
  Nucleus-Nucleus Collisions (Quark Matter 2015): Kobe, Japan, September
  27-October 3, 2015}}},\ }\href {\doibase 10.1016/j.nuclphysa.2016.02.010}
  {\bibfield  {journal} {\bibinfo  {journal} {Nucl. Phys.}\ }\textbf {\bibinfo
  {volume} {A956}},\ \bibinfo {pages} {797} (\bibinfo {year} {2016})},\ \Eprint
  {http://arxiv.org/abs/1601.00466} {arXiv:1601.00466 [nucl-th]} \BibitemShut
  {NoStop}%
\bibitem [{\citenamefont {Ivanov}\ \emph {et~al.}(2006)\citenamefont {Ivanov},
  \citenamefont {Russkikh},\ and\ \citenamefont {Toneev}}]{Ivanov:2005yw}%
  \BibitemOpen
  \bibfield  {author} {\bibinfo {author} {\bibfnamefont {Y.}~\bibnamefont
  {Ivanov}}, \bibinfo {author} {\bibfnamefont {V.}~\bibnamefont {Russkikh}}, \
  and\ \bibinfo {author} {\bibfnamefont {V.}~\bibnamefont {Toneev}},\ }\href
  {\doibase 10.1103/PhysRevC.73.044904} {\bibfield  {journal} {\bibinfo
  {journal} {Phys. Rev. C}\ }\textbf {\bibinfo {volume} {73}},\ \bibinfo
  {pages} {044904} (\bibinfo {year} {2006})},\ \Eprint
  {http://arxiv.org/abs/nucl-th/0503088} {arXiv:nucl-th/0503088} \BibitemShut
  {NoStop}%
\bibitem [{\citenamefont {Shen}\ and\ \citenamefont
  {Schenke}(2019)}]{Shen:2018pty}%
  \BibitemOpen
  \bibfield  {author} {\bibinfo {author} {\bibfnamefont {C.}~\bibnamefont
  {Shen}}\ and\ \bibinfo {author} {\bibfnamefont {B.}~\bibnamefont {Schenke}},\
  }\bibfield  {booktitle} {\emph {\bibinfo {booktitle} {{Proceedings, 27th
  International Conference on Ultrarelativistic Nucleus-Nucleus Collisions
  (Quark Matter 2018): Venice, Italy, May 14-19, 2018}}},\ }\href {\doibase
  10.1016/j.nuclphysa.2018.08.007} {\bibfield  {journal} {\bibinfo  {journal}
  {Nucl. Phys.}\ }\textbf {\bibinfo {volume} {A982}},\ \bibinfo {pages} {411}
  (\bibinfo {year} {2019})},\ \Eprint {http://arxiv.org/abs/1807.05141}
  {arXiv:1807.05141 [nucl-th]} \BibitemShut {NoStop}%
\bibitem [{\citenamefont {Becattini}\ \emph {et~al.}(2006)\citenamefont
  {Becattini}, \citenamefont {Manninen},\ and\ \citenamefont
  {Gazdzicki}}]{Becattini:2005xt}%
  \BibitemOpen
  \bibfield  {author} {\bibinfo {author} {\bibfnamefont {F.}~\bibnamefont
  {Becattini}}, \bibinfo {author} {\bibfnamefont {J.}~\bibnamefont {Manninen}},
  \ and\ \bibinfo {author} {\bibfnamefont {M.}~\bibnamefont {Gazdzicki}},\
  }\href {\doibase 10.1103/PhysRevC.73.044905} {\bibfield  {journal} {\bibinfo
  {journal} {Phys. Rev.}\ }\textbf {\bibinfo {volume} {C73}},\ \bibinfo {pages}
  {044905} (\bibinfo {year} {2006})},\ \Eprint
  {http://arxiv.org/abs/hep-ph/0511092} {arXiv:hep-ph/0511092 [hep-ph]}
  \BibitemShut {NoStop}%
\bibitem [{\citenamefont {Becattini}\ \emph {et~al.}(2013)\citenamefont
  {Becattini}, \citenamefont {Bleicher}, \citenamefont {Kollegger},
  \citenamefont {Schuster}, \citenamefont {Steinheimer},\ and\ \citenamefont
  {Stock}}]{Becattini:2012xb}%
  \BibitemOpen
  \bibfield  {author} {\bibinfo {author} {\bibfnamefont {F.}~\bibnamefont
  {Becattini}}, \bibinfo {author} {\bibfnamefont {M.}~\bibnamefont {Bleicher}},
  \bibinfo {author} {\bibfnamefont {T.}~\bibnamefont {Kollegger}}, \bibinfo
  {author} {\bibfnamefont {T.}~\bibnamefont {Schuster}}, \bibinfo {author}
  {\bibfnamefont {J.}~\bibnamefont {Steinheimer}}, \ and\ \bibinfo {author}
  {\bibfnamefont {R.}~\bibnamefont {Stock}},\ }\href {\doibase
  10.1103/PhysRevLett.111.082302} {\bibfield  {journal} {\bibinfo  {journal}
  {Phys. Rev. Lett.}\ }\textbf {\bibinfo {volume} {111}},\ \bibinfo {pages}
  {082302} (\bibinfo {year} {2013})},\ \Eprint {http://arxiv.org/abs/1212.2431}
  {arXiv:1212.2431 [nucl-th]} \BibitemShut {NoStop}%
\bibitem [{\citenamefont {Cleymans}\ \emph {et~al.}(2006)\citenamefont
  {Cleymans}, \citenamefont {Oeschler}, \citenamefont {Redlich},\ and\
  \citenamefont {Wheaton}}]{Cleymans:2005xv}%
  \BibitemOpen
  \bibfield  {author} {\bibinfo {author} {\bibfnamefont {J.}~\bibnamefont
  {Cleymans}}, \bibinfo {author} {\bibfnamefont {H.}~\bibnamefont {Oeschler}},
  \bibinfo {author} {\bibfnamefont {K.}~\bibnamefont {Redlich}}, \ and\
  \bibinfo {author} {\bibfnamefont {S.}~\bibnamefont {Wheaton}},\ }\href
  {\doibase 10.1103/PhysRevC.73.034905} {\bibfield  {journal} {\bibinfo
  {journal} {Phys. Rev.}\ }\textbf {\bibinfo {volume} {C73}},\ \bibinfo {pages}
  {034905} (\bibinfo {year} {2006})},\ \Eprint
  {http://arxiv.org/abs/hep-ph/0511094} {arXiv:hep-ph/0511094 [hep-ph]}
  \BibitemShut {NoStop}%
\bibitem [{\citenamefont {Torrieri}\ and\ \citenamefont
  {Rafelski}(2007)}]{Torrieri:2006yb}%
  \BibitemOpen
  \bibfield  {author} {\bibinfo {author} {\bibfnamefont {G.}~\bibnamefont
  {Torrieri}}\ and\ \bibinfo {author} {\bibfnamefont {J.}~\bibnamefont
  {Rafelski}},\ }\href {\doibase 10.1103/PhysRevC.75.024902} {\bibfield
  {journal} {\bibinfo  {journal} {Phys. Rev.}\ }\textbf {\bibinfo {volume}
  {C75}},\ \bibinfo {pages} {024902} (\bibinfo {year} {2007})},\ \Eprint
  {http://arxiv.org/abs/nucl-th/0608061} {arXiv:nucl-th/0608061 [nucl-th]}
  \BibitemShut {NoStop}%
\bibitem [{\citenamefont {Andronic}\ \emph {et~al.}(2011)\citenamefont
  {Andronic}, \citenamefont {Braun-Munzinger}, \citenamefont {Redlich},\ and\
  \citenamefont {Stachel}}]{Andronic:2011yq}%
  \BibitemOpen
  \bibfield  {author} {\bibinfo {author} {\bibfnamefont {A.}~\bibnamefont
  {Andronic}}, \bibinfo {author} {\bibfnamefont {P.}~\bibnamefont
  {Braun-Munzinger}}, \bibinfo {author} {\bibfnamefont {K.}~\bibnamefont
  {Redlich}}, \ and\ \bibinfo {author} {\bibfnamefont {J.}~\bibnamefont
  {Stachel}},\ }\bibfield  {booktitle} {\emph {\bibinfo {booktitle} {{Quark
  matter. Proceedings, 22nd International Conference on Ultra-Relativistic
  Nucleus-Nucleus Collisions, Quark Matter 2011, Annecy, France, May 23-28,
  2011}}},\ }\href {\doibase 10.1088/0954-3899/38/12/124081} {\bibfield
  {journal} {\bibinfo  {journal} {J. Phys.}\ }\textbf {\bibinfo {volume}
  {G38}},\ \bibinfo {pages} {124081} (\bibinfo {year} {2011})},\ \Eprint
  {http://arxiv.org/abs/1106.6321} {arXiv:1106.6321 [nucl-th]} \BibitemShut
  {NoStop}%
\bibitem [{\citenamefont {Braun-Munzinger}\ \emph {et~al.}(2003)\citenamefont
  {Braun-Munzinger}, \citenamefont {Redlich},\ and\ \citenamefont
  {Stachel}}]{BraunMunzinger:2003zd}%
  \BibitemOpen
  \bibfield  {author} {\bibinfo {author} {\bibfnamefont {P.}~\bibnamefont
  {Braun-Munzinger}}, \bibinfo {author} {\bibfnamefont {K.}~\bibnamefont
  {Redlich}}, \ and\ \bibinfo {author} {\bibfnamefont {J.}~\bibnamefont
  {Stachel}},\ }\href {\doibase 10.1142/9789812795533_0008} {\ ,\ \bibinfo
  {pages} {491} (\bibinfo {year} {2003})},\ \Eprint
  {http://arxiv.org/abs/nucl-th/0304013} {arXiv:nucl-th/0304013 [nucl-th]}
  \BibitemShut {NoStop}%
\bibitem [{\citenamefont {Andronic}\ \emph {et~al.}(2006)\citenamefont
  {Andronic}, \citenamefont {Braun-Munzinger},\ and\ \citenamefont
  {Stachel}}]{Andronic:2005yp}%
  \BibitemOpen
  \bibfield  {author} {\bibinfo {author} {\bibfnamefont {A.}~\bibnamefont
  {Andronic}}, \bibinfo {author} {\bibfnamefont {P.}~\bibnamefont
  {Braun-Munzinger}}, \ and\ \bibinfo {author} {\bibfnamefont {J.}~\bibnamefont
  {Stachel}},\ }\href {\doibase 10.1016/j.nuclphysa.2006.03.012} {\bibfield
  {journal} {\bibinfo  {journal} {Nucl. Phys.}\ }\textbf {\bibinfo {volume}
  {A772}},\ \bibinfo {pages} {167} (\bibinfo {year} {2006})},\ \Eprint
  {http://arxiv.org/abs/nucl-th/0511071} {arXiv:nucl-th/0511071 [nucl-th]}
  \BibitemShut {NoStop}%
\bibitem [{\citenamefont {Vovchenko}\ \emph {et~al.}(2016)\citenamefont
  {Vovchenko}, \citenamefont {Begun},\ and\ \citenamefont
  {Gorenstein}}]{Vovchenko:2015idt}%
  \BibitemOpen
  \bibfield  {author} {\bibinfo {author} {\bibfnamefont {V.}~\bibnamefont
  {Vovchenko}}, \bibinfo {author} {\bibfnamefont {V.}~\bibnamefont {Begun}}, \
  and\ \bibinfo {author} {\bibfnamefont {M.}~\bibnamefont {Gorenstein}},\
  }\href {\doibase 10.1103/PhysRevC.93.064906} {\bibfield  {journal} {\bibinfo
  {journal} {Phys. Rev. C}\ }\textbf {\bibinfo {volume} {93}},\ \bibinfo
  {pages} {064906} (\bibinfo {year} {2016})},\ \Eprint
  {http://arxiv.org/abs/1512.08025} {arXiv:1512.08025 [nucl-th]} \BibitemShut
  {NoStop}%
\bibitem [{\citenamefont {Alba}\ \emph {et~al.}(2014)\citenamefont {Alba},
  \citenamefont {Alberico}, \citenamefont {Bellwied}, \citenamefont {Bluhm},
  \citenamefont {Mantovani~Sarti}, \citenamefont {Nahrgang},\ and\
  \citenamefont {Ratti}}]{Alba:2014eba}%
  \BibitemOpen
  \bibfield  {author} {\bibinfo {author} {\bibfnamefont {P.}~\bibnamefont
  {Alba}}, \bibinfo {author} {\bibfnamefont {W.}~\bibnamefont {Alberico}},
  \bibinfo {author} {\bibfnamefont {R.}~\bibnamefont {Bellwied}}, \bibinfo
  {author} {\bibfnamefont {M.}~\bibnamefont {Bluhm}}, \bibinfo {author}
  {\bibfnamefont {V.}~\bibnamefont {Mantovani~Sarti}}, \bibinfo {author}
  {\bibfnamefont {M.}~\bibnamefont {Nahrgang}}, \ and\ \bibinfo {author}
  {\bibfnamefont {C.}~\bibnamefont {Ratti}},\ }\href {\doibase
  10.1016/j.physletb.2014.09.052} {\bibfield  {journal} {\bibinfo  {journal}
  {Phys. Lett.}\ }\textbf {\bibinfo {volume} {B738}},\ \bibinfo {pages} {305}
  (\bibinfo {year} {2014})},\ \Eprint {http://arxiv.org/abs/1403.4903}
  {arXiv:1403.4903 [hep-ph]} \BibitemShut {NoStop}%
\bibitem [{\citenamefont {Karsch}\ and\ \citenamefont
  {Redlich}(2011)}]{Karsch:2010ck}%
  \BibitemOpen
  \bibfield  {author} {\bibinfo {author} {\bibfnamefont {F.}~\bibnamefont
  {Karsch}}\ and\ \bibinfo {author} {\bibfnamefont {K.}~\bibnamefont
  {Redlich}},\ }\href {\doibase 10.1016/j.physletb.2010.10.046} {\bibfield
  {journal} {\bibinfo  {journal} {Phys. Lett.}\ }\textbf {\bibinfo {volume}
  {B695}},\ \bibinfo {pages} {136} (\bibinfo {year} {2011})},\ \Eprint
  {http://arxiv.org/abs/1007.2581} {arXiv:1007.2581 [hep-ph]} \BibitemShut
  {NoStop}%
\bibitem [{\citenamefont {Garg}\ \emph {et~al.}(2013)\citenamefont {Garg},
  \citenamefont {Mishra}, \citenamefont {Netrakanti}, \citenamefont {Mohanty},
  \citenamefont {Mohanty}, \citenamefont {Singh},\ and\ \citenamefont
  {Xu}}]{Garg:2013ata}%
  \BibitemOpen
  \bibfield  {author} {\bibinfo {author} {\bibfnamefont {P.}~\bibnamefont
  {Garg}}, \bibinfo {author} {\bibfnamefont {D.~K.}\ \bibnamefont {Mishra}},
  \bibinfo {author} {\bibfnamefont {P.~K.}\ \bibnamefont {Netrakanti}},
  \bibinfo {author} {\bibfnamefont {B.}~\bibnamefont {Mohanty}}, \bibinfo
  {author} {\bibfnamefont {A.~K.}\ \bibnamefont {Mohanty}}, \bibinfo {author}
  {\bibfnamefont {B.~K.}\ \bibnamefont {Singh}}, \ and\ \bibinfo {author}
  {\bibfnamefont {N.}~\bibnamefont {Xu}},\ }\href {\doibase
  10.1016/j.physletb.2013.09.019} {\bibfield  {journal} {\bibinfo  {journal}
  {Phys. Lett.}\ }\textbf {\bibinfo {volume} {B726}},\ \bibinfo {pages} {691}
  (\bibinfo {year} {2013})},\ \Eprint {http://arxiv.org/abs/1304.7133}
  {arXiv:1304.7133 [nucl-ex]} \BibitemShut {NoStop}%
\bibitem [{\citenamefont {Borsanyi}\ \emph {et~al.}(2013)\citenamefont
  {Borsanyi}, \citenamefont {Fodor}, \citenamefont {Katz}, \citenamefont
  {Krieg}, \citenamefont {Ratti},\ and\ \citenamefont
  {Szabo}}]{Borsanyi:2013hza}%
  \BibitemOpen
  \bibfield  {author} {\bibinfo {author} {\bibfnamefont {S.}~\bibnamefont
  {Borsanyi}}, \bibinfo {author} {\bibfnamefont {Z.}~\bibnamefont {Fodor}},
  \bibinfo {author} {\bibfnamefont {S.~D.}\ \bibnamefont {Katz}}, \bibinfo
  {author} {\bibfnamefont {S.}~\bibnamefont {Krieg}}, \bibinfo {author}
  {\bibfnamefont {C.}~\bibnamefont {Ratti}}, \ and\ \bibinfo {author}
  {\bibfnamefont {K.~K.}\ \bibnamefont {Szabo}},\ }\href {\doibase
  10.1103/PhysRevLett.111.062005} {\bibfield  {journal} {\bibinfo  {journal}
  {Phys. Rev. Lett.}\ }\textbf {\bibinfo {volume} {111}},\ \bibinfo {pages}
  {062005} (\bibinfo {year} {2013})},\ \Eprint {http://arxiv.org/abs/1305.5161}
  {arXiv:1305.5161 [hep-lat]} \BibitemShut {NoStop}%
\bibitem [{\citenamefont {Borsanyi}\ \emph {et~al.}(2014)\citenamefont
  {Borsanyi}, \citenamefont {Fodor}, \citenamefont {Katz}, \citenamefont
  {Krieg}, \citenamefont {Ratti},\ and\ \citenamefont
  {Szabo}}]{Borsanyi:2014ewa}%
  \BibitemOpen
  \bibfield  {author} {\bibinfo {author} {\bibfnamefont {S.}~\bibnamefont
  {Borsanyi}}, \bibinfo {author} {\bibfnamefont {Z.}~\bibnamefont {Fodor}},
  \bibinfo {author} {\bibfnamefont {S.~D.}\ \bibnamefont {Katz}}, \bibinfo
  {author} {\bibfnamefont {S.}~\bibnamefont {Krieg}}, \bibinfo {author}
  {\bibfnamefont {C.}~\bibnamefont {Ratti}}, \ and\ \bibinfo {author}
  {\bibfnamefont {K.~K.}\ \bibnamefont {Szabo}},\ }\href {\doibase
  10.1103/PhysRevLett.113.052301} {\bibfield  {journal} {\bibinfo  {journal}
  {Phys. Rev. Lett.}\ }\textbf {\bibinfo {volume} {113}},\ \bibinfo {pages}
  {052301} (\bibinfo {year} {2014})},\ \Eprint {http://arxiv.org/abs/1403.4576}
  {arXiv:1403.4576 [hep-lat]} \BibitemShut {NoStop}%
\bibitem [{\citenamefont {Bellwied}\ \emph {et~al.}(2020)\citenamefont
  {Bellwied}, \citenamefont {Borsanyi}, \citenamefont {Fodor}, \citenamefont
  {Guenther}, \citenamefont {Noronha-Hostler}, \citenamefont {Parotto},
  \citenamefont {Pasztor}, \citenamefont {Ratti},\ and\ \citenamefont
  {Stafford}}]{Bellwied:2019pxh}%
  \BibitemOpen
  \bibfield  {author} {\bibinfo {author} {\bibfnamefont {R.}~\bibnamefont
  {Bellwied}}, \bibinfo {author} {\bibfnamefont {S.}~\bibnamefont {Borsanyi}},
  \bibinfo {author} {\bibfnamefont {Z.}~\bibnamefont {Fodor}}, \bibinfo
  {author} {\bibfnamefont {J.~N.}\ \bibnamefont {Guenther}}, \bibinfo {author}
  {\bibfnamefont {J.}~\bibnamefont {Noronha-Hostler}}, \bibinfo {author}
  {\bibfnamefont {P.}~\bibnamefont {Parotto}}, \bibinfo {author} {\bibfnamefont
  {A.}~\bibnamefont {Pasztor}}, \bibinfo {author} {\bibfnamefont
  {C.}~\bibnamefont {Ratti}}, \ and\ \bibinfo {author} {\bibfnamefont {J.~M.}\
  \bibnamefont {Stafford}},\ }\href {\doibase 10.1103/PhysRevD.101.034506}
  {\bibfield  {journal} {\bibinfo  {journal} {Phys. Rev.}\ }\textbf {\bibinfo
  {volume} {D101}},\ \bibinfo {pages} {034506} (\bibinfo {year} {2020})},\
  \Eprint {http://arxiv.org/abs/1910.14592} {arXiv:1910.14592 [hep-lat]}
  \BibitemShut {NoStop}%
\bibitem [{\citenamefont {Braun-Munzinger}\ \emph {et~al.}(2020)\citenamefont
  {Braun-Munzinger}, \citenamefont {Friman}, \citenamefont {Redlich},
  \citenamefont {Rustamov},\ and\ \citenamefont
  {Stachel}}]{Braun-Munzinger:2020jbk}%
  \BibitemOpen
  \bibfield  {author} {\bibinfo {author} {\bibfnamefont {P.}~\bibnamefont
  {Braun-Munzinger}}, \bibinfo {author} {\bibfnamefont {B.}~\bibnamefont
  {Friman}}, \bibinfo {author} {\bibfnamefont {K.}~\bibnamefont {Redlich}},
  \bibinfo {author} {\bibfnamefont {A.}~\bibnamefont {Rustamov}}, \ and\
  \bibinfo {author} {\bibfnamefont {J.}~\bibnamefont {Stachel}},\ }\href@noop
  {} {\  (\bibinfo {year} {2020})},\ \Eprint {http://arxiv.org/abs/2007.02463}
  {arXiv:2007.02463 [nucl-th]} \BibitemShut {NoStop}%
\bibitem [{\citenamefont {Alba}\ \emph {et~al.}(2020)\citenamefont {Alba},
  \citenamefont {Sarti}, \citenamefont {Noronha-Hostler}, \citenamefont
  {Parotto}, \citenamefont {Portillo-Vazquez}, \citenamefont {Ratti},\ and\
  \citenamefont {Stafford}}]{Alba:2020jir}%
  \BibitemOpen
  \bibfield  {author} {\bibinfo {author} {\bibfnamefont {P.}~\bibnamefont
  {Alba}}, \bibinfo {author} {\bibfnamefont {V.~M.}\ \bibnamefont {Sarti}},
  \bibinfo {author} {\bibfnamefont {J.}~\bibnamefont {Noronha-Hostler}},
  \bibinfo {author} {\bibfnamefont {P.}~\bibnamefont {Parotto}}, \bibinfo
  {author} {\bibfnamefont {I.}~\bibnamefont {Portillo-Vazquez}}, \bibinfo
  {author} {\bibfnamefont {C.}~\bibnamefont {Ratti}}, \ and\ \bibinfo {author}
  {\bibfnamefont {J.~M.}\ \bibnamefont {Stafford}},\ }\href@noop {} {\
  (\bibinfo {year} {2020})},\ \Eprint {http://arxiv.org/abs/2002.12395}
  {arXiv:2002.12395 [hep-ph]} \BibitemShut {NoStop}%
\bibitem [{\citenamefont {Schofield}\ \emph {et~al.}(1969)\citenamefont
  {Schofield}, \citenamefont {Litster},\ and\ \citenamefont
  {Ho}}]{Schofield:1969}%
  \BibitemOpen
  \bibfield  {author} {\bibinfo {author} {\bibfnamefont {P.}~\bibnamefont
  {Schofield}}, \bibinfo {author} {\bibfnamefont {J.~D.}\ \bibnamefont
  {Litster}}, \ and\ \bibinfo {author} {\bibfnamefont {J.~T.}\ \bibnamefont
  {Ho}},\ }\href {\doibase 10.1103/PhysRevLett.23.1098} {\bibfield  {journal}
  {\bibinfo  {journal} {Phys. Rev. Lett.}\ }\textbf {\bibinfo {volume} {23}},\
  \bibinfo {pages} {1098} (\bibinfo {year} {1969})}\BibitemShut {NoStop}%
\bibitem [{\citenamefont {Zinn-Justin}(2001)}]{Zinn:2001}%
  \BibitemOpen
  \bibfield  {author} {\bibinfo {author} {\bibfnamefont {J.}~\bibnamefont
  {Zinn-Justin}},\ }\href {\doibase
  https://doi.org/10.1016/S0370-1573(00)00126-5} {\bibfield  {journal}
  {\bibinfo  {journal} {Physics Reports}\ }\textbf {\bibinfo {volume} {344}},\
  \bibinfo {pages} {159 } (\bibinfo {year} {2001})},\ \bibinfo {note}
  {renormalization group theory in the new millennium}\BibitemShut {NoStop}%
\bibitem [{\citenamefont {Chattopadhyay}\ and\ \citenamefont
  {Heinz}(2020)}]{Chattopadhyay:2019jqj}%
  \BibitemOpen
  \bibfield  {author} {\bibinfo {author} {\bibfnamefont {C.}~\bibnamefont
  {Chattopadhyay}}\ and\ \bibinfo {author} {\bibfnamefont {U.~W.}\ \bibnamefont
  {Heinz}},\ }\href {\doibase 10.1016/j.physletb.2019.135158} {\bibfield
  {journal} {\bibinfo  {journal} {Phys. Lett. B}\ }\textbf {\bibinfo {volume}
  {801}},\ \bibinfo {pages} {135158} (\bibinfo {year} {2020})},\ \Eprint
  {http://arxiv.org/abs/1911.07765} {arXiv:1911.07765 [nucl-th]} \BibitemShut
  {NoStop}%
\bibitem [{\citenamefont {Janik}\ and\ \citenamefont
  {Peschanski}(2006)}]{Janik:2005zt}%
  \BibitemOpen
  \bibfield  {author} {\bibinfo {author} {\bibfnamefont {R.~A.}\ \bibnamefont
  {Janik}}\ and\ \bibinfo {author} {\bibfnamefont {R.~B.}\ \bibnamefont
  {Peschanski}},\ }\href {\doibase 10.1103/PhysRevD.73.045013} {\bibfield
  {journal} {\bibinfo  {journal} {Phys. Rev. D}\ }\textbf {\bibinfo {volume}
  {73}},\ \bibinfo {pages} {045013} (\bibinfo {year} {2006})},\ \Eprint
  {http://arxiv.org/abs/hep-th/0512162} {arXiv:hep-th/0512162} \BibitemShut
  {NoStop}%
\bibitem [{\citenamefont {Brewer}\ \emph {et~al.}(2018)\citenamefont {Brewer},
  \citenamefont {Mukherjee}, \citenamefont {Rajagopal},\ and\ \citenamefont
  {Yin}}]{Brewer:2018abr}%
  \BibitemOpen
  \bibfield  {author} {\bibinfo {author} {\bibfnamefont {J.}~\bibnamefont
  {Brewer}}, \bibinfo {author} {\bibfnamefont {S.}~\bibnamefont {Mukherjee}},
  \bibinfo {author} {\bibfnamefont {K.}~\bibnamefont {Rajagopal}}, \ and\
  \bibinfo {author} {\bibfnamefont {Y.}~\bibnamefont {Yin}},\ }\href {\doibase
  10.1103/PhysRevC.98.061901} {\bibfield  {journal} {\bibinfo  {journal} {Phys.
  Rev.}\ }\textbf {\bibinfo {volume} {C98}},\ \bibinfo {pages} {061901}
  (\bibinfo {year} {2018})},\ \Eprint {http://arxiv.org/abs/1804.10215}
  {arXiv:1804.10215 [hep-ph]} \BibitemShut {NoStop}%
\bibitem [{\citenamefont {Li}\ and\ \citenamefont
  {Kapusta}(2019)}]{Li:2018ini}%
  \BibitemOpen
  \bibfield  {author} {\bibinfo {author} {\bibfnamefont {M.}~\bibnamefont
  {Li}}\ and\ \bibinfo {author} {\bibfnamefont {J.~I.}\ \bibnamefont
  {Kapusta}},\ }\href {\doibase 10.1103/PhysRevC.99.014906} {\bibfield
  {journal} {\bibinfo  {journal} {Phys. Rev.}\ }\textbf {\bibinfo {volume}
  {C99}},\ \bibinfo {pages} {014906} (\bibinfo {year} {2019})},\ \Eprint
  {http://arxiv.org/abs/1808.05751} {arXiv:1808.05751 [nucl-th]} \BibitemShut
  {NoStop}%
\bibitem [{\citenamefont {Danielewicz}\ \emph {et~al.}(2002)\citenamefont
  {Danielewicz}, \citenamefont {Lacey},\ and\ \citenamefont
  {Lynch}}]{Danielewicz:2002pu}%
  \BibitemOpen
  \bibfield  {author} {\bibinfo {author} {\bibfnamefont {P.}~\bibnamefont
  {Danielewicz}}, \bibinfo {author} {\bibfnamefont {R.}~\bibnamefont {Lacey}},
  \ and\ \bibinfo {author} {\bibfnamefont {W.~G.}\ \bibnamefont {Lynch}},\
  }\href {\doibase 10.1126/science.1078070} {\bibfield  {journal} {\bibinfo
  {journal} {Science}\ }\textbf {\bibinfo {volume} {298}},\ \bibinfo {pages}
  {1592} (\bibinfo {year} {2002})},\ \Eprint
  {http://arxiv.org/abs/nucl-th/0208016} {arXiv:nucl-th/0208016} \BibitemShut
  {NoStop}%
\bibitem [{\citenamefont {Adamczewski-Musch}\ \emph {et~al.}(2019)\citenamefont
  {Adamczewski-Musch} \emph {et~al.}}]{Adamczewski-Musch:2019byl}%
  \BibitemOpen
  \bibfield  {author} {\bibinfo {author} {\bibfnamefont {J.}~\bibnamefont
  {Adamczewski-Musch}} \emph {et~al.} (\bibinfo {collaboration} {HADES}),\
  }\href {\doibase 10.1038/s41567-019-0583-8} {\bibfield  {journal} {\bibinfo
  {journal} {Nature Phys.}\ }\textbf {\bibinfo {volume} {15}},\ \bibinfo
  {pages} {1040} (\bibinfo {year} {2019})}\BibitemShut {NoStop}%
\bibitem [{\citenamefont {Horowitz}\ \emph {et~al.}(2020)\citenamefont
  {Horowitz}, \citenamefont {Piekarewicz},\ and\ \citenamefont
  {Reed}}]{Horowitz:2020evx}%
  \BibitemOpen
  \bibfield  {author} {\bibinfo {author} {\bibfnamefont {C.}~\bibnamefont
  {Horowitz}}, \bibinfo {author} {\bibfnamefont {J.}~\bibnamefont
  {Piekarewicz}}, \ and\ \bibinfo {author} {\bibfnamefont {B.}~\bibnamefont
  {Reed}},\ }\href@noop {} {\  (\bibinfo {year} {2020})},\ \Eprint
  {http://arxiv.org/abs/2007.07117} {arXiv:2007.07117 [nucl-th]} \BibitemShut
  {NoStop}%
\bibitem [{\citenamefont {Baym}\ \emph {et~al.}(2018)\citenamefont {Baym},
  \citenamefont {Hatsuda}, \citenamefont {Kojo}, \citenamefont {Powell},
  \citenamefont {Song},\ and\ \citenamefont {Takatsuka}}]{Baym:2017whm}%
  \BibitemOpen
  \bibfield  {author} {\bibinfo {author} {\bibfnamefont {G.}~\bibnamefont
  {Baym}}, \bibinfo {author} {\bibfnamefont {T.}~\bibnamefont {Hatsuda}},
  \bibinfo {author} {\bibfnamefont {T.}~\bibnamefont {Kojo}}, \bibinfo {author}
  {\bibfnamefont {P.~D.}\ \bibnamefont {Powell}}, \bibinfo {author}
  {\bibfnamefont {Y.}~\bibnamefont {Song}}, \ and\ \bibinfo {author}
  {\bibfnamefont {T.}~\bibnamefont {Takatsuka}},\ }\href {\doibase
  10.1088/1361-6633/aaae14} {\bibfield  {journal} {\bibinfo  {journal} {Rept.
  Prog. Phys.}\ }\textbf {\bibinfo {volume} {81}},\ \bibinfo {pages} {056902}
  (\bibinfo {year} {2018})},\ \Eprint {http://arxiv.org/abs/1707.04966}
  {arXiv:1707.04966 [astro-ph.HE]} \BibitemShut {NoStop}%
\bibitem [{\citenamefont {Kolomeitsev}\ and\ \citenamefont
  {Voskresensky}(2015)}]{Kolomeitsev:2014gfa}%
  \BibitemOpen
  \bibfield  {author} {\bibinfo {author} {\bibfnamefont {E.}~\bibnamefont
  {Kolomeitsev}}\ and\ \bibinfo {author} {\bibfnamefont {D.}~\bibnamefont
  {Voskresensky}},\ }\href {\doibase 10.1103/PhysRevC.91.025805} {\bibfield
  {journal} {\bibinfo  {journal} {Phys. Rev. C}\ }\textbf {\bibinfo {volume}
  {91}},\ \bibinfo {pages} {025805} (\bibinfo {year} {2015})},\ \Eprint
  {http://arxiv.org/abs/1412.0314} {arXiv:1412.0314 [nucl-th]} \BibitemShut
  {NoStop}%
\bibitem [{\citenamefont {Ofengeim}\ \emph {et~al.}(2019)\citenamefont
  {Ofengeim}, \citenamefont {Gusakov}, \citenamefont {Haensel},\ and\
  \citenamefont {Fortin}}]{Ofengeim:2019fjy}%
  \BibitemOpen
  \bibfield  {author} {\bibinfo {author} {\bibfnamefont {D.}~\bibnamefont
  {Ofengeim}}, \bibinfo {author} {\bibfnamefont {M.}~\bibnamefont {Gusakov}},
  \bibinfo {author} {\bibfnamefont {P.}~\bibnamefont {Haensel}}, \ and\
  \bibinfo {author} {\bibfnamefont {M.}~\bibnamefont {Fortin}},\ }\href
  {\doibase 10.1103/PhysRevD.100.103017} {\bibfield  {journal} {\bibinfo
  {journal} {Phys. Rev. D}\ }\textbf {\bibinfo {volume} {100}},\ \bibinfo
  {pages} {103017} (\bibinfo {year} {2019})},\ \Eprint
  {http://arxiv.org/abs/1911.08407} {arXiv:1911.08407 [astro-ph.HE]}
  \BibitemShut {NoStop}%
\bibitem [{\citenamefont {Alford}\ \emph {et~al.}(2019)\citenamefont {Alford},
  \citenamefont {Harutyunyan},\ and\ \citenamefont
  {Sedrakian}}]{Alford:2019kdw}%
  \BibitemOpen
  \bibfield  {author} {\bibinfo {author} {\bibfnamefont {M.}~\bibnamefont
  {Alford}}, \bibinfo {author} {\bibfnamefont {A.}~\bibnamefont {Harutyunyan}},
  \ and\ \bibinfo {author} {\bibfnamefont {A.}~\bibnamefont {Sedrakian}},\
  }\href {\doibase 10.1103/PhysRevD.100.103021} {\bibfield  {journal} {\bibinfo
   {journal} {Phys. Rev. D}\ }\textbf {\bibinfo {volume} {100}},\ \bibinfo
  {pages} {103021} (\bibinfo {year} {2019})},\ \Eprint
  {http://arxiv.org/abs/1907.04192} {arXiv:1907.04192 [astro-ph.HE]}
  \BibitemShut {NoStop}%
\bibitem [{\citenamefont {Bemfica}\ \emph
  {et~al.}(2019{\natexlab{b}})\citenamefont {Bemfica}, \citenamefont
  {Disconzi},\ and\ \citenamefont {Noronha}}]{Bemfica:2019knx}%
  \BibitemOpen
  \bibfield  {author} {\bibinfo {author} {\bibfnamefont {F.~S.}\ \bibnamefont
  {Bemfica}}, \bibinfo {author} {\bibfnamefont {M.~M.}\ \bibnamefont
  {Disconzi}}, \ and\ \bibinfo {author} {\bibfnamefont {J.}~\bibnamefont
  {Noronha}},\ }\href {\doibase 10.1103/PhysRevD.100.104020} {\bibfield
  {journal} {\bibinfo  {journal} {Phys. Rev. D}\ }\textbf {\bibinfo {volume}
  {100}},\ \bibinfo {pages} {104020} (\bibinfo {year} {2019}{\natexlab{b}})},\
  \Eprint {http://arxiv.org/abs/1907.12695} {arXiv:1907.12695 [gr-qc]}
  \BibitemShut {NoStop}%
\end{thebibliography}%

\end{document}